\documentclass[MOOR,nonblindrev]{informs3} 

\usepackage{dsfont}
\usepackage{amssymb, amsmath}
\usepackage{enumitem}
\usepackage{booktabs}
\usepackage{siunitx}
\usepackage{xcolor}
\usepackage[colorinlistoftodos]{todonotes}
\usepackage{soul}

\newcommand{\A}{\mathsf{A}}
\newcommand{\M}{\mathcal{M}}

\newcommand{\F}{\mathfrak{F}}
\newcommand{\ssalg}{\mathfrak{A}}
\newcommand{\salg}{\mathfrak{F}}
\newcommand{\I}{\mathcal{I}}

\newcommand{\Id}{{\mathds{1}}}

\newcommand{\E}{{\mathbb{E}}}
\newcommand{\N}{{\mathbb{N}}}
\renewcommand{\P}{{\mathbb{P}}}

\newcommand{\R}{{\mathbb{R}}}

\newcommand{\x}{{\boldsymbol{x}}}

\newcommand{\X}{{\boldsymbol{X}}}
\newcommand{\W}{{\boldsymbol{W}}}

\newcommand{\Y}{{Y}}

\renewcommand{\I}{\mathcal{I}}

\newcommand{\Var}{{\mathrm{Var}}}
\newcommand{\VaR}{{\mathrm{VaR}}}
\newcommand{\ES}{{\mathrm{ES}}}
\newcommand{\RVaR}{{\mathrm{RVaR}}}

\newcommand{\g}{\mathsf{g}}

\newcommand{\sensA}{\xi_S(Y;\ssalg)}
\newcommand{\sens}{\xi_S(Y;}

\usepackage[plainpages=false,hyperfootnotes=false]{hyperref}
\hypersetup{
  colorlinks   = true, 
  urlcolor     = blue, 
  linkcolor    = red, 
  citecolor   = blue 
}

\usepackage{natbib}
 \bibpunct[, ]{(}{)}{,}{a}{}{,}%
 \def\bibfont{\small}%
\TheoremsNumberedThrough     
\ECRepeatTheorems

\EquationsNumberedBySection 

\begin{document}

\VOLUME{}%
\NO{}%
\MONTH{}
\YEAR{}
\FIRSTPAGE{}%
\LASTPAGE{}%
\SHORTYEAR{}
\ISSUE{} %
\LONGFIRSTPAGE{} %
\DOI{}%

\RUNAUTHOR{Fissler \& Pesenti}

\RUNTITLE{Sensitivity Measures Based on Scoring Functions}

\TITLE{Sensitivity Measures Based on Scoring Functions}

\ARTICLEAUTHORS{%
\AUTHOR{Tobias Fissler}
\AFF{Institute for Statistics and Mathematics, Vienna University of Economics and Business (WU), Vienna, Austria\thanks{\EMAIL{tobias.fissler@wu.ac.at}}
\AUTHOR{Silvana M. Pesenti}
\AFF{Department of Statistical Sciences, University of Toronto, Toronto, Canada\thanks{\EMAIL{silvana.pesenti@utoronto.ca}}}} 
\vspace{1.5em}
 \today
\vspace{1.5em}
} 

\ABSTRACT{%
We propose a holistic framework for constructing sensitivity measures for any elicitable functional $T$ of a response variable. The sensitivity measures, termed score-based sensitivities, are constructed via scoring functions that are (strictly) consistent for $T$. These score-based sensitivities quantify the relative improvement in predictive accuracy when available information, e.g.,~from explanatory variables, is used ideally.
We establish intuitive and desirable properties of these sensitivities and discuss advantageous choices of scoring functions leading to scale-invariant sensitivities.   

Since elicitable functionals typically possess rich classes of (strictly) consistent scoring functions, we demonstrate how Murphy diagrams can provide a picture of all score-based sensitivity measures. We discuss the family of score-based sensitivities for the mean functional (of which the Sobol indices are a special case) and risk functionals such as Value-at-Risk, and the pair Value-at-Risk and Expected Shortfall. The sensitivity measures are illustrated using numerous examples, including the Ishigami--Homma test function. In a simulation study, estimation of score-based sensitivities for a non-linear insurance portfolio is performed using neural nets.
}%


\maketitle

%


\section{Introduction}

We consider the context of quantitative risk management where $\Y$ describes a random variable of interest, e.g.,  an (insurance) portfolio. The vector $\X = (X_1, \ldots , X_n)$, $n \in \N$, describes the risk factors or risk factor changes which determine $Y$ via a mapping or aggregation function $\g \colon \R^n \to \R$, such that $\Y  = \g(\X)$, see e.g.~\cite{QRM2015}. Typical in applications is that not all risk factors $\X$ are observable, e.g., only functions of $\X$ are observable, or their observation is costly. In such situations, the available information may be expressed via a random vector $\W = (W_1, \ldots, W_d)$, $d\in \N$, and in the absence of knowing what factors determine $Y$, $\W$ may contain information irrelevant for modelling $Y$.

The key question we aim to address is: How sensitive is $Y$ with respect to $\W$? 
Or more specifically: What is the information value of $\W$ for $Y$?
This can be made more precise by assessing by how much the uncertainty of $Y$ is reduced when knowing/learning $\W$.
We will see that this latter question can be equivalently rephrased as: What is the gain in predictive accuracy for $Y$ when knowing $\W$?
To answer these questions one needs to first clarify some imminent ones. 
How is uncertainty of $Y$ measured? And what is the predictive target?
As for the latter, one distinguishes between probabilistic predictions, that is, specifying the full conditional distribution of $Y$ given $\W$, and point predictions. For point predictions the conditional distribution of $Y$ given $\W$ is summarised by a functional $T$ such as the mean or a law-determined risk measure, e.g., Value-at-Risk (VaR) or Expected Shortfall (ES). 
Then, truthful prediction amounts to specifying the correct conditional distribution $F_{Y|\W}$ or a functional thereof, $T(F_{Y|\W})$.

The above raised question is typically addressed using sensitivity analysis, and in particular, via sensitivity measures or importance measures \citep{Saltelli2008Wiley}. Sensitivity measures associate the uncertainty in $Y$ with the uncertainties in the risk factors in a way that allows for e.g., importance ranking of risk factors. 
Such an importance ranking may inform where to direct scarce resources to collect more data. Besides, it may lead to model simplifications when a factor is deemed to have irrelevant explanatory power.
The literature on sensitivity measures is vast and we refer to \cite{Borgonovo2016EJOR-review} and \cite{Razavi2021FutureSA} for an extensive review. Examples of sensitivity measures include variance-based \citep{Saltelli2002JASA}, moment-independent \citep{Borgonovo2007RESS}, and quantile-based sensitivities \citep{Browne2017WP}. Alternative approaches include those based on divergence measures \citep{Gamboa2018SIAMUQ, Pesenti2019EJOR, Fort2021SIAMUQ, Pesenti2021SSRN-Wasser} and differential sensitivity measures, see \cite{Tanakas2016RA} and \cite{Pesenti2021RA} in a risk management context. However, as argued in \cite{borgonovo2021EJOR}, the choice of a sensitivity measure should be intimately tied to the functional of interest $T$ via the notion of strictly consistent scoring functions and moreover reflect the information value of risk factors. 

A scoring function or scoring rule $S(\cdot,\cdot)$ maps the tuple $(z,y)$, consisting of a prediction $z$ and an observation $y$ of $Y$, to the non-negative real number $S(z,y)$ with the convention that smaller values of $S$ reflect more accurate predictions of $Y$. 
Here, the prediction $z$ may be a point prediction, an interval, or the entire probability distribution or density of $Y$.
As forcefully argued in \cite{MurphyDaan1985, EngelbergManskiETAL2009}, and \cite{ Gneiting2011}, a score $S$ should be strictly consistent for a functional $T$. A strictly consistent score incentivises truthful predictions in the sense that the expected score is strictly minimised by the target at hand, which in our setup is the conditional distribution $F_{Y|\W}$ or a functional thereof, $T(F_{Y|\W})$. Examples for strictly consistent scoring functions are the squared loss for the mean functional and the negative log-likelihood or the continuous ranked probability score for the conditional distribution $F_{Y|\W}$ 
\citep{GneitingRaftery2007}.
If a functional $T$ admits a strictly consistent scoring function it is called elicitable \citep{LambertETAL2008}.
Many functionals of interest are elicitable, e.g.~the mean via the squared loss $S(z,y) = (z-y)^2$ and the median via the absolute loss $S(z,y) = |z-y|$. As for risk measures, entropic risk measures, VaR, and expectiles are elicitable, while the variance, standard deviation, and ES fail to be elicitable \citep{Osband1985, Weber2006, Gneiting2011}. 
However, the pair consisting of the mean and variance (or standard deviation) and the pair VaR and ES admit strictly consistent scoring functions \citep{FisslerZiegel2016}.

If the functional of interest $T$ is elicitable, the information value of $\W$ for $Y$ can be measured by the potential reduction of uncertainty when knowing $\W$ -- expressed by a  strictly consistent scoring function $S$ -- that is by $
\E[S(T(F_Y),Y)] - \E[S(T(F_{Y|\W}),Y)]$; a notion
first proposed in the sensitivity literature by \cite{borgonovo2021EJOR}. 
We build on their suggestion by defining for a strictly consistent scoring function $S$ the score-based sensitivity measure of $Y$ to information $\W$ as
\begin{equation*}\label{eq:sens intro}
\xi_S(Y;\W) = \frac{\E[S(T(F_Y),Y)] - \E[S(T(F_{Y|\W}),Y)]}{\E[S(T(F_Y),Y)]}\,,
\end{equation*}
see Definition \ref{def:sensitivity-measure} for a precise statement of the involved assumptions.
By construction, the score-based sensitivity measure attains values in $[0,1]$ and is unitless, which allows for comparison of sensitivities to risk factors that live on different scales. Sensitivity measures constructed via strictly consistent scoring functions include the Sobol indices (and extensions thereof), corresponding to the mean functional and the squared loss \citep{borgonovo2021EJOR}, and the contrast index studied in \cite{Maume2018SPL}, corresponding to the pinball loss and the VaR functional. \cite{Fort2016ML} consider a related concept using contrast functions in the machine learning literature.
Sensitivity concepts that reflect value of information  (which typically are intimately connected to scoring functions) have been studied by \cite{Felli1998MDM} and more recently by \cite{Borgonovo2017RA} in the context of probability safety assessment, and by \cite{Straub2021RES} in applications to reliability analysis.

In this paper, we provide a comprehensive framework for constructing score-based sensitivity measures for any elicitable functional $T$. Moreover, we establish universal properties of score-based sensitivity measures that are inherited by any elicitable functional with any strictly consistent scoring function. In particular, we argue that sensitivity measures should possess the \textit{zero information gain} property, a property significantly weaker than \cite{borgonovo2021EJOR}'s nullity-implies-independence  property. 
Indeed, while the nullity-implies-independence property means that a zero sensitivity implies that $\W$ and $Y$ are independent, we only require that a sensitivity of  zero is equivalent to $\W$ being irrelevant for modelling $T(F_Y)$. Furthermore, we show that a score-based sensitivity measure is equal to 1 if and only if $\W$ contains \emph{all relevant} information to model $T(F_Y)$. We additionally define an interaction sensitivity measure -- termed interaction information -- that quantifies the information value of interactions of risk factors. In the special case of the squared loss scoring function we recover the well-known Sobol interaction terms.

One imminent challenge of score-based sensitivity measures is the choice of scoring function. An elicitable functionals $T$ typically admits infinitely many strictly consistent scoring functions. Furthermore, as we illustrate in examples, different scoring functions may lead to sensitivity measures that rank information, and thus risk factors, differently. To overcome these difficulties, we advocate for scoring functions that lead to  scale-invariant sensitivities. We moreover promote and illustrate the usage of Murphy diagrams, which has been impressively demonstrated in the context of scoring functions in \cite{EhmETAL2016}.
Another challenge is the estimation of the score-based sensitivity, in particular the term $\E[S(T(F_{Y|\W}),Y)]$. One way to address this is with neural nets from machine learning, which we do in Section \ref{sec:applications}.

This paper is organised as follows. Section \ref{sec:senitivity} motivates and defines score-based sensitivity measures. In Section \ref{sec:properties} we discuss their universal properties and introduce a sensitivity that quantifies the value of information of interactions of risk factors; termed interaction information. Section \ref{sec:choice} discusses the choice of strictly consistent scoring functions and defines score-based Murphy diagrams for sensitivity measures. We illustrate the score-based sensitivities on the Ishigami--Homma test function and a non-linear insurance portfolio in Section \ref{sec:applications}.

\section{From Scoring Functions to Sensitivity Measures}
\label{sec:senitivity}

Let $(\Omega, \F, \P)$ be a complete probability space on which we identify random elements which almost surely coincide.
Moreover, if not stated explicitly, all events such as equalities, inequalities, etc., are to be understood in an almost sure sense.
We use the decision-theoretic setup and notation of \cite{Gneiting2011} and \cite{FFHR2021}.
For this let $\M^0(\R)$ be the class of all Borel probability measures on $\R$ and let $\M \subseteq \M' \subseteq \M^0(\R)$ be two sub-classes. 
We equip $\M^0(\R)$ with the $\sigma$-algebra generated by the family of evaluation maps $\{\pi_B\}_{B\in\salg}$ given by $\pi_B\colon \M^0(\R) \to [0,1]$, which map $ \mu\mapsto \pi_B(\mu) = \mu(B)$.
We moreover identify any probability measure $\mu\in \M^0(\R)$ with its cumulative distribution function $F(x) = \mu((-\infty,x])$, $x\in\R$. 
Further, let $\A$ be an action domain -- in our context, this is typically the interval $(0,\infty)$, $\R$, or $\R^k$ --
equipped with with the Borel $\sigma$-algebra. 
The predictive goal is then described by the measurable functional  $T \colon \M'\to \A$ and we refer to \cite{FisslerHolzmann2022} for measurability results for functionals of interest such as the mean, expectiles, Value-at-Risk (VaR), and Expected Shortfall (ES).

For a random variable $Y$ in some class $\mathcal Y \subseteq L^0(\Omega, \F, \P)$ we denote its cumulative distribution function by $F_Y(\cdot) = \P(Y \le \cdot)$. For a sub-$\sigma$-algebra $\ssalg\subseteq \F$ -- often referred to as \emph{information set} -- we denote by $F_{Y|\ssalg}$ a regular version of the conditional distribution of $Y$ given $\ssalg$. (Recall that $F_{Y|\ssalg}$ is a measurable map from $\Omega$ to $\M^0(\R)$ \citep{FisslerHolzmann2022}.)
If $F_Y\in\M'$ and $F_{Y|\ssalg}\in\M'$ (almost surely, which will be suppressed in the sequel), 
we can consider the (measurable) random variables
$T(Y)=T(F_Y)$ and $T(Y|\ssalg)= T(F_{Y|\ssalg})$.
If $\ssalg$ is generated by some random vector $\W$, i.e. $\ssalg = \sigma(\W)$, then we simply write $F_{Y|\W} = F_{Y|\ssalg}$ and use the shorthand $T(Y|\W)=T(Y|\sigma(\W))$.

\begin{definition}[Consistency \& Elicitability]
A scoring function (or score) is a measurable map $S\colon\A\times\R\to[0,\infty]$. For a functional $T\colon\M'\to \A$ and a sub-class $\M\subseteq \M'$ the scoring function $S$ may satisfy the following properties:
\begin{enumerate}[label = $(\roman*)$]
    \item 
    $S$ is \emph{$\M$-consistent} for $T$, if
    for all $F\in\M$ and for all $z\in\A$
    \begin{equation}
    \label{eq:consistency}
    \int S(T(F),y)\,\mathrm{d}F(y) \le \int S(z,y)\,\mathrm{d}F(y)\,.
    \end{equation}

    \item
    $S$ is \emph{strictly} $\M$-consistent for $T$, if it is $\M$-consistent for $T$ and if the inequality in \eqref{eq:consistency} is strict for all $z\neq T(F)$.
    
    \item 
    $T$ is \textit{elicitable} on $\M$, if there exists a strictly $\M$-consistent scoring function for $T$.

\end{enumerate}
\end{definition}
Elicitability of a functional is equivalent to the fact that it is a Bayes act or the minimiser of an expected score \citep{Gneiting2011}.

We will work with the following set of assumptions throughout the rest of the paper.

\begin{assumption}\label{asm:dirac-score}
Let $S$ be an $\M$-consistent scoring function for $T\colon \M'\to \A$ and $\M\subseteq\M'$. Suppose that $F_Y\in\M$ for all $Y\in\mathcal Y$ and that $\delta_y\in\M'$ for all $y\in\R$.\footnote{For $y\in\R$, the point measure $\delta_y$ is a probability measure such that for any $B\subseteq \R$, $\delta_y(B) = 1$ if $y\in B$ and $\delta_y(B)=0$ otherwise. Note that $F_{Y|\salg} = F_{Y|Y} = \delta_Y$ is a random probability measure on $\R$.} 
Let the following hold:
\begin{enumerate}[label = $(\roman*)$]
    \item 
    $S(T(\delta_y),y)=0$ for all $y\in\R$.
    \item
    $S(T(\delta_y),y)<S(z,y)$ for all $y\in\R$, $z\neq T(\delta_y)$.
    \item
    For all $F\in\M$, $\int S(T(F),y)\,\mathrm{d}F(y) <\infty$.
    \item
    For all $Y\in \mathcal Y$, $\int S(T(F_Y),y)\,\mathrm{d}F_Y(y) >0$.
\end{enumerate}
\end{assumption}
Assumption \ref{asm:dirac-score} part (ii) amounts to strict consistency for $T$ on the class of point measures $\{\delta_y\colon y\in\R\}$. Part (i) is a normalisation condition and can be achieved under (ii) by considering the normalised score $\tilde S(z,y) = S(z,y) - S(T(\delta_y),y)$.
The finiteness in (iii) is implied if $S$ is strictly $\M$-consistent. Indeed, if $\int S(T(F),y)\,\mathrm{d}F(y)=\infty$, it cannot be \emph{strictly} smaller than $\int S(z,y)\,\mathrm{d}F(y)$ for $z\neq T(F)$.
Condition (iv) usually holds if $S$ is strictly $\M$-consistent and $Y$ is not constant almost surely.

It is well-known that consistent scoring functions respect increasing information sets \cite[Theorem 1]{HolzmannEulert2014}. This 
means that if $S$ is $\M$-consistent for $T$ and if $\ssalg\subseteq \ssalg'\subseteq \F$ are two sub-$\sigma$-algebras such that $F_{Y|\ssalg}, F_{Y|\ssalg'}\in \M$, then
    \begin{equation}\label{eq:HE2014}
        \E[S(T(Y|\ssalg'), Y)]\le \E[S(T(Y|\ssalg), Y)].
    \end{equation}
If $S$ is strictly $\M$-consistent for $T$, then equality in \eqref{eq:HE2014} holds only if  $T(Y|\ssalg') = T(Y|\ssalg)$. (The main argument behind this result is an application of the definition of (strict) consistency in combination with the tower property for the conditional expectation.)
This provides a motivation for considering, for $\ssalg \subseteq \F$, the term
    \begin{equation}
    \label{eq:sensitivity}
        \E[S(T(Y),Y)] - \E[S(T(Y|\ssalg),Y)]\ge0,
    \end{equation}
which corresponds to the \emph{resolution term} of the score decomposition in \cite{Pohle2020}, the \emph{discrimination} in \cite{GneitingResin2021}, 
and it is a special instance for a \emph{score divergence} \citep{ThorarinsdottirGneitingGissibl2013, GneitingRaftery2007, Dawid2007}, 
which is related to the \emph{cost of uncertainty} discussed in \cite{FrankelKamenica2019}.
The resolution term \eqref{eq:sensitivity} quantifies how helpful the information $\ssalg$ is to improve the correct baseline model $T(Y)$, when used ideally. 
This improvement is naturally measured in terms of a consistent score \citep{Gneiting2011}. The next definition quantifies this improvement relative to the so-called \textit{oracle improvement} of full information
\begin{equation*}
    \label{eq:inequality}
\E[S(T(Y),Y)] - \E[S(T(Y|\salg),Y)] = \E[S(T(Y),Y)]  >0,
\end{equation*}
where equality follows from Assumption \ref{asm:dirac-score} (i) and the fact that $T(Y|\salg) = T(\delta_Y)$ and strict positivity from Assumption \ref{asm:dirac-score} (iv).

The resolution term normalised by the oracle improvement motivates to consider the following notion of sensitivity measure. The sensitivity measure is inspired by \cite{borgonovo2021EJOR} who introduced sensitivity measures based on scoring functions to the sensitivity literature.

\begin{definition}[Sensitivities based on scoring functions]
\label{def:sensitivity-measure}
Let $S$ be an $\M$-consistent scoring function for $T\colon \M'\to\A$ with $\M \subseteq \M'$ satisfying Assumption \ref{asm:dirac-score}, 
$Y$ a random variable, and 
$\ssalg\subseteq \salg$ an information set with $F_Y, F_{Y|\ssalg}\in\M$.
Then the \emph{sensitivity} of $\Y$ to $\ssalg$ based on $S$ is given by
\begin{equation}
\label{eq:sensitivity-measure}
\sensA = 
\frac{\E[S(T(Y),Y)] - \E[S(T(Y|\ssalg),Y)]}{\E[S(T(Y),Y)]}\,.
\end{equation}
\end{definition}

The larger the value of $\sensA$, the larger is the information value of $\ssalg$, measured by the scoring function $S$.
While \cite{borgonovo2021EJOR} consider sensitivity measures based solely on the numerator of \eqref{eq:sensitivity-measure}, our proposed sensitivity measure is normalised to lie between 0 and 1; see Theorem \ref{thm:properties} for details.
This normalisation is achieved by the division with the oracle improvement
$\E[S(T(Y),Y)]$; a quantity  called \emph{uncertainty} in the forecasting literature \citep{GneitingResin2021}. $\E[S(T(Y),Y)]$ is a generalisation of the variance which arises when $S$ is the squared loss.
Our sensitivity measures closely resemble the \emph{universal coefficient of determination} which was recently introduced by \cite{GneitingResin2021} and which generalises the classical coefficient of determination R$^2$.
The subtle difference is that in this work, the term $T(Y|\ssalg)$ in \eqref{eq:sensitivity-measure} is conditionally calibrated with respect to the information $\ssalg$, whereas \cite{GneitingResin2021} require the notion of a ``conditionally $T$-calibrated forecast'', which is generally a weaker assumption.
Moreover, \eqref{eq:sensitivity-measure} is a special instance of a \emph{skill score} due to \cite{Murhpy1973}; see also \cite{GneitingRaftery2007} for a discussion.

Exploiting the consistency of $S$ for $T$ in Definition \ref{def:sensitivity-measure}, we could replace $\E[S(T(Y),Y)]$ by $\min_{z\in\R} \E[S(z,Y)]$ and $\E[S(T(Y|\ssalg),Y)]$ by $\E\big[\min_{Z\in\ssalg} \E[S(Z,Y)|\ssalg]\big]$, where the latter minimum is taken over all $\ssalg$-measurable random variables $Z$. Thus, the score-based sensitivity only depends on the functional $T$ through its consistent score.

In the sequel, we shall use the shorthand $\sens \W)=\sens \sigma(\W))$ to indicate the sensitivity of $Y$ with respect to the information generated by a random vector $\W$.

\section{Properties of Score-based Sensitivities}
\label{sec:properties}

Theorem \ref{thm:properties} provides the core properties of our score-based sensitivities.

\begin{theorem}\label{thm:properties}
Let $S$ be an $\M$-consistent scoring function for $T\colon \M'\to\A$ with $\M\subseteq \M'$ satisfying Assumption \ref{asm:dirac-score}, 
$Y$ a random variable, and 
$\ssalg$ an information set with $F_Y, F_{Y|\ssalg}\in\M$. Then,
 $\sensA$ satisfies the following properties.
\begin{enumerate}[label = (\alph*)]
    \item 
    \label{thm:normalisation}
    \emph{Normalisation:} $\sensA \in [0,1]$.
    \item 
    \label{thm:zero-info}
    \emph{Zero information gain:} $T(Y|\ssalg) = T(Y)$ implies $\sensA=0$. If $S$ is strictly $\M$-consistent then $\sensA=0$  only if $T(Y|\ssalg) = T(Y)$.
    \item    
    \label{thm:full-info}
    \emph{Full information gain:}
    $T(Y|\ssalg) = T(\delta_Y)$ implies that $\sensA=1$.  If $S$ is strictly $\M$-consistent then $\sensA=1$ only if $T(Y|\ssalg) =  T(\delta_Y)$.
    \item 
    \label{thm:monotone}
    \emph{Monotonicity with respect to nested information:} For any information set $\ssalg'$ with $\ssalg \subseteq \ssalg'$ 
    and $F_{Y|\ssalg'}\in\M$
    it holds that $\sensA\le \sens \ssalg')$. If $S$ is strictly $\M$-consistent then $\sensA = \sens \ssalg')$ only if $T(Y|\ssalg) = T(Y|\ssalg')$.
\end{enumerate}
\end{theorem}

\textit{Proof of Theorem \ref{thm:properties}.}
The normalisation follows from the non-negativity of the resolution term \eqref{eq:sensitivity}. The remaining assertions are a corollary of Theorem 1 in \cite{HolzmannEulert2014}.
\hfill\Halmos

The normalisation of the proposed sensitivity has the obvious advantage of rendering the sensitivity unitless. This facilitates comparison of sensitivities to risk factors that live on different scales in a straightforward manner. 
In the subsequent subsections we discuss the properties of the proposed sensitivity measures, including the ones established in Theorem \ref{thm:properties}, and provide illustrating examples.
Interestingly, these properties (almost) correspond to four of the five axioms stipulated in \cite{GriessenbergerETAL2022} for a dependence measure between $Y$ and $\ssalg$, which yields an alternative angle on our sensitivity measures.

\subsection{Zero Information Gain}
\label{subsec:zero}

We first start with an obvious, yet important, corollary of the zero information gain property stated in Theorem \ref{thm:properties} \ref{thm:zero-info}.

\begin{corollary}\label{cor:independence}
Under the assumptions of Theorem \ref{thm:properties}, if $Y$ and $\ssalg$ are independent, then $\sensA=0$.
\end{corollary}

Corollary \ref{cor:independence} follows from Theorem \ref{thm:properties} since independence of $Y$ and $\ssalg$ is equivalent to $F_{Y|\ssalg} = F_Y$, which implies that $ T(Y|\ssalg) = T(Y)$. 

Clearly, if $Y$ and $\ssalg$ are stochastically independent, $\ssalg$ does not contain any information about $Y$. 
The perspective of the zero information gain property, however, is more nuanced than independence. Assume that $\ssalg$ contains information about $Y$, i.e., they are not independent, but the information is not relevant for modelling $T$, i.e., $T(Y|\ssalg) = T(Y)$, then the score-based sensitivity measure $\sensA$ is equal to 0. Thus, the sensitivity measure being equal to 0 implies that $\ssalg$ contains no relevant information for modelling $T(Y)$.
Examples \ref{example:zero information gain 1} and \ref{example:zero information gain 2} below illustrate such situations.
For the mean functional $T$, the identity $T(Y|\ssalg) = T(Y)$ is also known as \emph{mean independence} \cite[p. 25]{Wooldridge} which is weaker than independence but stronger than uncorrelatedness.  
Following this terminology $T(Y|\ssalg) = T(Y)$ can be called \textit{$T$-independence} of $Y$ from $\ssalg$.

\cite{borgonovo2021EJOR} advocate that sensitivity measures should satisfy the so-called  \emph{nullity-implies-independence} (n.i.i.) property, see Definition 4 in \cite{borgonovo2021EJOR}; a property inspired by R\'eny's postulate D for measures of dependence \citep{Renyi1959}. A (sensitivity) measure satisfies the n.i.i. property -- R\'eny's postulate D -- if the measure is zero if and only if $\ssalg$ and $Y$ are independent. While R\'enyi considers measures of dependence between any two random variables, a sensitivity measure in our manuscript (and the extant literature) is a measure between the unconditional distribution $F_Y$ and the conditional distribution $F_{Y|\ssalg}$  of the output $Y$. Therefore, in our context, the n.i.i.\ property means that the sensitivity measure is 0 if and only if $F_Y$ and $F_{Y|\ssalg}$ coincide (almost surely).
Since the conditional distribution $F_{Y|\ssalg}$ enters the sensitivity measure \eqref{eq:sensitivity-measure} only via the functional $T$, one cannot hope for the n.i.i.\ property unless $T$ is the identity functional. 
If a modeller is interested in the information value of $\ssalg$ for $T(Y)$ only, the  n.i.i.\ property needs to be replaced by an achievable desideratum which is the zero information gain property. 
In this spirit, we may call the zero information gain property the ``nullity-implies-$T$-independence'', which we illustrate in Examples \ref{example:zero information gain 1} and \ref{example:zero information gain 2}. We emphasise that for the special case of probabilistic predictions, i.e.~when $T$ is the identity map, the zero information gain property coincides with the n.i.i.\ property; see Proposition 6 in  \cite{borgonovo2021EJOR}.

\begin{example}
\label{example:zero information gain 1}
Consider the output $Y = X_1 X_2 + X_3$, where $X_1, X_2, X_3$ are independent and non-deterministic, $X_1>0$, and $\E[X_2]=0$. Clearly, $Y$ and $X_1$ are dependent. 
For $T$ the mean functional, we obtain that $T(Y|X_1) = \E[Y|X_1] = \E[X_3] = \E[Y]$. Thus, knowing $X_1$ does not help to make a mean-model more precise. 
Therefore, $\sens X_1)=0$ for any consistent scoring function $S$ for the mean functional.
\end{example}

For the next example, we recall the definition of Value-at-Risk (VaR) of a random variable $Y$ at level $\alpha\in(0,1)$ which is given by
\[
\VaR_\alpha(Y)= \inf\{t\in\R\mid F_Y(t)\ge\alpha\}\,,
\]
corresponding to the left-continuous $\alpha$-quantile of the distribution $F_Y$ of $Y$. Since VaR is a law-determined functional, we may write $\VaR_\alpha(F_Y)= \VaR_\alpha(Y)$.

\begin{example}
\label{example:zero information gain 2}
Let $X_1,X_2,X_3$ be independent, $X_1$ has a Bernoulli distribution with $p = \P(X_1=0) = 1- \P(X_1=1) \in(0, \alpha)$, and $X_2<C<X_3$ almost surely for a constant $C>0$. The output $Y = \Id_{\{X_1=0\}}X_2 + \Id_{\{X_1=1\}}X_3$ may be viewed as an insurance portfolio consisting of small claims -- smaller than $C$ -- which occur with probability $p$, and large claims -- larger than $C$ -- occurring with probability $1-p$. 
Then, 
\[
\VaR_{\alpha}(Y|X_3) = \VaR_{\alpha}(Y|X_2,X_3) = X_3.
\]
Moreover, 
\begin{equation*}
\VaR_{\alpha}(Y|X_2) = \VaR_{\alpha}(Y) =
\VaR_{(\alpha-p)/(1-p)}(X_3).
\end{equation*}
Thus, knowing $X_2$ does not improve the modelling of the $\VaR$ functional even though $Y$ and $X_2$ are dependent.
The reason is that $X_2$ only affects the distribution of $Y$ in the left tail below the $p$-quantile.
As a consequence, for any consistent score $S$ for $\VaR_\alpha$ it holds
$\sens X_2)=0$.
\end{example}

\subsection{Full Information Gain}

The counterpart of zero information gain is full information gain. We start with a straightforward and important observation.

\begin{corollary}\label{cor:full info}
Under the assumptions of Theorem \ref{thm:properties}, if $Y$ is $\ssalg$-measurable, then $\sensA=1$.
\end{corollary}

Again, the proof of Corollary \ref{cor:full info} is simple. If $Y$ is $\ssalg$-measurable, the conditional distribution of $Y$ given $\ssalg$ reduces to a random point measure in $Y$, that is $F_{Y|\ssalg} = \delta_Y$. This in turn implies that $T(F_{Y|\ssalg}) = T(\delta_Y)$ and therefore $S(T(F_{Y|\ssalg}) ,Y) =0$ by Assumption \ref{asm:dirac-score}(i).

Corollary \ref{cor:full info} for example applies if $\ssalg$ is generated by a real-valued random variable $W$ such that the pair $(W, Y)$ is co- or countermonotonic. Another example of $Y$ being $\ssalg$-measurable is if $\ssalg$ is generated by all risk factors, i.e., $\ssalg = \sigma(\X)$ and $Y = \g(\X)$. 
Similar comments about the relevance of information $\ssalg$ with respect to only $T(Y)$ can be made corresponding to the ones in Subsection \ref{subsec:zero}.
In particular, the below example shows that we can obtain a sensitivity of 1, i.e.~we have all relevant information for $T$, even if $Y$ itself cannot be fully explained by $\ssalg$.  

\begin{example}[Example \ref{example:zero information gain 2} continued] \label{ex:full-information-gain}
We consider $X_1,X_2,X_3$ independent, $X_1$ Bernoulli distributed with $p = \P(X_1=0) = 1- \P(X_1=1) \in(0, \alpha)$, and $X_2<C<X_3$ almost surely, for some constant $C>0$, where $X_3$ has finite mean. Then, the output $Y = \Id_{\{X_1=0\}}X_2 + \Id_{\{X_1=1\}}X_3$ is not measurable with respect to $\sigma(X_1,  X_3)$. However, for the functional $T(Y) =\E[ Y\Id_{\{Y >C\}}]$, we have that
$T(Y|Y) = Y\Id_{\{Y >C\}}= X_3\Id_{\{X_1=1\}}= T(Y|X_1,X_3)$.
By part c) of Theorem \ref{thm:properties}, for any consistent scoring function $S$ for $T$ we get that $\xi_S(Y;Y) = \xi_S(Y;X_1,X_3)=1$.
Example \ref{ex:full-information-gain} shows that knowing $(X_1, X_3)$ fully explains $T(Y) = \E[Y\Id_{\{Y > C\}}]$, which implies that $\sens X_1, X_3) = 1$, however, $Y$ is not measurable with respect to  $\sigma(X_1, X_3)$. 
\end{example}

\subsection{Monotonicity with respect to Nested Information}

It is difficult to compare arbitrary sets of information. Put mathematically, there is no canonical total order on the class of all sub-$\sigma$-algebras of $\salg$. However, the subset relation is a sensible partial order on this class and Theorem \ref{thm:properties} \ref{thm:monotone} asserts that our sensitivities are monotone with respect to this partial order.
Some direct consequences are immediate, for example in regard to transformations of explanatory variables.

\begin{corollary}\label{cor:transformations}
Under the assumptions of Theorem \ref{thm:properties} \ref{thm:monotone}, suppose that $\ssalg'$ is generated by a $d$-dimensional random vector $\W$. Then for any measurable function $h\colon \R^d \to\R ^m$, $m\le d$, it holds that $\xi_S \big(Y;h(\W)\big)\le \sens \W)$.
If $h$ is injective, then $\xi_S \big(Y;h(\W)\big)= \sens \W)$.
\end{corollary}

The proof follows by invoking that $\sigma\big(h(\W)\big) \subseteq \sigma(\W)$. The inclusion becomes an equality if $h$ is an injection.

Corollary \ref{cor:transformations} implies that injective transformations of risk factors do not affect the sensitivities. Examples include affine transformations or (component-wise) logarithmic scaling of explanatory variables. If the transformation is not injective, then the transformation may induce a loss of information, e.g., a projection to components of $\W$ (reduction of dimensionality) implies that $\sens \W_\I)\le \sens \W)$, for any subvector $\W_\I$ of $\W$.

Again, we stress that monotonicity with respect to nested information sets only takes relevant information for modelling $T(Y)$ into account. We illustrate this with the following example.

\begin{example}
\label{example:non-additive0}
Let $Y = X_1 + X_2$, where $X_1,X_2$ are independent standard normal.
The target functional $T$ is the mean and and $S(z,y) = (z-y)^2$ the squared loss.
Then, we obtain that
\[
\sens X_1) = \sens X_2) = \frac{1}{2}\,.
\]
Next, we consider the transformation $h(x) = |x|$, then $T\left(Y | h(X_1)\right) = \E\big[Y\big||X_1|\big] =T\left(Y | h(X_2)\right) =  \E\big[Y\big| |X_2|\big] = \E[Y]=0$. Using part \ref{thm:zero-info} of Theorem \ref{thm:properties}, the sensitivity to $|X_i|$, $i = 1,2$, are
\[
\sens |X_1|) = \sens |X_2|) = 0.
\]
This also constitutes another example for Theorem \ref{thm:properties} \ref{thm:zero-info}, where the information $|X_1|$ is not independent of $Y$ but irrelevant for prediction purposes.
\end{example}

\subsection{Interaction Information}
\label{sec:Interaction}
Important in sensitivity analysis, particularly for non-additive models, is the calculation of \textit{interaction indices}, which quantify the effects that multiple risk factors have jointly on the functional of interest minus their individual effects, see e.g. Chapter 4 in \cite{Saltelli2008Wiley}. 
In an information value setting, the relevant question that quantifies \emph{interaction information} between two information sets $\ssalg_1$ and $\ssalg_2$ is: ``How much do we learn from knowing both $\ssalg_1, \ssalg_2$ jointly, once we already know $\ssalg_1$ and $\ssalg_2$ individually''. We formalise this in the next definition.

\begin{definition}[Interaction Information]
The sensitivity of $Y$ with respect to the interaction information 
of $\ssalg_1, \ssalg_2$ based on $S$ is given by
\begin{equation*}
\label{eq:interaction}
\sens \ssalg_1 \wedge \ssalg_2) = \max\big\{\sens \ssalg_1 \vee \ssalg_2)  - \sens \ssalg_1) - \sens \ssalg_2) \,,\; 0\big\},
\end{equation*}
where $\ssalg_1 \vee \ssalg_2$ is the smallest $\sigma$-algebra which contains all elements of $\ssalg_1$ and $\ssalg_2$, i.e., $\ssalg_1 \vee \ssalg_2=\sigma(\ssalg_1 \cup \ssalg_2)$.
\end{definition}

\begin{remark}
The sensitivity with respect to the interaction information $\sens \ssalg_1 \wedge \ssalg_2)$ does generally not coincide with the sensitivity with respect to the intersection of the two information sets, $\sens \ssalg_1 \cap \ssalg_2)$.\footnote{For example, consider the probability space $(\R^2, \mathfrak{B}(\R^2), \mathcal N(0,1)^{\otimes 2})$, where $\mathfrak{B}(\R^2)$ is the Borel $\sigma$-algebra on $\R^2$ and $\mathcal N(0,1)^{\otimes 2}$ is the measure of a standard normal distribution on $\R^2$. For $\boldsymbol{t}=(t_1,t_2)\in\R^2$ define the random variables $X_1(\boldsymbol{t})=t_1$ and $X_2(\boldsymbol{t}) = t_2$. Then $X_1,X_2$ are two independent standard normal random variables and $\sigma(X_1)\cap\sigma(X_2) $ is the trivial $\sigma$-algebra $\{\emptyset, \R^2\}$. If $X_3$ is a linear combination of $X_1$ and $X_2$ such that $X_1$ and $X_3$ are not perfectly correlated, then $\sigma(X_1)\cap\sigma(X_3) $ is still trivial.}
\end{remark}

\begin{example}[Variance-based Interactions]
For variance-based sensitivity measures, that is, when $T$ is the mean and $S$ the squared loss, 
we recover the Sobol interaction term if the information sets $\ssalg_1$ and $\ssalg_2$ are independent.
Indeed, the sensitivity of $Y$ with respect to the interaction information of $\ssalg_1$ and $\ssalg_2$ is 
\begin{equation}\label{ex:interaction-vairance}
    \sens \ssalg_1 \wedge \ssalg_2) 
    = \frac{\Var \left(\E\left[Y|\ssalg_1 \vee \ssalg_2\right] \right)}{\Var(Y)}
    - \frac{\Var \left(\E\left[Y|\ssalg_1\right] \right)}{\Var(Y)}
    - \frac{\Var \left(\E\left[Y|\ssalg_2\right] \right)}{\Var(Y)}\;\ge\; 0\,.
\end{equation}
The first summand in \eqref{ex:interaction-vairance} is the normalised \textit{joint effect} of $\ssalg_1$ and $\ssalg_2$, while the second and third term are the normalised first-order effects to $\ssalg_1$ and $\ssalg_2$, respectively; see e.g., \cite{Saltelli2002JASA}.
\end{example}

While for the special case of Sobol indices the interaction terms are always non-negative, this need not be the case for general scoring functions and prediction functionals, as is illustrated in the next proposition and example.

\begin{proposition}[Information is Not Additive]
\label{prop:additivity}
Sensitivities are generally neither subadditive nor superadditive, in the sense that the term $\sens \ssalg_1 \vee \ssalg_2)  - \sens \ssalg_1) - \sens \ssalg_2)$ can generally take any sign.
\end{proposition}

Proposition \ref{prop:additivity} follows from the following example.

\begin{example}
\label{example:non-additive}
Let $Y = X_1 + X_2$, where $X_1,X_2$ are jointly normal with means 0, variances 1 and correlation $\rho \in (-1,1]$.
 The target functional $T$ is the mean and we consider the scoring function $S(z,y) = (z - y)^2$.
Moreover, let $\ssalg_i = \sigma(X_i)$, $i=1,2$.
In this context $\sens \ssalg_1 \vee \ssalg_2) = \sens X_1,X_2) = \sens Y) = 1$.
On the other hand,
\[
\sens X_1) = \sens X_2) = \frac{1+\rho}{2}.
\]
\end{example}
Interestingly, we obtain additivity in Example \ref{example:non-additive} for the case when $\ssalg_1$ and $\ssalg_2$ are independent, i.e., $\sens \ssalg_1 \vee \ssalg_2) = \sens \ssalg_1) + \sens \ssalg_2)$ if $\rho = 0$.
The following modification of Example \ref{example:zero information gain 1} shows, however, that independence of $\ssalg_1$ and $\ssalg_2$ does not imply additivity.
\begin{example}
\label{example:non-additive2}
Let $X_1$, $X_2$ be independent random variables with mean 0 and $Y=X_1X_2$. Since $\E[Y|X_1] = \E[Y|X_2] = \E[Y]=0$, part \ref{thm:zero-info} of Theorem \ref{thm:properties} implies that $\sens X_1) = \sens X_2) = 0$ for any consistent score for the mean functional. But clearly $\sens X_1,X_2)=1$.
The setup of Example \ref{example:non-additive2} can be rephrased in that $Y$ is pairwise mean-independent from $X_1$ and $X_2$, but not mean-independent from $(X_1,X_2)$.
\end{example}

\section{Choice of Score-based Sensitivities}
\label{sec:choice}
So far we did not discuss the choice of (strictly) consistent scoring function for a functional $T$ to compute score-based sensitivities. This is, however, an important issue since almost all elicitable functionals possess rich classes of (strictly) consistent scoring functions each of which could lead to a different score-based sensitivity measure. In this section, we discuss potential choices of scoring function and their implication on the corresponding score-based sensitivities. We first discuss scale-invariant score-based sensitivity measures which leads to the subclass of homogeneous scores (Section \ref{subsec:scale-invariant}). We further introduce score-based Murphy diagrams which allow for graphical illustrations of score-based sensitivities (Sections \ref{subsec:Murphy elementary} and \ref{subsec:Murphy homogeneous}).

Throughout this section, the concepts are illustrated on the mean functional and the $\alpha$-quantile (or $\VaR_\alpha$). 
For this, we first recall their family of consistent scoring functions in the next proposition, referring to \cite{Gneiting2011}. 
For an overview of characterisation results of other important functionals such as entropic risk measures, expectiles, the mode, the pairs (mean, variance), ($\VaR_\alpha$, $\ES_\alpha$), and Range Value-at-Risk (RVaR) together with its VaR-components, we refer to Appendix \ref{app:scores}.

\begin{proposition}[\cite{Gneiting2011}]
\label{prop:score_mean_quantile}
\begin{enumerate}[label = $\roman*)$]
\item
Let $\M$ be the class of distributions with finite mean. If $\phi$ is (strictly) convex with subgradient $\phi'$, then 
\begin{equation}
    \label{eq:Bregman}
    S_{\phi}(z,y) = \phi(y)- \phi(z) + \phi'(z)(z-y)
    \,,\qquad z,y\in\R\,,
\end{equation}
is (strictly) $\M$-consistent for the mean, if 
$\int |\phi(y)|\,\mathrm{d}F(y) <\infty$ for all $F\in\M$. 
Moreover, on the class of compactly supported measures, any (strictly) consistent scoring function for the mean which is continuously differentiable in its first argument and which satisfies $S(y,y)=0$ is necessarily of the form \eqref{eq:Bregman}.

\item
If $g$ is increasing, the score 
\begin{equation}
    \label{eq:GPL}
    S_g(z,y) = \big(\Id_{\{y\le z\}} - \alpha \big)\big(g(z) - g(y)\big)
    \,,\qquad z,y\in\R\,,
\end{equation}
is $\M$-consistent for $\VaR_\alpha$ if $\int |g(y)|\,\mathrm{d}F(y) <\infty$ for all $F\in\M$. 
If $g$ is strictly increasing and if for all $F\in\M$, $F(\VaR_\alpha(F)+\epsilon)>\alpha$ for all $\epsilon>0$, then \eqref{eq:GPL} is strictly $\M$-consistent.
Moreover, on the class of compactly supported measures, any consistent scoring function for $\VaR_\alpha$ which is continuous in its first argument, which admits a continuous derivative for all $z\neq y$ and which satisfies $S(y,y)=0$ is necessarily of the form \eqref{eq:GPL}.

\end{enumerate}
\end{proposition}
Scoring functions of the form \eqref{eq:Bregman} are called \emph{Bregman scores} and the scores in \eqref{eq:GPL} are termed \emph{generalised piecewise linear scores}. Almost all different choices of $\phi$ in \eqref{eq:Bregman} and $g$ in \eqref{eq:GPL} lead to different score-based sensitivity measures for the mean and VaR, respectively. Indeed, the next proposition states the family of score-based sensitivity measures for the mean and VaR arising from their scoring functions given in Proposition 
\ref{prop:score_mean_quantile}.
To that end, recall the definition of the Expected Shortfall (ES) of a random variable $Y$ at level $\alpha\in(0,1)$ \citep{Acerbi2002}
\begin{align}\label{eq:def ES}
    \ES_\alpha(Y)
    &= \frac{1}{1-\alpha} \int_{\alpha}^1 \VaR_{\beta}(Y)\,\mathrm{d}\beta \\[0.5em] \nonumber
&= \frac{1}{1-\alpha} \E[Y \Id_{\{Y>\VaR_\alpha(Y)\}}] + \frac{1}{1-\alpha}\VaR_\alpha(Y)\big(1-\alpha - \P(Y> \VaR_\alpha(Y))\big).
\end{align}
Since ES is law-determined, we may write $\ES_\alpha(F_Y) $ instead of $\ES_\alpha(Y)$.

\begin{proposition}
\label{prop:sens mean quantile}
\begin{enumerate}[label = $\roman*)$]
\item
For any strictly convex function $\phi\colon\R\to\R$ and any random variable $Y$ such that $\phi(Y)$ is integrable and such that $\E\left[\phi(Y)\right] \neq \phi\left(\E[Y]\right)$, the score-based sensitivity for $Y$ induced by the Bregman score $S_\phi$ \eqref{eq:Bregman} is
\begin{equation}\label{eq:sens-mean}
    \xi_{S_\phi}(Y;\ssalg)
    = 
    \frac{\E\left[\phi\left(\E[Y|\ssalg]\right)\right] - \phi\left(\E[Y]\right)}{\E\left[\phi(Y)\right] - \phi\left(\E[Y]\right)}\,.
\end{equation}
\item
For any strictly increasing function $g\colon\R\to\R$ and any random variable $Y$ such that $g(Y)$ is integrable and such that $\ES_\alpha(g(Y)) \neq \E[g(Y)]$, the score-based sensitivity for the $\alpha$-quantile, $\alpha\in(0,1)$, induced by the generalised piecewise linear score $S_g$ \eqref{eq:GPL} is 
\begin{equation}\label{eq:sens-quantile}
    \xi_{S_g}(Y;\ssalg)
    = 
    \frac{\ES_\alpha(g(Y)) - \E\big[\ES_\alpha(g(Y)|\ssalg)]}{\ES_\alpha(g(Y)) - \E[g(Y)]}\,.
\end{equation}
\end{enumerate}
\end{proposition}
\textit{Proof of Proposition  \ref{prop:sens mean quantile}.}
First, we prove \eqref{eq:sens-mean}. For this, let $\ssalg \subseteq \F$, then 
\begin{align*}
    \E\big[ S_\phi(\E[Y|\ssalg], Y )\big]
    &= 
    \E\big[\phi(Y) - 
    \phi\left(\E[Y|\ssalg]\right)
     + \phi'\left(\E[Y|\ssalg]\right)(\E[Y|\ssalg] - Y) \big]
     \\
     &= 
    \E\big[\phi(Y)\big] - 
    \E\big[\phi\left(\E[Y|\ssalg]\right)\big]
     + \E\big[\E\big[\phi'\left(\E[Y|\ssalg]\right)(\E[Y|\ssalg] - Y) |\ssalg\big]\big]
     \\
    &= 
    \E\big[\phi(Y)\big] - 
    \E\big[\phi\left(\E[Y|\ssalg]\right)\big]\,,
\end{align*}
which shows \eqref{eq:sens-mean}. To prove \eqref{eq:sens-quantile}, let $\ssalg \subseteq \F$, and note that since $g$ is strictly increasing $\VaR_\alpha(g(Y)|\ssalg) = g\left(\VaR_\alpha(Y|\ssalg)\right)$. Next, we use the second identity in \eqref{eq:def ES} to obtain
\begin{align*}
    \E\big[ S_g(\VaR_\alpha(Y), Y )\big]
    &= \E\big[\left(\Id_{\{Y \le \VaR_\alpha(Y)\}} - \alpha\right)
    \left(
    g\big(\VaR_\alpha(Y)\big) - g(Y)
    \right) \big]\\
    &= 
    \E\big[\left(1  - \alpha - \Id_{\{g(Y) > \VaR_\alpha(g(Y))\}}\right)
    \big(
    \VaR_\alpha(g(Y)) - g(Y)
    \big) \big] \\
    &=
    \big(1  - \alpha - \P\big(g(Y) > \VaR_\alpha(g(Y))\big)\big)\VaR_\alpha(g(Y)) \\
    &\qquad + \E\big[\Id_{\{g(Y) > \VaR_\alpha(g(Y))\}} g(Y)\big] - (1-\alpha)\E[g(Y)]\\
    &= (1-\alpha)\ES_\alpha(g(Y)) - (1-\alpha)\E[g(Y)]\,.
\end{align*}
Similarly, 
\begin{align*}
    \E\big[ S_g(\VaR_\alpha(Y|\ssalg), Y )\big] 
    &= \E\big[ \E[S_g(\VaR_\alpha(Y|\ssalg), Y )|\ssalg]\big] \\
    &= \E\big[ (1-\alpha)\ES_\alpha(g(Y)|\ssalg) - (1-\alpha)\E[g(Y)|\ssalg] \big] \\
    &= (1-\alpha) \E\big[\ES_\alpha(g(Y)|\ssalg)] - (1-\alpha)\E[g(Y)]\,.
\end{align*}
Inserting the above into the formula for the sensitivity measures completes the proof.
\hfill\Halmos

Obviously, scaling the scoring function leaves the value of the sensitivity measure unaffected,
i.e., $\xi_{cS} = \xi_{S}$, for any $c>0$.
But otherwise, different choices of scoring functions lead to different sensitivities. In particular, these different sensitivities may induce different rankings of information value,
see Example \ref{ex:zero-info-homo} in Section \ref{subsec:Murphy homogeneous} for an illustration. Proposition \ref{prop:sens mean quantile}, however, offers some further insight on the ordering of sensitivity measures for the mean and quantile. Specifically, for the mean functional, Proposition \ref{prop:sens mean quantile} (case i)) states an if and only if condition for the ordering of sensitivities. Indeed, $\xi_{S_\phi}(Y;\ssalg)\ge \xi_{S_\phi}(Y;\ssalg')$, if and only if, $\E\left[\phi\left(\E[Y|\ssalg]\right)\right]\ge
\E\left[\phi\left(\E[Y|\ssalg']\right)\right]$ for $\phi$ convex and two (not necessarily nested) information sets $\ssalg$, $\ssalg'\subseteq \F$, see also Theorem 3.1 in \cite{KruegerZiegel2021}. This condition thus establishes monotonicity of the sensitivity measure for the mean functional for situations beyond nested information sets. 
An Example of interest is $Y = X_1 + X_2$, where $X_1, X_2$ are independent and normally distributed with mean 0 and variances $\sigma_1^2, \sigma_2^2$, respectively. Then it holds for all convex functions $\phi$ that
$\xi_{S_\phi}(Y;X_1) > \xi_{S_\phi}(Y;X_2)$, if and only if, $\sigma_1^2>\sigma_2^2$, see also Example 3.2. in \cite{KruegerZiegel2021}. A similar argument can be made for the quantile, i.e. part ii) of Proposition \ref{prop:sens mean quantile}. Indeed the sensitivities for the $\VaR_\alpha$ are ordered
$\xi_{S_g}(Y;\ssalg)\ge \xi_{S_g}(Y;\ssalg')$,
if and only if, 
$\E\big[\ES_\alpha(g(Y)|\ssalg)] \le \E\big[\ES_\alpha(g(Y)|\ssalg')]$, for $g$ increasing and two not necessarily nested information sets $\ssalg$, $\ssalg' \subseteq \F$. For further discussion, we refer the interested reader to Theorem C.1 in the online appendix of \cite{KruegerZiegel2021}.
\vspace{1em}

There are many ways to choose a scoring function. 
One can use a scoring function motivated by tradition and interpretability. For the mean functional, for example, the traditional choice is the squared loss which results in the Sobol indices.
We would like to promote two alternative choices. First, in Subsection \ref{subsec:scale-invariant} we consider scale-invariant sensitivities, i.e., sensitivities that are unaffected by scaling $Y$, which are induced by the important subclass of positively homogeneous scores.
Second, in Subsection \ref{subsec:Murphy elementary}, we use Murphy diagrams to illustrate score-based sensitivities \emph{simultaneously} for basically the entire class of consistent scoring functions for a functional. 
Subsection \ref{subsec:Murphy homogeneous} combines the ideas of its two previous subsections and promotes a way to assess sensitivities simultaneously for all positively homogeneous scores.

\subsection{Scale-invariant Sensitivities}
\label{subsec:scale-invariant}

\cite{BaucellsBorgonovo2013} introduce and advocate for the use of sensitivities which are invariant under (strictly) monotone transformations of the output $Y$. A sensitivity $\xi_S$ based on $S$ is invariant under monotone transformations if for any random variable $Y$ and any information set $\ssalg \subseteq \F$
\[
\xi_S\big(u(Y);\ssalg\big) = \sensA
\]
for all strictly increasing functions $u\colon\R\to\R$.
The only score-based sensitivity reported in Table 6 of \cite{borgonovo2021EJOR} which is invariant under monotone transformations of the output $Y$ is based on the log-score, $S(f,y) = -\log(f(y))$ where $f$ is a predictive density. The log-score is a strictly proper scoring rule which is tailored to evaluate probabilistic predictions (corresponding to the situation where $T$ is the identity map).
To the best of our knowledge, we are not aware of any other score-based sensitivity, in particular also for point predictions, which is invariant under monotone transformations of the output $Y$. 
There are, however, examples of sensitivity measures that, while not score-based sensitivities, are transformation invariant under monotone transformation of the output random variables. We refer the interested reader to \cite{Plischke2019copula} and \cite{BaucellsBorgonovo2013}.

We suggest to consider a weaker, but still very relevant, invariance criterion. 
\begin{definition}[Scale-invariance]
A sensitivity $\xi_S$ based on $S$ is \emph{scale-invariant} if
for any random variable $Y$ and any information set $\ssalg \subseteq \F$
\[
\xi_S(cY;\ssalg) = \sensA\, \quad \text{for all}\quad c>0\,.
\]
\end{definition}
A scale-invariant sensitivity measure is a sensitivity measure that takes on the same value, independent of the units in which $Y$ is reported. We will see that a score-based sensitivity measure is scale-invariant if the employed scoring function is homogeneous, which in turn implies that the considered functional $T$ is homogeneous of degree 1. Thus, to discuss scale-invariant score-based sensitivity measures, we first define positive homogeneity for scoring functions.

\begin{definition}[Homogeneity]
Let $D$ be the positive half-line, the negative half-line or the whole $\R$.
A scoring function $S\colon D\times D \to\R$ is positively homogeneous of degree $b\in\R$, if 
$S(cz,cy) = c^b S(z,y)$ for all $z,y\in D$ and for all $c>0$.
We call a scoring function positively homogeneous if there exists a $b\in\R$ such that it is positively homogeneous of degree $b$.
\end{definition}

Next, we establish a sufficient condition for an elicitable functional to be homogeneous of degree 1.

\begin{proposition}[Homogeneous Elicitable Functionals]\label{prop:hom-elicitable}
Let $D$ be the positive half-line, the negative half-line, or the whole $\R$.
If there is a positively homogeneous score $S\colon D\times D \to\R$ which is strictly $\M$-consistent for $T\colon\M\to D$, then $T$ is positively homogeneous of degree 1.
\end{proposition}

\textit{Proof of Proposition \ref{prop:hom-elicitable}}.
The assertion is a special instance of Lemma 4.3 in \cite{FisslerZiegel2019}.
For completeness, we provide a short proof.
Let $S\colon D\times D \to\R$ be strictly $\M$-consistent for $T$ and positively homogeneous of degree $b\in\R$. 
Then, for all $Y$ with $F_Y\in\M$ and for all $c>0$, it holds that if $z \in D$, then setting $z^\prime c = z$ implies that $z^\prime\in D$, and we obtain that
\begin{align*}
    T(cY) 
    &= \argmin_{z \in D} \E\left[S(z, cY)\right]
    = c\, \argmin_{z^\prime \in D} \E\left[S(c z^\prime, cY)\right]
    = c\, \argmin_{z \in D} \;c^b  \;\E\left[S(z, Y)\right]
    = c\, T(Y)\,.
\end{align*}
\hfill\Halmos

Many functionals of interests such as monetary risk measures, which include the entropic risk measures, VaR, ES, and RVaR, are by definition homogeneous of degree 1 (scale-\emph{equivariant}), i.e., $T(cY) = cT(Y)$ for all $ c>0$. Moreover, these elicitable functional possess homogeneous scores. Note that 
positive homogeneity of $T$ implies that $T(cY|\ssalg) = cT(Y|\ssalg)$ for any information set $\ssalg$. The next proposition provides a sufficient condition for a score-based sensitivity to be scale-invariant.

\begin{proposition}
\label{prop:scale-invariance}
Let $D$ be the positive half-line, the negative half-line, or the whole $\R$.
If there is a positively homogeneous score $S\colon D\times D \to\R$ which is strictly $\M$-consistent for $T\colon\M\to D$, 
then, the score-based sensitivity $\xi_S$ is scale-invariant.
\end{proposition}

The proof of Proposition \ref{prop:scale-invariance} is a direct consequence of Proposition \ref{prop:hom-elicitable} and that $T(cY|\ssalg) = cT(Y|\ssalg)$, for all $c >0$ and $\ssalg \subseteq \F$.
\hfill \Halmos

The above proposition implies that if scale-invariance is a desired property a homogeneous score should be chosen for the score-based sensitivity measures.
Many elicitable and positively homogeneous functionals admit positively homogeneous scoring functions, which enables to build scale-invariant score-based sensitivities.
We recall the family of homogeneous scoring functions for the mean and VaR functional from the literature.
\begin{proposition}[Homogeneous Scores for the Mean \citep{NoldeZiegel2017, Patton2011}]
\label{prop: hom mean}
The class of strictly consistent and $b$-homogeneous scores for the mean satisfying $S(y,y)=0$ are given by any positive multiple of a member of the Patton family
\begin{equation}
    \label{eq:Patton}
    S_b(z,y) = 
    \begin{cases}
    \frac{y^b - z^b}{b(b-1)}- \frac{z^{b-1}}{b-1}(y-z), & b\in\R\setminus \{0,1\},\\[0.5em]
    \frac{y}{z} - \log\left(\frac{y}{z}\right) - 1, & b=0,\\[0.5em]
    y\log\left(\frac{y}{z}\right) - (y-z), & b=1,
    \end{cases}
\end{equation}
where we require that $z,y>0$. 
\end{proposition}
Note that homogeneous scores of a given degree are unique up to positive scaling and that scaling a score does not affect the sensitivity. The positively homogeneous scores in \eqref{eq:Patton} arise from \eqref{eq:Bregman} upon choosing the strictly convex functions $\phi_b(y) = y^b/(b(b-1))$ for $b\in\R\setminus\{0,1\}$, $\phi_0(y)= -\log(y)$ for $b = 0$, and $\phi_1(y) = -y\log(y)$ for $b = 1$. 
Note that in \eqref{eq:Patton} we require $z,y$ to be strictly positive. 
There are no strictly consistent and $b$-homogeneous scores for the mean on $\R\times\R$, since there are no convex and $b$-homogeneous functions on $\R$ for $b\le 1$.
For $b>1$, however, one may choose
\[
\phi_b(y) = d_1y^b\Id_{{\{y>0\}}} + d_2|y|^b\Id_{{\{y<0\}}}, \qquad y\in\R
\]
for positive constants $d_1,d_2>0$.
We refer to the supplement of \cite{NoldeZiegel2017} which contains the characterisation results of positively homogeneous and consistent scoring functions for the $\tau$-expectile; for which the mean arises for $\tau=1/2$. Moreover, this supplement contains similar characterisation results for VaR and the pair consisting of VaR and ES. We state the former.

\begin{proposition}[Homogeneous Scores for VaR \citep{NoldeZiegel2017}]
\label{prop: hom-score-var}
The class of strictly consistent and $b$-homogeneous scores for $\VaR_\alpha$ satisfying $S(y,y)=0$ are given by any positive multiple of
\begin{equation*} 
    S_b(z,y) = \left(\Id_{\{y \le z\}} - \alpha\right) \big(g(z) - g(y)  \big) \,,
\end{equation*}
where 
\begin{equation*}
g(y) = 
\begin{cases}
y^b\Id_{\{y>0\}} - d|y|^b\Id_{\{y<0\}}  \qquad\qquad &\text{if} \quad b >0 \text{ and } y \in \R,
\\[0.5em]
\log(y)  &\text{if} \quad b = 0 \text{ and } y >0,
\\[0.5em]
-y^b   &\text{if} \quad b <0 \text{ and } y >0\,,
\end{cases}%
\end{equation*}%
for a positive constant $d>0$.
\end{proposition}

Again, if $y>0$, homogeneous scores of a certain degree are unique up to positive scaling.
We could also consider transformations other than scaling, such as e.g., translations. Then, if the score $S$ is invariant and the functional $T$ is equivariant for that transformation, we directly retrieve the corresponding invariance of the score-based sensitivity for that transformation.
If we consider transformations with respect to which the functional $T$ of interest is not equivariant or the score $S$ is not invariant, we do not obtain an invariance property of the resulting sensitivity measure $\xi_S$.
In case of the mean and the coefficient of determination, R$^2$, this fact is very well known and described in standard Econometrics textbooks.\footnote{E.g. \citet[p.~194]{Wooldridge} writes ``\ldots it is \emph{not} legitimate to compare $R$-squareds from models where $y$ is the dependent variable in one case and $\log(y)$ is the dependent variable in the other. These measures explain variations in different variables.'' (The emphasis in the quote is original.)}
We further illustrate this in the following example.

\begin{example}
\label{ex:transform-y}
Let $Y = \exp(X_1 + X_2)$, where $X_1$ $X_2$ are independent and standard normally distributed. 
Consider the mean functional with the squared loss score. Then the sensitivity of $\log(Y)$ with respect to $X_1$ is 
\begin{equation*}
\xi_S(\log(Y);X_1)
=
\frac{\Var(\E[\log(Y)|X_1])}{\Var(\log(Y))}
=
\frac{\Var(X_1 + \E[X_2])}{\Var(\log(Y))}
= \frac{1}{2}\,.
\end{equation*}
In contrast, the sensitivity of 
$Y$ with respect to $X_1$ based on the squared loss is
\begin{equation*}
\xi_S(Y;X_1)
= 
\frac{\Var\left(\E\left[Y|X_1\right]\right)}{\Var\left(Y\right)}
= 
\frac{\Var\left(e^{X_1}\E\left[e^{X_2}\right]\right)}{\Var\left(Y\right)}
= 
\frac{1}{e+1} \approx 0.27\,.
\end{equation*}
We emphasise that for a function $h$, the sensitivity of $h(Y)$ with respect to $X_1$ is in general not equal to the sensitivity of $Y$ with respect to $X_1$ if $h$ is not affine.
\end{example}

\subsection{Murphy Diagrams Based on Elementary Scores}
\label{subsec:Murphy elementary}

In contrast to Subsection \ref{subsec:scale-invariant}, where we discuss the choice of score-based sensitivity measure by imposing additional restrictions on the scoring functions, here, we pursue a different strategy.
Since different choices of scoring functions may lead to different rankings of information in terms of score-based sensitivities, we suggest to \emph{simultaneously} consider  (ideally) \emph{all} sensitivity measures that arise by (strictly) consistent scoring functions.
While there are many characterisation results for consistent scoring functions available in the spirit of Proposition \ref{prop:score_mean_quantile},
these classes are typically indexed by an infinite dimensional parameter space, e.g., the space of convex functions $\phi$ for the mean or the space of increasing functions $g$ for the quantile.
This fact renders a computation and comparison of all score-based sensitivities practically infeasible at first glance.

\cite{EhmETAL2016} establish so-called \emph{mixture representations} of classes of all consistent scoring functions for the mean\footnote{And more generally also the $\tau$-expectile.} and the $\alpha$-quantile; subject to mild regularity conditions.
That is, they show that $S$ is a consistent scoring function for the mean (or the $\alpha$-quantile)
if and only if there is a non-negative and $\sigma$-finite measure $H$ on $\R$ such that 
\begin{equation}\label{eq:mixture}
S(z,y) = \int_{\R} S_{\theta}(z,y)\, \mathrm{d} H(\theta),
\end{equation}
where the class $\{S_{\theta}, \, \theta\in\R\}$ consists of co-called \emph{elementary scores}, which are consistent scores for the mean (or the $\alpha$-quantile).
Moreover, the measure $H$ is uniquely determined by $S$. We recall the formal statements.

\begin{proposition}[Elementary Scores for the Mean and VaR \citep{EhmETAL2016}]
\label{prop:mixture}\hfill
\begin{enumerate}[label = $\roman*)$]
\item
Let $S_{\phi}$ be a Bregman score \eqref{eq:Bregman} such that the subgradient $\phi'$ is left-continuous.
Then
\begin{equation*}
    \label{eq:Bregman mixing}
    S_{\phi}(z,y) = \int_\R S_\theta(z,y)\, \mathrm{d}\phi'(\theta)\,,
\end{equation*}
where for $\theta\in\R$, the \emph{elementary scores} for the mean functional are given by
\begin{equation}
    \label{eq:Bregman elementary}
    S_\theta(z,y) = \begin{cases}
|y-\theta|, & \text{if } \theta \in\big[\min(z,y), \max(z,y)\big) \\
0, & \text{otherwise}.
\end{cases}
\end{equation}
\item
Let $S_{g}^{\alpha}$ be a generalised piecewise linear score \eqref{eq:GPL} such that $g$ is left-continuous. 
Then
\begin{equation*}
    \label{eq:GPL mixing}
    S_{g}^{\alpha}(z,y) = \int_\R S^{\alpha}_\theta(z,y)\, \mathrm{d}g(\theta)\,,
\end{equation*}
where for $\theta\in\R$, the \emph{elementary scores} for the $\VaR_\alpha$ are given by
\begin{equation*}
    \label{eq:GPL elementary}
    S^{\alpha}_\theta(z,y) = 
    \begin{cases}
    1 - \alpha, \qquad &\text{if}\quad \theta \in [y, z)\\
    \alpha, &\text{if}\quad \theta \in [z, y)\\
    0, & \text{otherwise}.
    \end{cases}
\end{equation*}
\end{enumerate}
\end{proposition}

These mixture representations of Proposition \ref{prop:mixture} open the door to assess prediction dominance with respect to almost any Bregman score or generalised piecewise linear score. I.e., for any (possibly random) forecasts $Z_1,Z_2$, we obtain that
$\E\big[S_{\phi}(Z_1,Y)\big]
\le \E\big[S_{\phi}(Z_2,Y)\big]
$ for all Bregman scores \eqref{eq:Bregman} with  left-continuous subgradient $\phi'$, if and only if, 
$\E\big[S_{\theta}(Z_1,Y)\big]
\le \E\big[S_{\theta}(Z_2,Y)\big]
$ for all elementary scores $S_{\theta}$ in \eqref{eq:Bregman elementary}. \cite{EhmETAL2016} demonstrate that this equivalence can be used to establish forecast dominance by inspecting the so called \textit{Murphy diagrams}. 
\begin{definition}[Murphy Diagrams; \cite{EhmETAL2016}]
Let $\{S_\theta, \ \theta\in\R\}$ be a class of scoring functions indexed by $\theta\in\R$. 
For an observation $Y$ and two (possibly random) forecasts $Z_1,Z_2$, the Murphy diagram of the score difference with respect to $\{S_\theta, \ \theta\in\R\}$ is the map
\[
\R\ni \theta \;\mapsto\; \E\big[S_{\theta}(Z_1,Y)\big] - \E\big[S_{\theta}(Z_2,Y)\big]\,.
\]
\end{definition}
Clearly, in applications one can approximate the Murphy diagram with the corresponding empirical counterparts of the expectations.
The following result shows that we can apply a similar rationale in our context of score-based sensitivities and therefore introduce Murphy diagrams for score-based sensitivity measures.

\begin{proposition}[Ordering of Score-based Sensitivities]
\label{prop:ordering}
Suppose $\mathcal S$ is a class of consistent scoring functions for a functional $T$ and $\{S_{\theta}, \, \theta\in\R\}$ a subclass of $\mathcal S$ such that for any $S\in\mathcal S$ there exists a non-negative $\sigma$-finite measure $H$ on $\R$ such that 
\begin{equation}
\label{eq:integral}
S(z,y) = \int_{\R}S_{\theta}(z,y)\,\mathrm{d}H(\theta).
\end{equation}
Then, for any response $Y$ and any information sets $\ssalg_1$ and $\ssalg_2$, the following are equivalent:
\begin{enumerate}[label = $\roman*)$]
    \item $\xi_S(Y;\ssalg_1)\le \xi_S(Y;\ssalg_2) \quad \text{for all } S\in\mathcal S$; \label{eq:equivalence-1}
    
    \item 
    $\xi_{S_\theta}(Y;\ssalg_1)\le \xi_{S_\theta}(Y;\ssalg_2) \quad \text{for all } \theta\in\R$.\label{eq:equivalence-2}
\end{enumerate}
\end{proposition}

\textit{Proof of Proposition \ref{prop:ordering}.}
The implication ``\ref{eq:equivalence-1} $\Rightarrow$ \ref{eq:equivalence-2}'' is obvious since by assumption $S_\theta\in\mathcal{S}$ for all $\theta\in\R$. For the other direction, suppose that \ref{eq:equivalence-2} holds. Then, for all $\theta\in\R$ it holds that 
$\xi_{S_{\theta}}(Y;\ssalg_1)\le \xi_{S_{\theta}}(Y;\ssalg_2)$
if and only if 
$\E[S_\theta(T(Y|\ssalg_1),Y)] - \E[S_\theta(T(Y|\ssalg_2),Y)]\ge0$. 
Hence, \ref{eq:equivalence-2} implies that for any $a,b>0$ and any $\theta_1,\theta_2\in\R$
\begin{equation}\label{eq:proof}
\E[aS_{\theta_1}(T(Y|\ssalg_1),Y) + bS_{\theta_2}(T(Y|\ssalg_1),Y)] - \E[aS_{\theta_1}(T(Y|\ssalg_2),Y) + bS_{\theta_2}(T(Y|\ssalg_2),Y)]\ge0.
\end{equation}
\eqref{eq:proof} in turn is equivalent to
$\xi_{aS_{\theta_1} + bS_{\theta_2}}(Y;\ssalg_1)\le \xi_{aS_{\theta_1} + bS_{\theta_2}}(Y;\ssalg_2)$.
Finally, \eqref{eq:proof}, the construction of the integral \eqref{eq:integral}, and Fubini's Theorem imply \ref{eq:equivalence-1}.
\hfill\Halmos

Proposition \ref{prop:ordering} implies that if we want to check whether information $\ssalg_2$ is more important for modelling $Y$ than $\ssalg_1$ with respect to all score-based sensitivities, it suffice to establish this ranking with respect to all elementary score-based sensitivities. This motivates to consider the following Murphy diagrams for score-based sensitivities.

\begin{definition}[Murphy Diagrams for Sensitivities]
\label{def:Murhpy sensitivity}
Let $\{S_\theta, \ \theta\in\R\}$ be a class of scoring functions indexed by $\theta\in\R$. 
The sensitivity of $Y$ to $\ssalg$ based on $\{S_\theta, \ \theta\in\R\}$ is given by the Murphy diagram for sensitivities
\[
\R\ni \theta \;\mapsto\; \xi_{S_{\theta}}(Y;\ssalg)\,.
\]
\end{definition}

\begin{example}[Example \ref{example:zero information gain 2} Continued]
\label{example:zero information gain 2 continued}
Let $X_1,X_2,X_3$ be independent, $X_1$ Bernoulli distributed with $p = \P(X_1=0) = 1- \P(X_1=1) \in(0, \alpha)$, and $X_2<C<X_3$ almost surely, for $C>0$, and consider the output $Y = \Id_{\{X_1=0\}}X_2 + \Id_{\{X_1=1\}}X_3$. We further choose $p = 0.8$, $C = 10$, $X_2$ uniformly distributed on $[0, C]$, and $X_3 = C + Z$, where $Z$ has a Gamma distribution with mean 20 and variance 10. 
\begin{table}[t]
    \centering
    \begin{tabular}{l l l}
    \toprule
    $T$     & $\E$ &  $\VaR_\alpha$\\[0.5em]
    \midrule
     $T(Y)$   & $p\E[X_2] + (1 - p)\E[X_3]$ &  $\VaR_{(\alpha-p)/(1-p)}(X_3)$
    \\[0.5em]
     $T(Y|X_1)$ & $\Id_{\{X_1 = 0\}}\E[X_2]  +  \Id_{\{X_1 = 1\}}\E[X_3]$ & $\Id_{\{X_1=0\}}\VaR_\alpha(X_2) + \Id_{\{X_1=1\}}\VaR_\alpha(X_3)$
     \\[0.5em]
     $T(Y|X_2)$ & $pX_2 + (1 - p)\E[X_3]$& $\VaR_{(\alpha-p)/(1-p)}(X_3)$
     \\[0.5em]
     $T(Y|X_3)$ & $p\E[X_2] + (1 - p)X_3$& $X_3$
     \\[0.5em]
     $T(Y|X_1, X_2)$ & $\Id_{\{X_1 = 0\}}X_2  +  \Id_{\{X_1 = 1\}}\E[X_3]$ & $\Id_{\{X_1=0\}}X_2 + \Id_{\{X_1=1\}}\VaR_\alpha(X_3)$
     \\[0.5em]
     $T(Y|X_1, X_3)$ & $\Id_{\{X_1 = 0\}}\E[X_2]  +  \Id_{\{X_1 = 1\}} X_3$ & $\Id_{\{X_1=0\}}\VaR_\alpha(X_2) + \Id_{\{X_1=1\}}X_3$
     \\[0.5em]
     $T(Y|X_2, X_3)$ & $pX_2  +  (1-p)X_3$ & $X_3$
     \\[0.5em]
    \bottomrule
    \end{tabular}
    \caption{Functional and conditional functional $T$ of Example \ref{example:zero information gain 2 continued}.}
    \label{tab:T-murphy}
\end{table}
Table \ref{tab:T-murphy} contains the conditional functionals for one and two risk factors for the mean functional and the $\VaR_{\alpha}$ used for calculating the score-based sensitivities. Figure \ref{fig:murphy-mean-var-ele} displays the corresponding Murphy diagrams of the elementary score-based sensitivities for the mean functional and the $\VaR_{0.9}$. All plots are based on $10^6$ simulations. We observe that the sensitivities of the mean to one risk factor are ordered, that is $\xi_{S_\theta}(Y;X_1)>\xi_{S_\theta}(Y;X_2)>\xi_{S_\theta}(Y;X_3)$ for all $\theta$, see top left panel of Figure \ref{fig:murphy-mean-var-ele}. This implies that the sensitivities to one risk factor are ordered for all strictly consistent scoring functions. This is in contrast to the sensitivity for $\VaR_{0.9}$. Comparing the mean functional with the $\VaR_{0.9}$, we observe that $X_1$ has a large sensitivity for both functionals, the sensitivity to $X_2$ is larger for the mean compared for the $\VaR_{0.9}$, and the sensitivity to $X_3$ is larger for the $\VaR_{0.9}$ compared for the mean. This reflects that $X_2$ influences the mean of $Y$ while $X_3$ influences the tail, hence the $\VaR_{0.9}$ of $Y$.
\begin{figure}[t]
    \centering
    \includegraphics[width=0.3\textwidth]{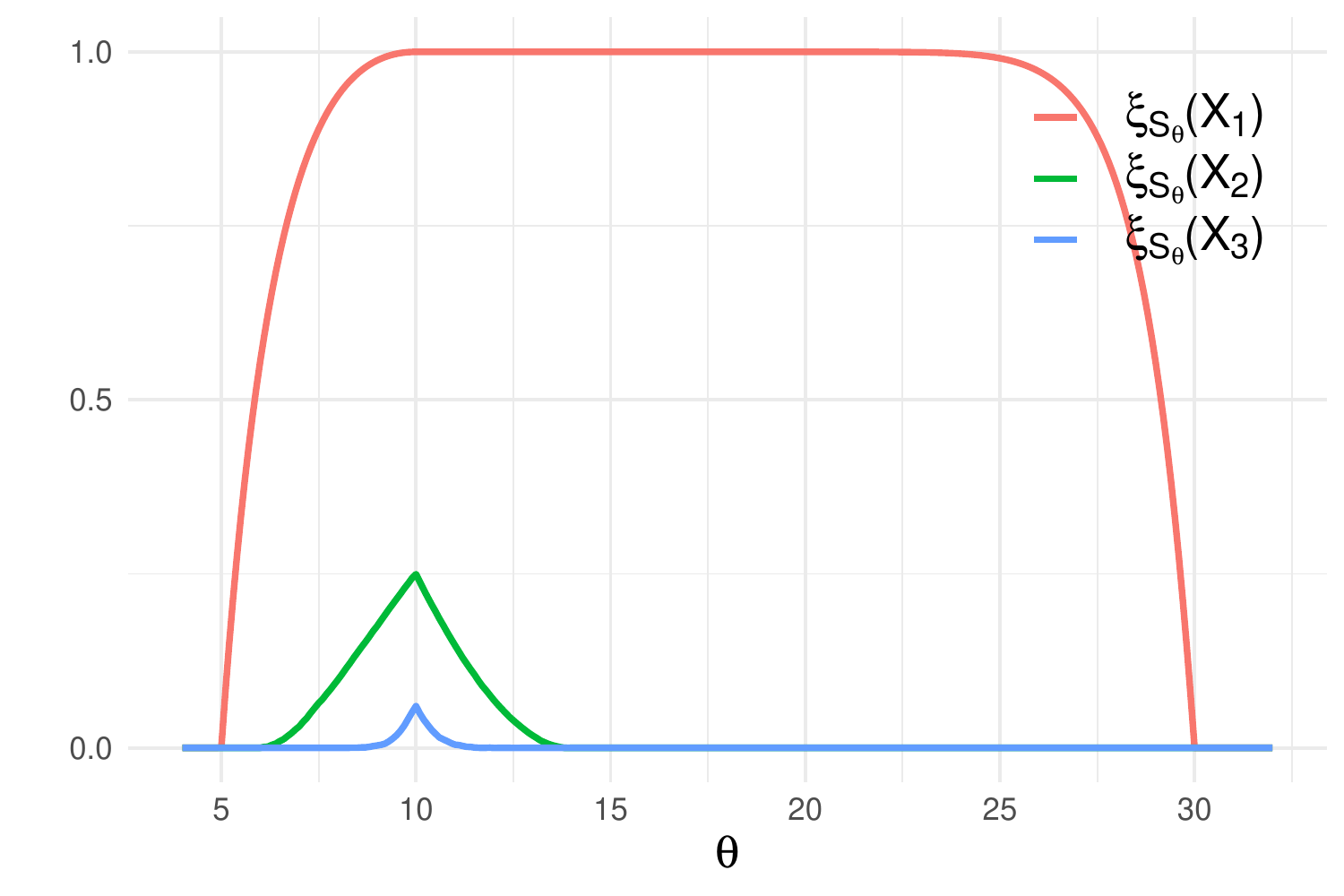}
    \hspace{0.5em}
    \includegraphics[width=0.3\textwidth]{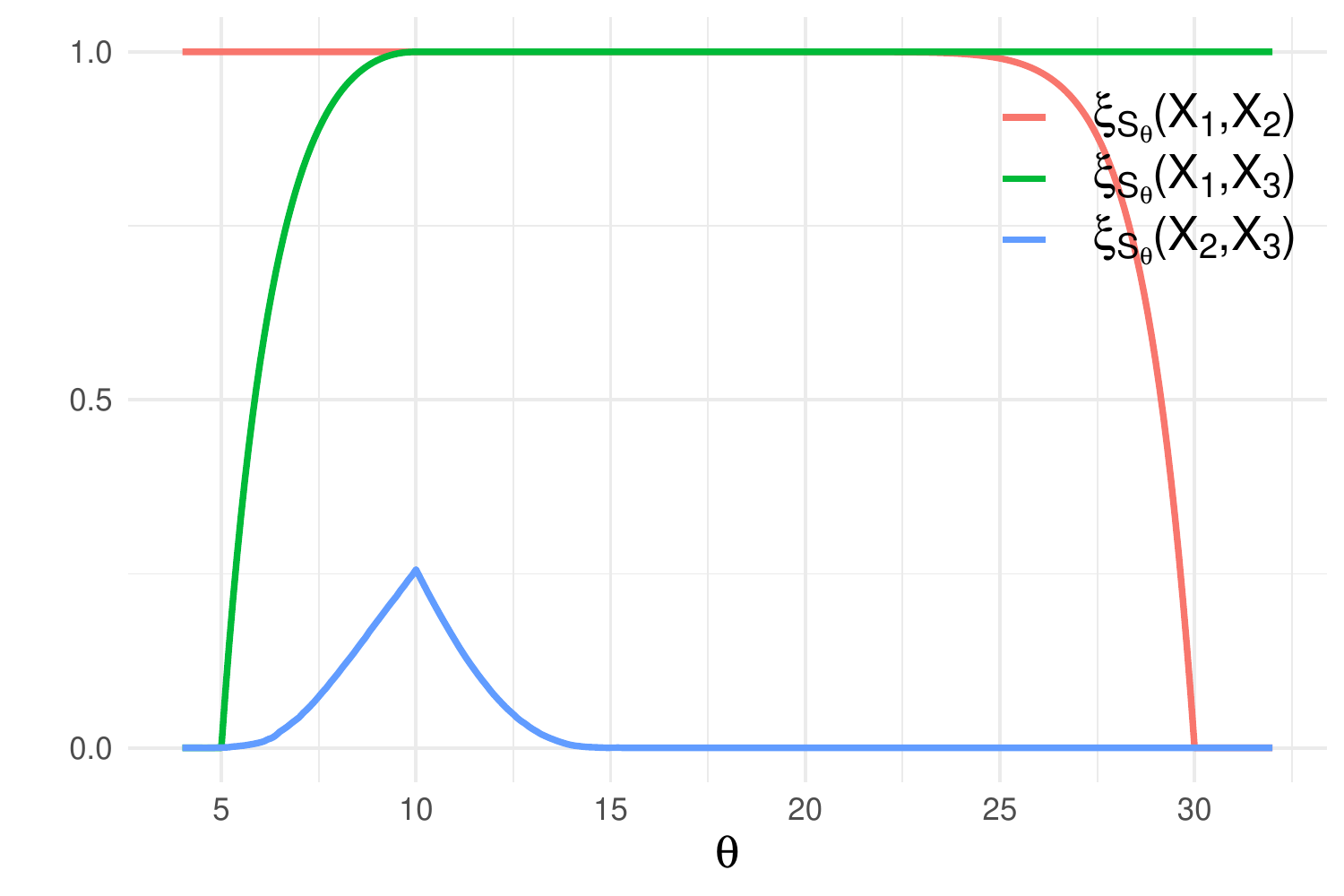}
    \hspace{0.5em}
    \includegraphics[width=0.3\textwidth]{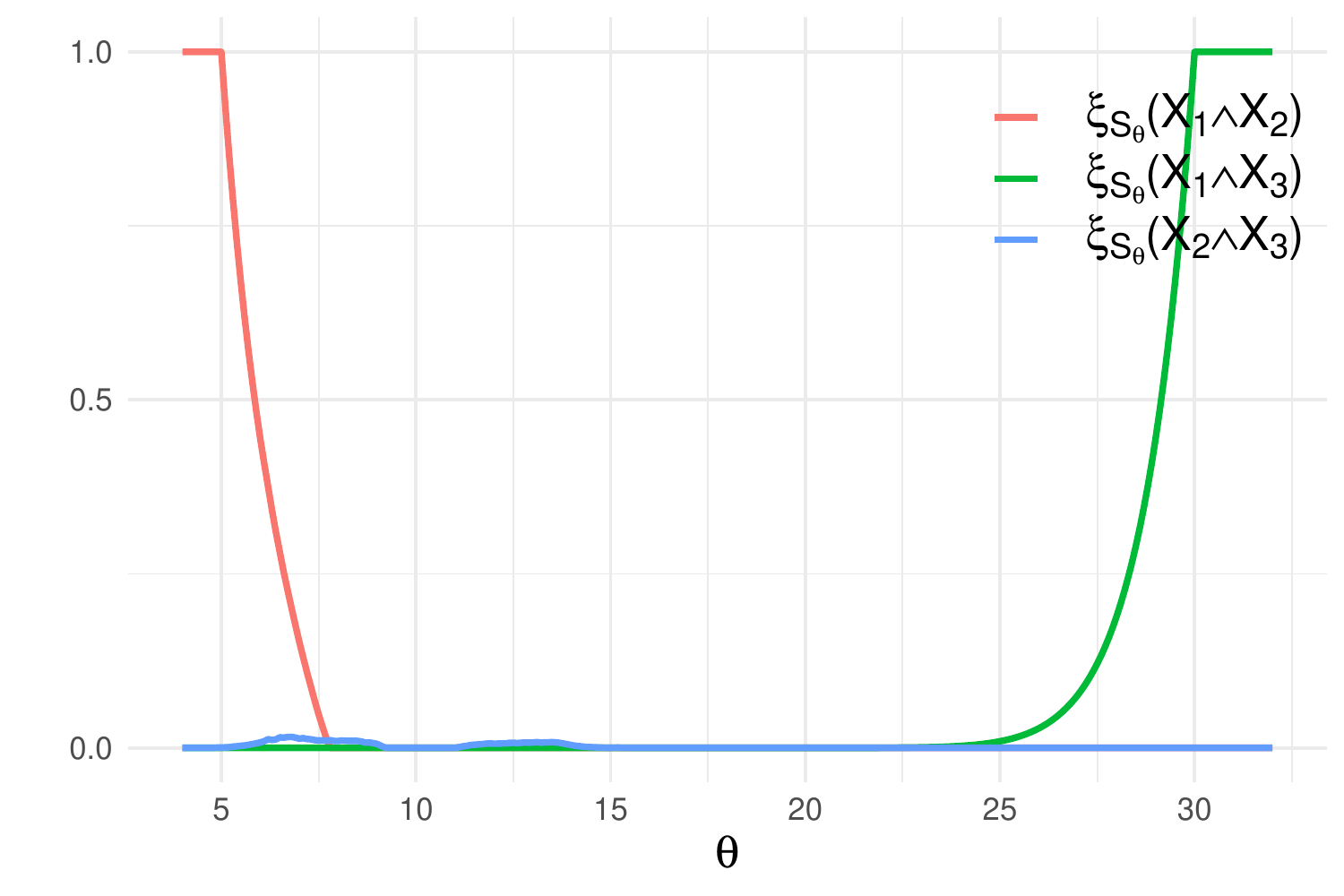}
    \\[0.5em]
    \includegraphics[width=0.3\textwidth]{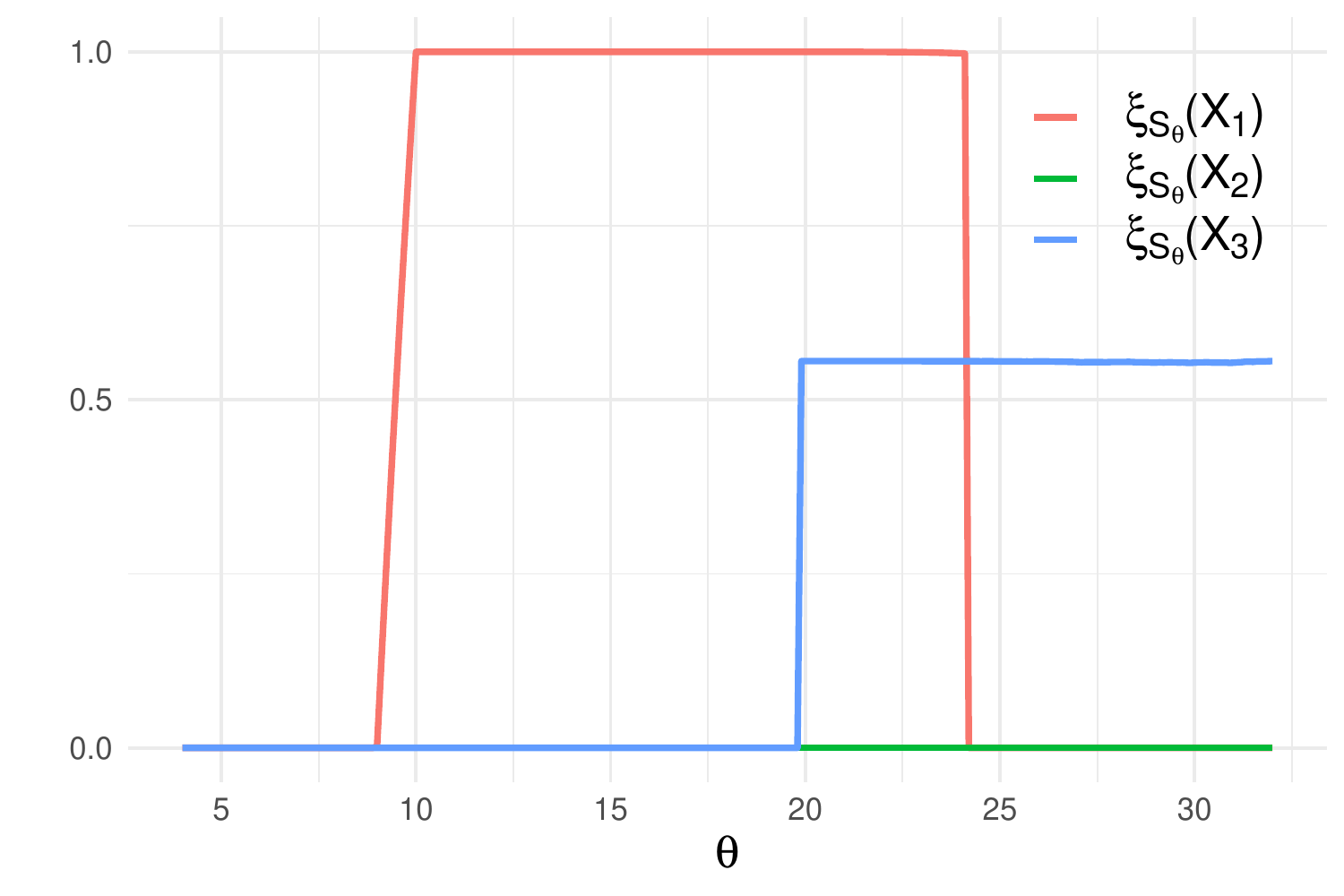}
    \hspace{0.5em}
    \includegraphics[width=0.3\textwidth]{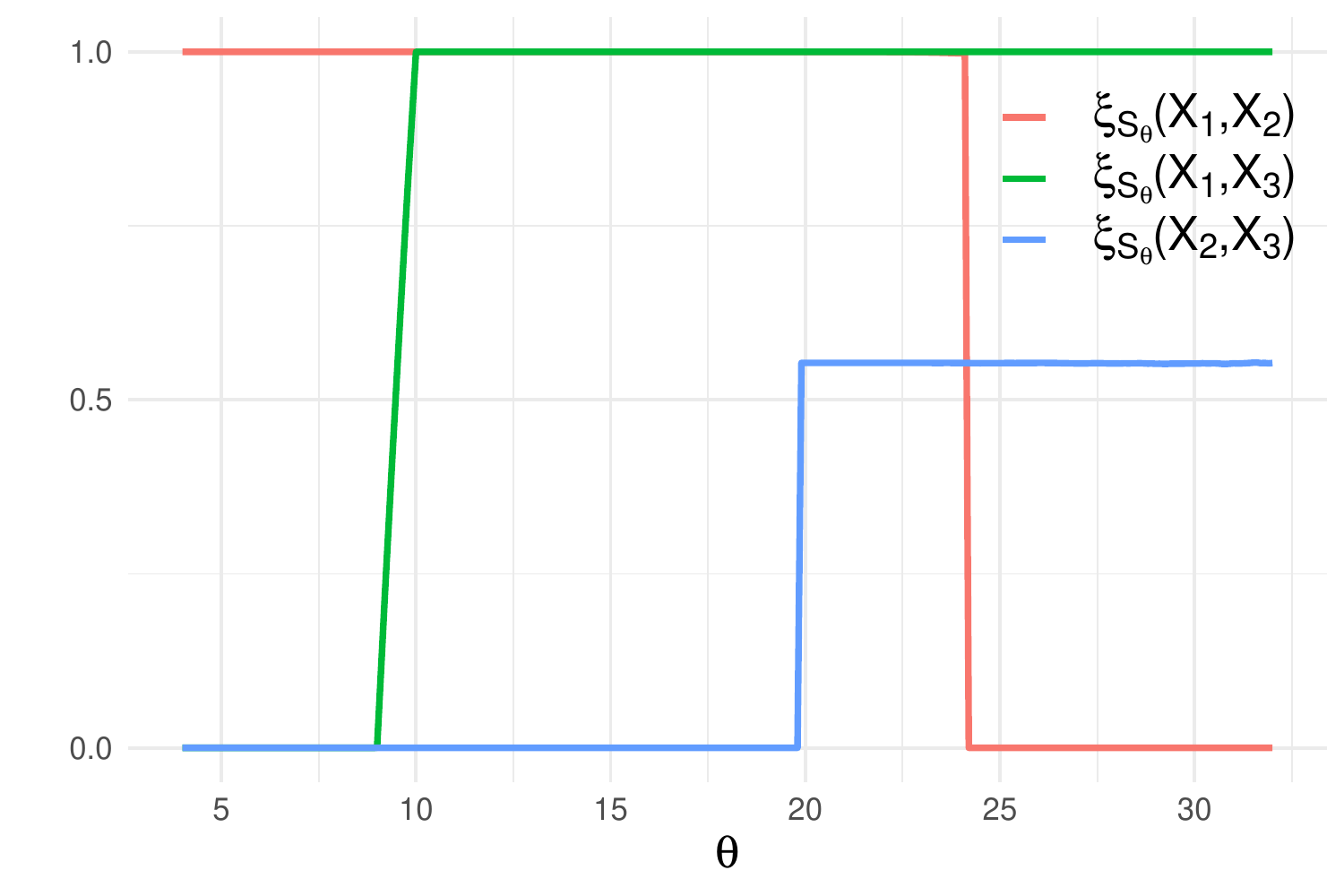}
    \hspace{0.5em}
    \includegraphics[width=0.3\textwidth]{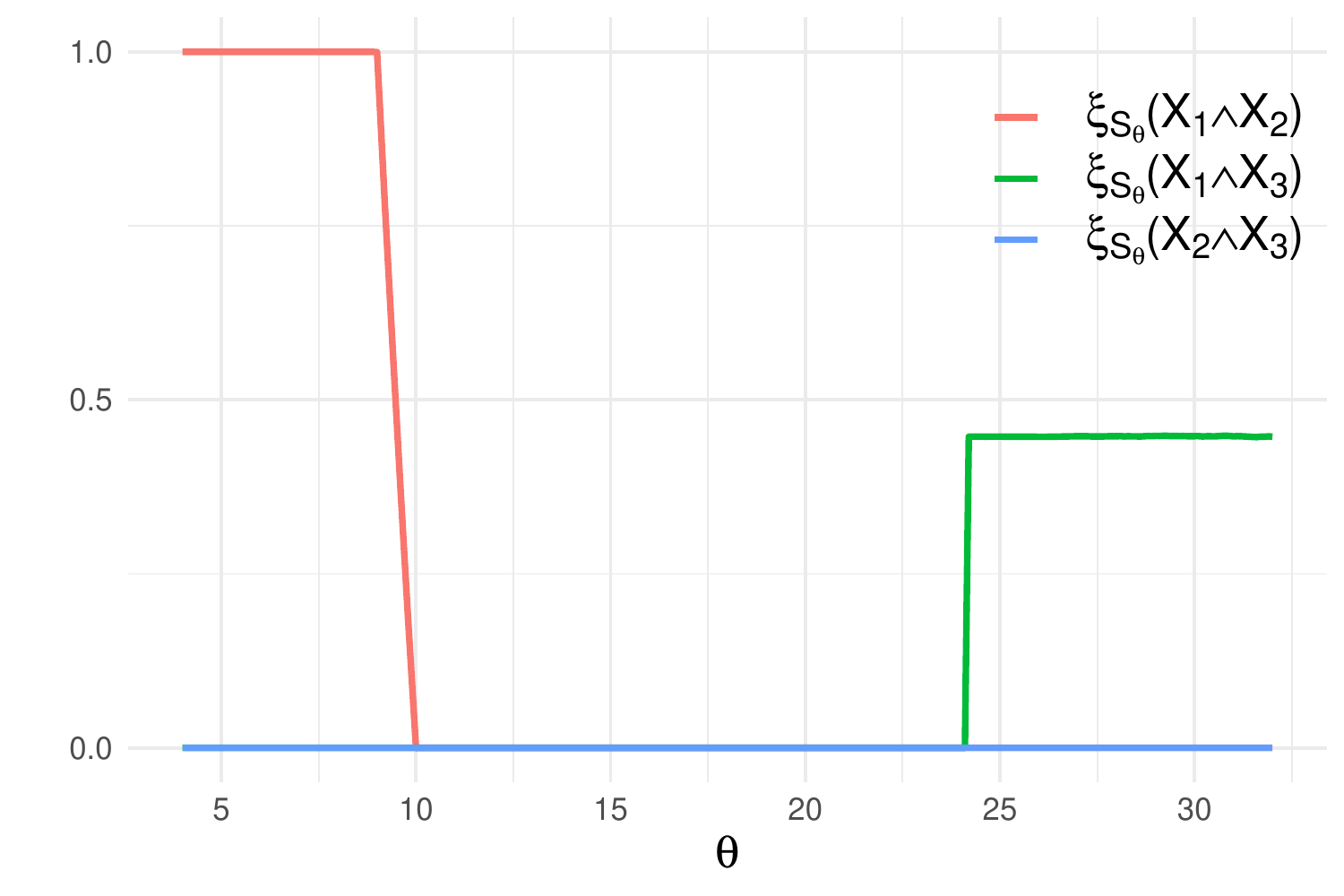}
    \caption{Elementary score-based sensitivities. Top panels display the mean functional and bottom panels the $\VaR_{0.9}$. Elementary score-based sensitivities to a single risk factor (left panel), to two risk factors (middle panel), and interaction sensitivities (right panels).}
    \label{fig:murphy-mean-var-ele}
\end{figure}
\end{example}

\subsection{Murphy Diagrams for Homogeneous Scores}
\label{subsec:Murphy homogeneous}

In Subsection \ref{subsec:scale-invariant} we have seen that a positively 1-homogeneous functional $T$ combined with a positively homogeneous scoring function renders the corresponding score-based sensitivity measure scale-invariant. 
Moreover, known characterisation results for positively homogeneous scores (Propositions \ref{prop: hom mean}, \ref{prop: hom-score-var}) state that, at least when $Y$ is positive, the $b$-homogeneous and strictly consistent scoring functions for the mean and the VaR are unique up to scaling. 
Since scaling the scores leaves the sensitivities unaffected, this means that for $Y>0$ and for the mean and the VaR, there is only a one-dimensional family of scores -- the $b$-homogeneous ones -- which render the sensitivity scale-invariant.

We propose to evaluate all of them jointly, making use of the Murphy diagram introduced in Definition \ref{def:Murhpy sensitivity}. The difference to the Murphy diagrams for elementary scores is that here the diagram is considered with respect to the parameter $b$, indicating the degree of homogeneity of the scoring function.
That is
\[
b\;\mapsto\; \xi_{S_b}(Y;\ssalg).
\]

We illustrate the homogeneous score-based Murphy diagrams for the mean and $\VaR_\alpha$ functional in the next example. For comparison with the Murphy diagrams for elementary scoring function, we illustrate the homogeneous score-based Murphy diagrams on the same Example \ref{example:zero information gain 2 continued}.

\begin{example}[Example \ref{example:zero information gain 2 continued} Continued]\label{ex:zero-info-homo}
Figure \ref{fig:murphy-mean-var-hom} displays the Murphy diagrams of the homogeneous score-based sensitivities. Comparing with the elementary score-based Murphy diagrams in Figure \ref{fig:murphy-mean-var-ele}, we observe a similar picture in that $X_1$ has a large sensitivity for both functionals, the sensitivity to $X_2$ is zero for the $\VaR_{0.9}$, and the sensitivity to $X_3$ is zero for the mean but large for the $\VaR_{0.9}$. The interaction sensitivities, right panels of Figure \ref{fig:murphy-mean-var-hom}, show that the interaction between $X_1$ and $X_3$ are non-negligible for both functionals. This is informative, as the sensitivity to $X_3$ for the mean is equal to zero for all choices of homogeneous scoring functions.
We obtain the Socol indices for $b=2$.

The sensitivities for the mean to one risk factor (top left panel in Figure \ref{fig:murphy-mean-var-hom}) are ordered: $\xi_{S_b}(Y; X_1) \ge \xi_{S_b}(Y; X_2)\ge \xi_{S_b}(Y; X_3)$. This is in contrast to the sensitivities for the $\VaR_{0.9}$ based on the homogeneous scores to one risk factor (bottom left panel in Figure \ref{fig:murphy-mean-var-hom}). Indeed, for $b = 0$ the sensitivities of $\VaR_{0.9}$ to $X_1$, $X_2$, and $X_3$ are respectively $0.45$, $0$, and $0.19$. For $b = 4$, the sensitivities of $\VaR_{0.9}$ to $X_1$ is equal to 0.32, the sensitivities of $\VaR_{0.9}$ to $X_2$ is $0$, and to $X_3$ is equal to $0.50$. Thus, different choices of strictly consistent scoring function for $\VaR$ can lead to different rankings of risk factors.

\begin{figure}
    \centering
    \includegraphics[width=0.3\textwidth]{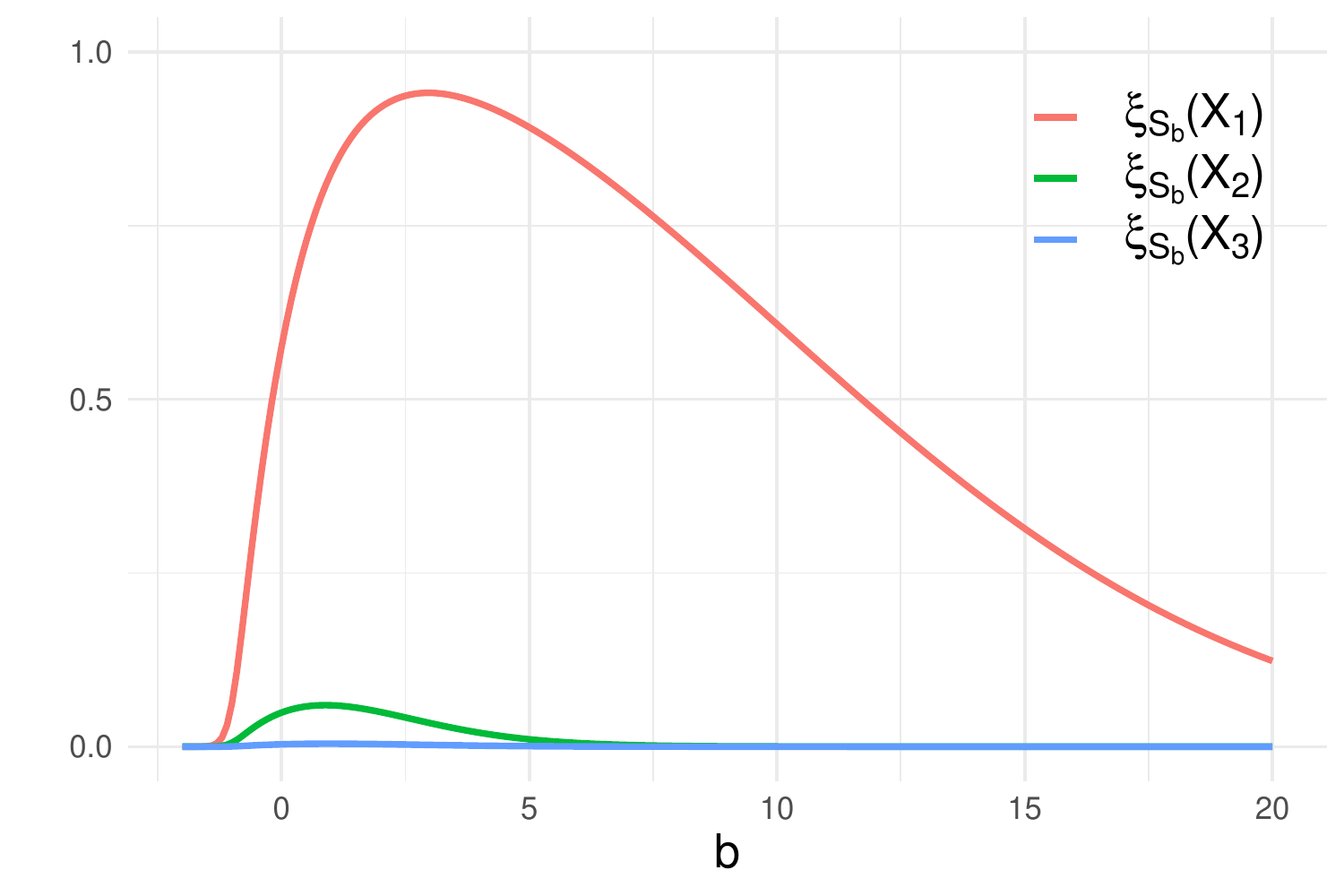}
    \hspace{0.5em}
    \includegraphics[width=0.3\textwidth]{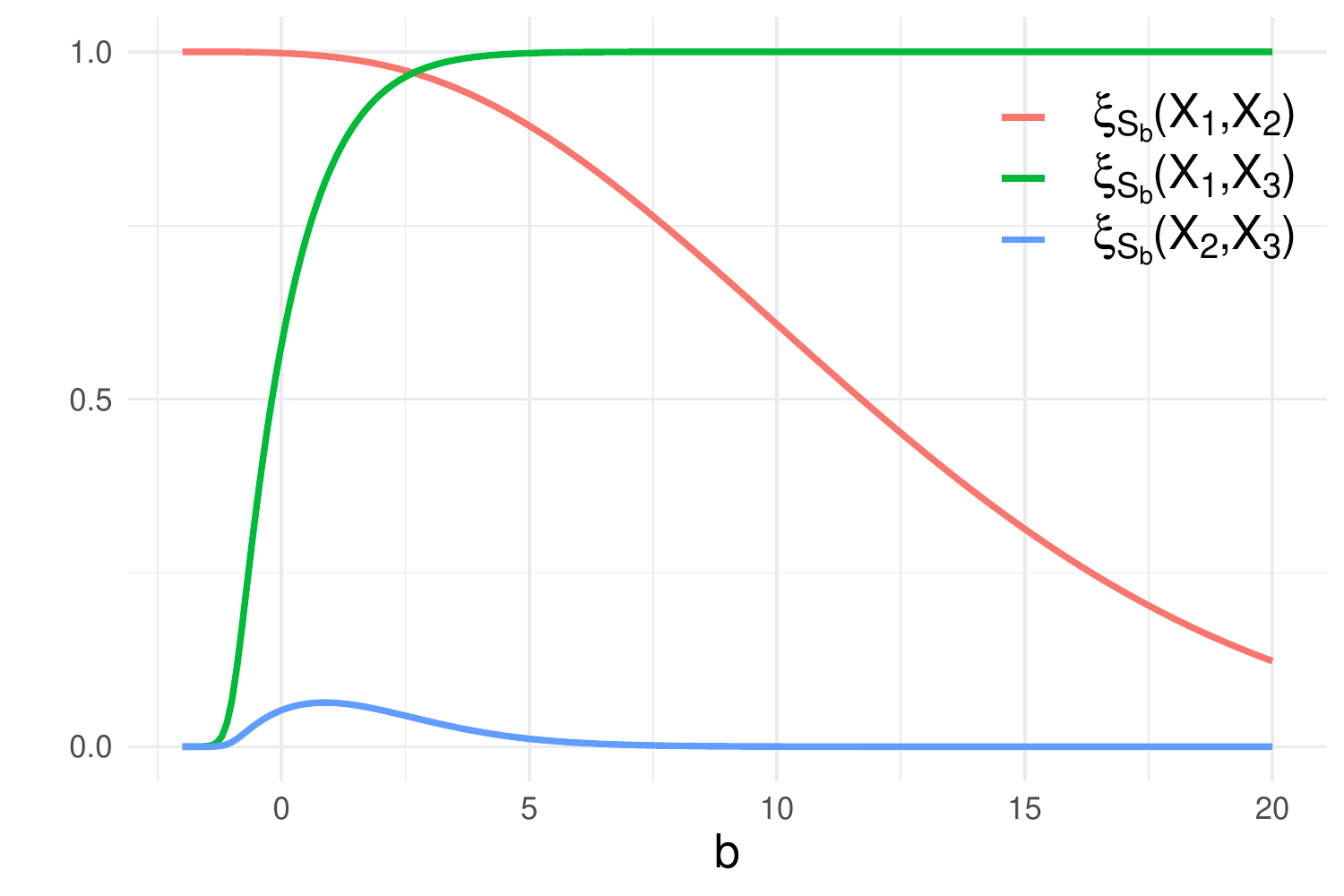}
    \hspace{0.5em}
    \includegraphics[width=0.3\textwidth]{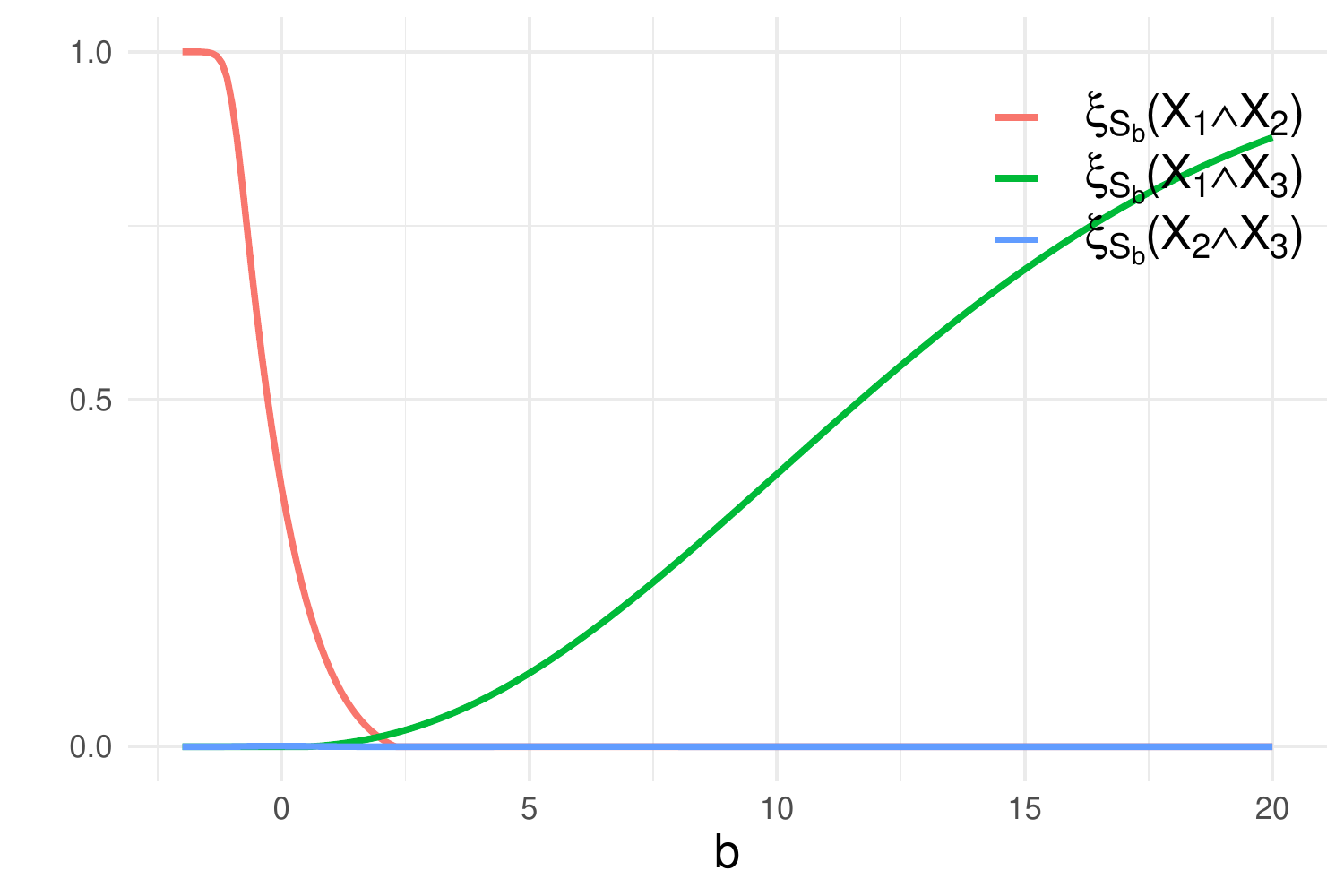}
    \\[0.5em]
    \includegraphics[width=0.3\textwidth]{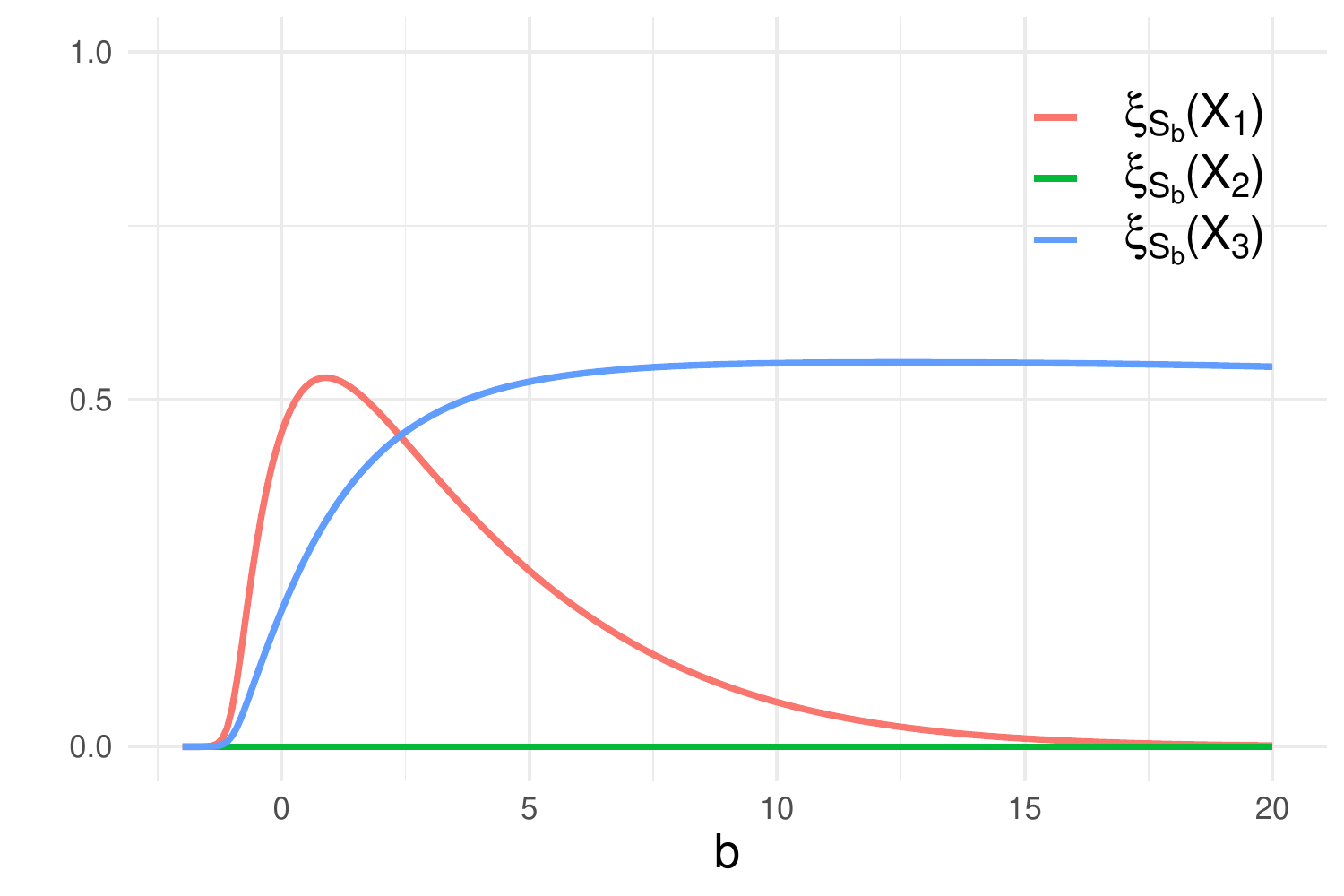}
    \hspace{0.5em}
    \includegraphics[width=0.3\textwidth]{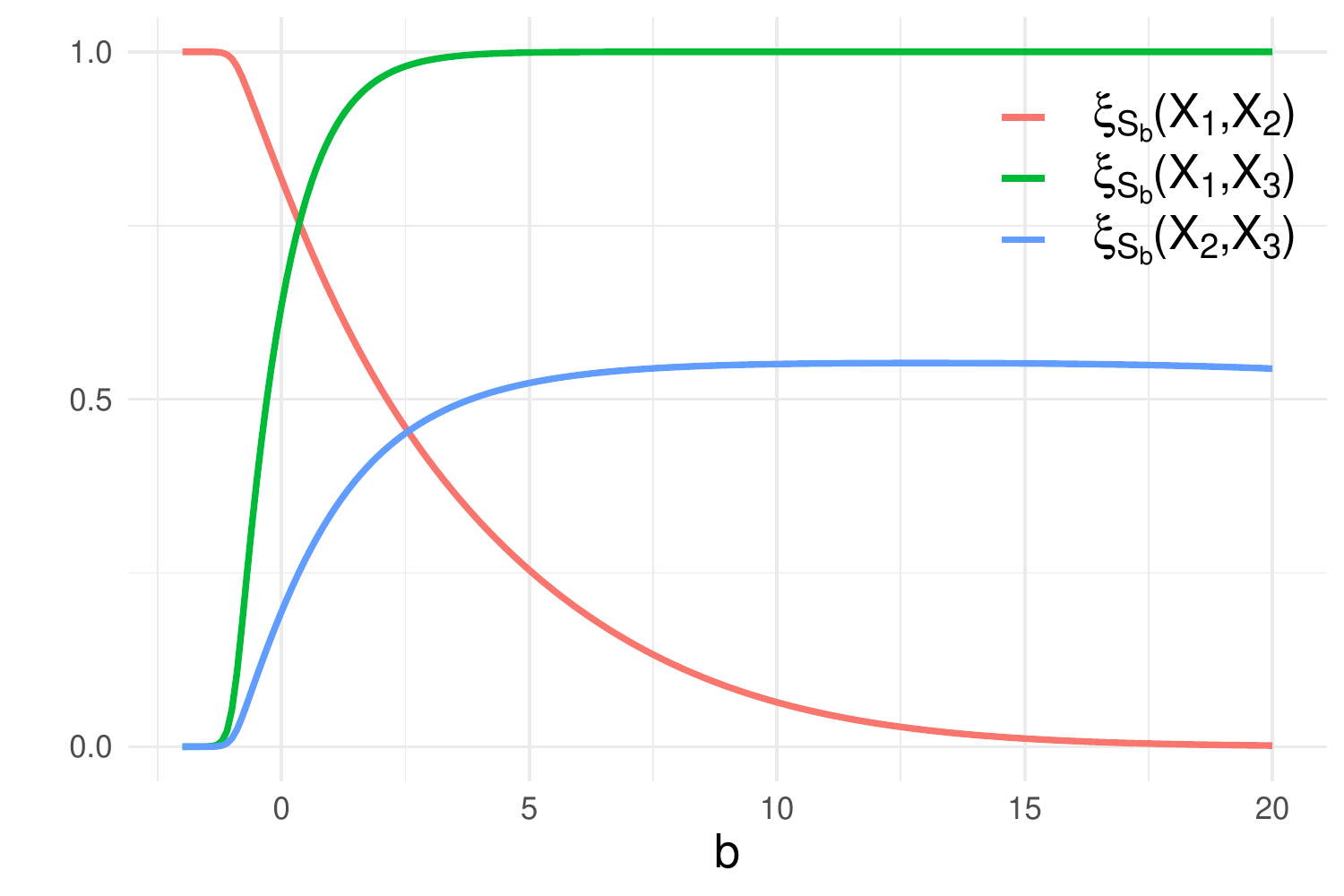}
    \hspace{0.5em}
    \includegraphics[width=0.3\textwidth]{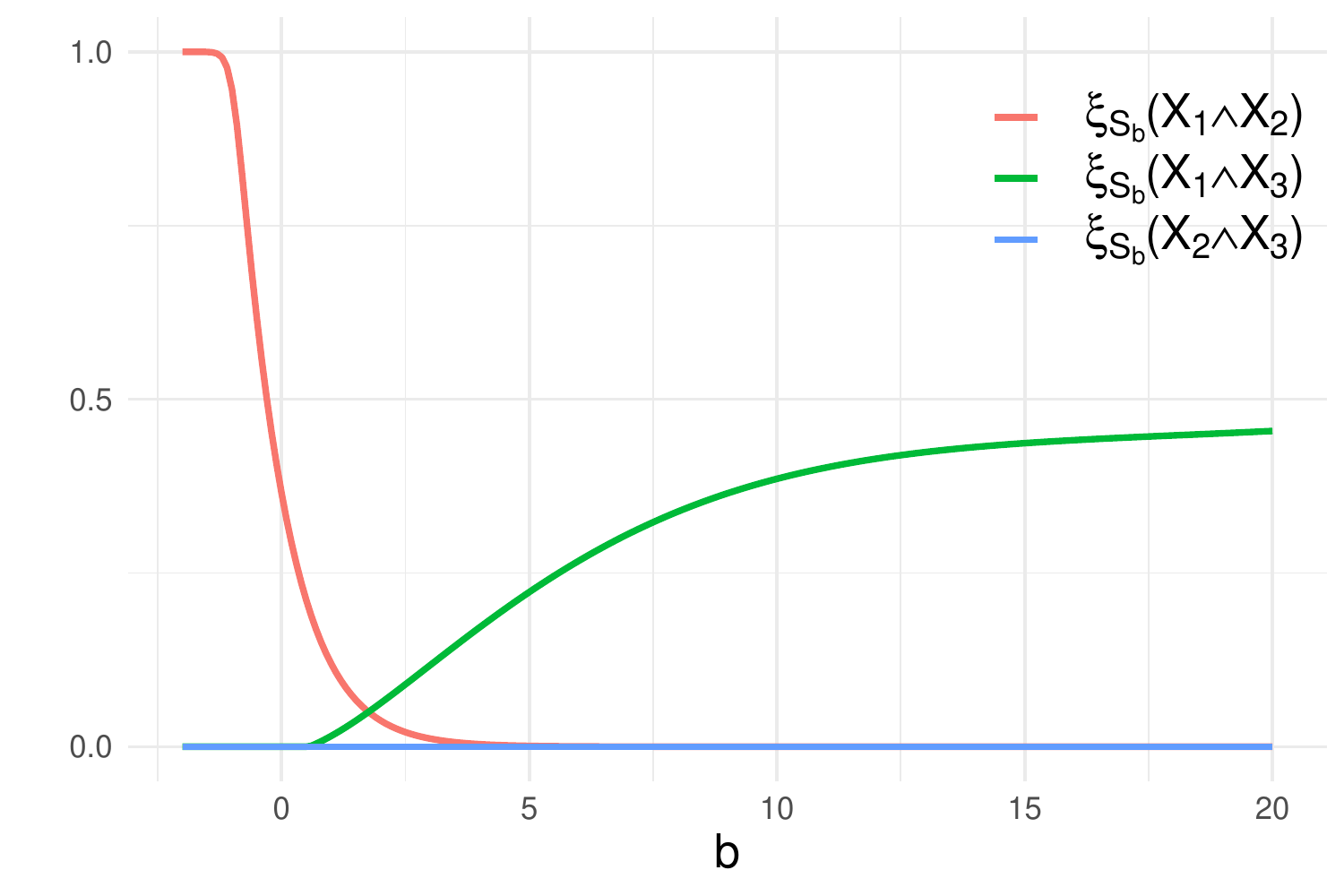}
    \caption{Homogeneous score-based sensitivities. Top panels display the mean functional and bottom panels the $\VaR_{0.9}$. The lines correspond to homogeneous score-based sensitivities to a single risk factor (left panel), to two risk factors (middle panel), and to interaction sensitivities (right panels).}
    \label{fig:murphy-mean-var-hom}
\end{figure}
\end{example}

Finally, we point out that the mixture representations of Proposition \ref{prop:mixture} hold for the corresponding elementary scores, but not for the positively homogeneous scores of Propositions \ref{prop: hom mean} and \ref{prop: hom-score-var}. Hence, while it is the case that an ordering of the sensitivities with respect to all elementary scores implies an ordering with respect to all positively homogeneous scores, the reverse does not hold.

\section{Applications}
\label{sec:applications}

In this section we illustrate the score-based sensitivities on the well-known Ishigami--Homma test function in sensitivity analysis and a non-linear insurance portfolio. 

The main challenge when calculating score-based sensitivities is the estimation of the conditional functionals $T(Y|X_i)$, for a risk factor $X_i$. For the  Ishigami--Homma test function the conditional mean functionals, i.e., the conditional expectations, are available in closed form for all risk factors of interest. Thus, estimating Murphy diagrams for elementary and homogeneous score-based sensitivities may be conducted straightforwardly using Monte Carlo approximations of the expectations. 
In the non-linear insurance portfolio, however, closed form conditional functionals are not available. Thus, we use neural nets to estimate the conditional VaR and ES.

\subsection{The Ishigami--Homma Test Function}
In this section we consider the Ishigami--Homma function given by \citep{Ishigami1990Proceedings}
\begin{equation*}
\label{eq:Ishigami--Homma}
    Y = \sin(X_1) + a_1 \sin(X_2)^2 + a_2 X_3^4 \sin(X_1)\,,
\end{equation*}
where $X_1, X_2, X_3$ are independent uniform random variables on $[-\pi, \pi]$. We consider the mean functional for comparison with the literature. 
\begin{figure}[t]
    \centering
    \includegraphics[width=0.3\textwidth]{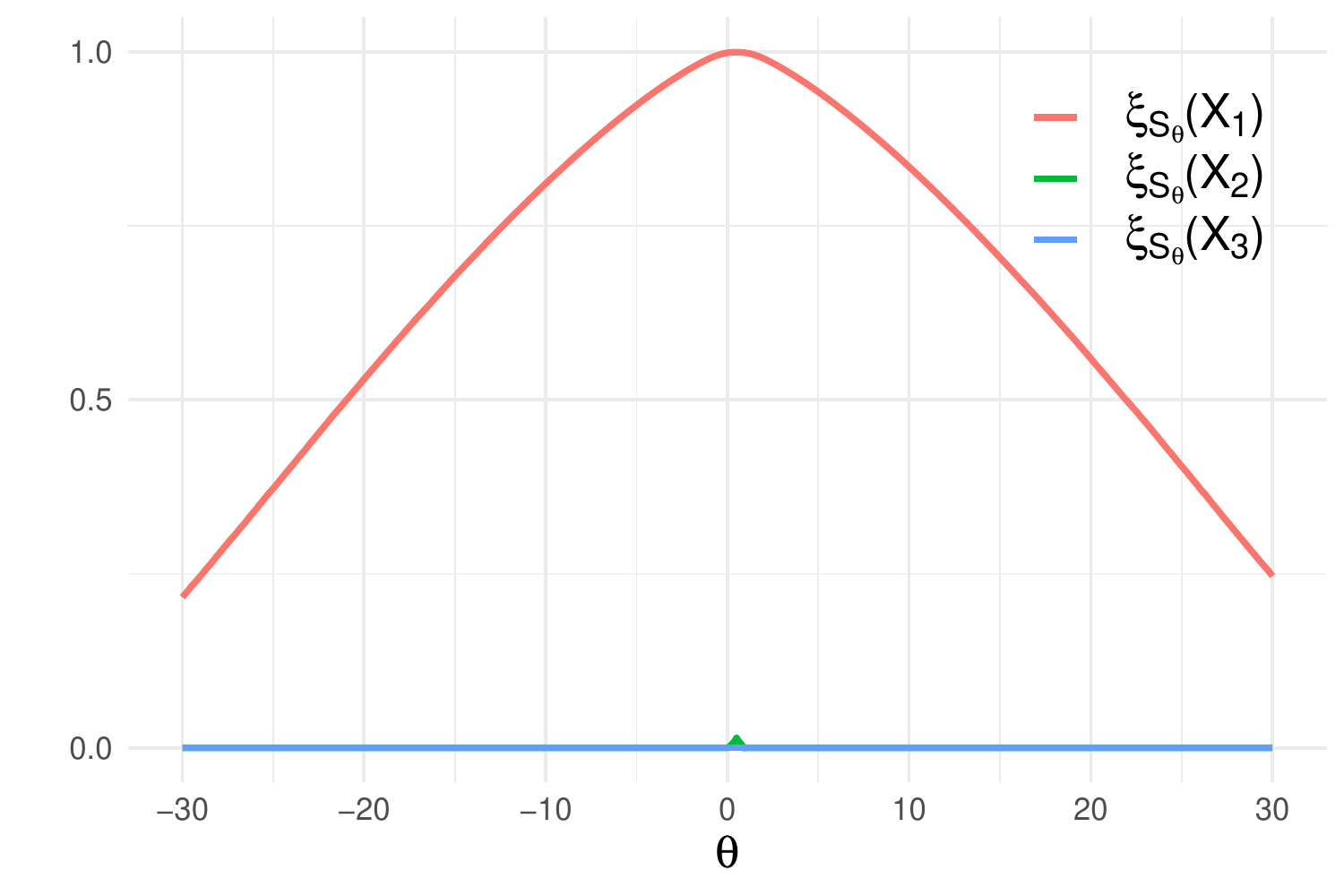}
    \hspace{0.5em}
    \includegraphics[width=0.3\textwidth]{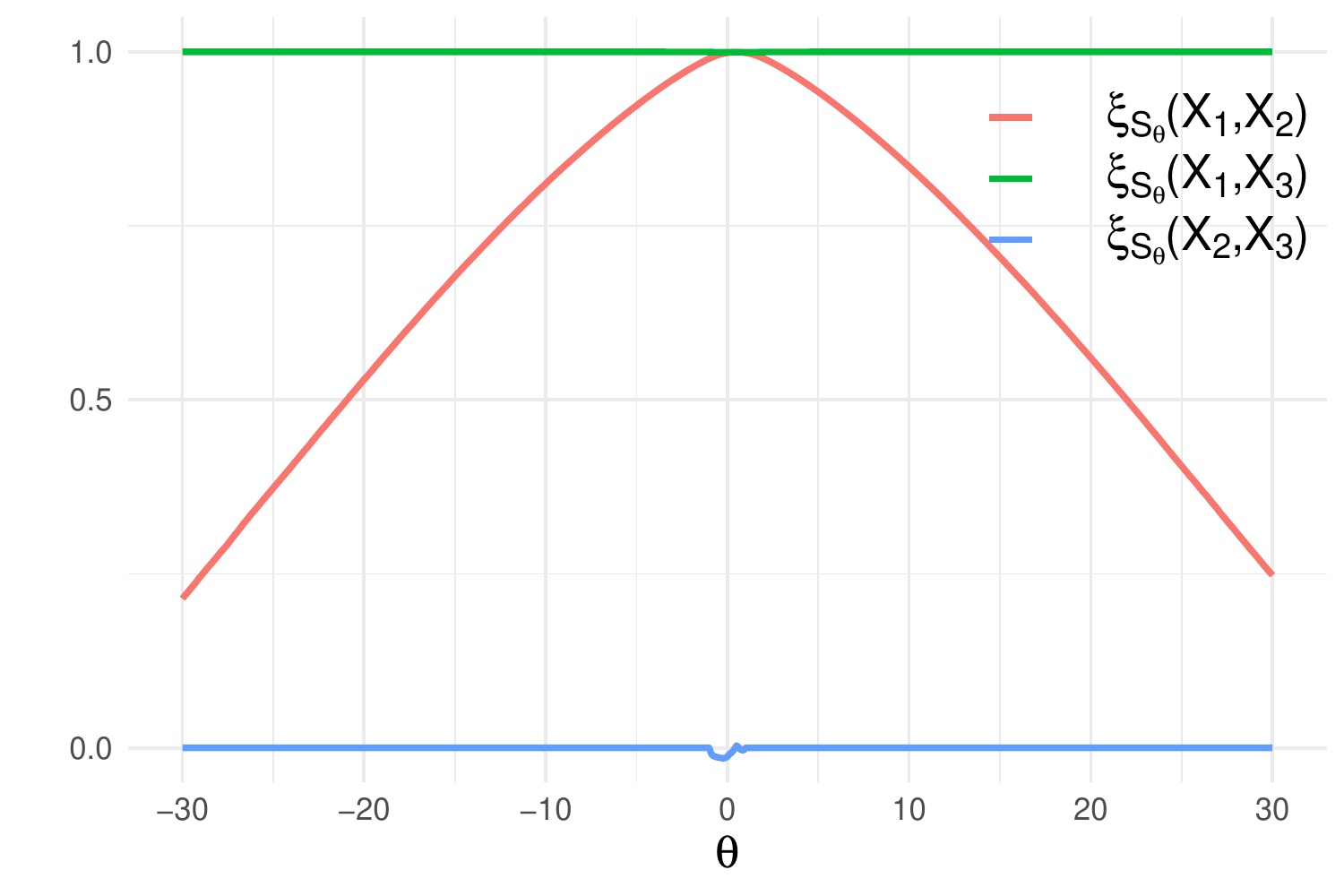}
    \hspace{0.5em}
    \includegraphics[width=0.3\textwidth]{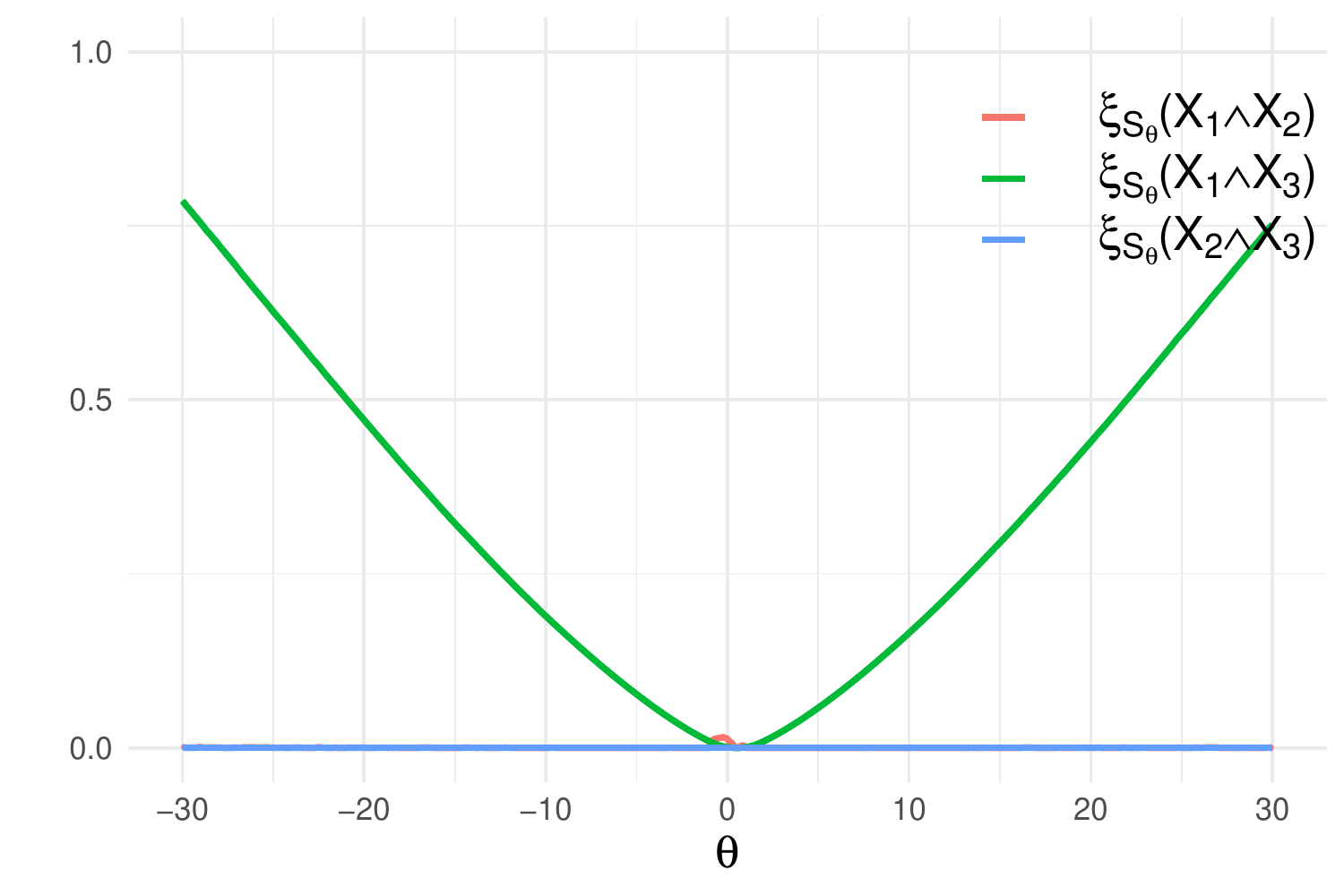}
    \\
    \includegraphics[width=0.3\textwidth]{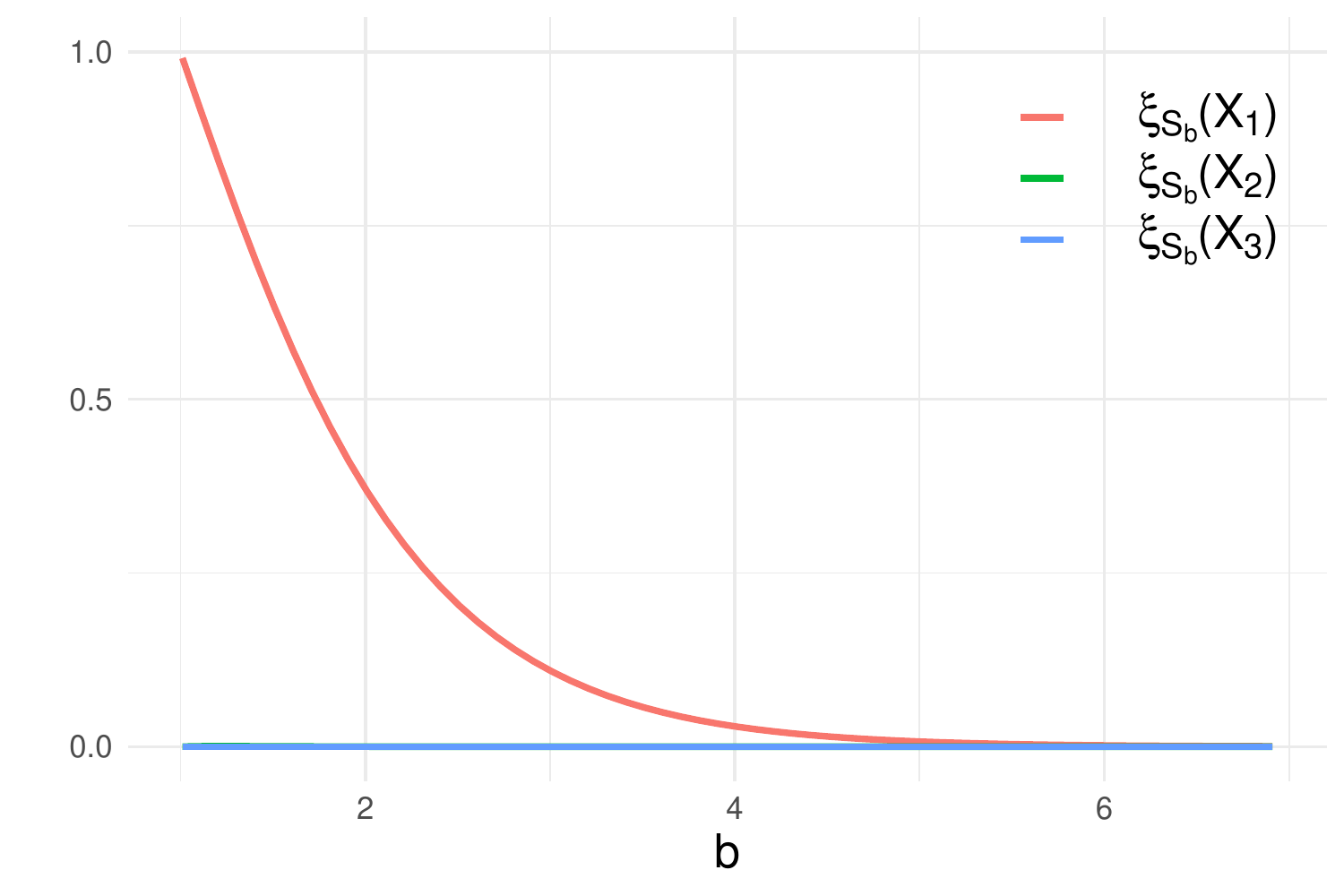}
    \hspace{0.5em}
    \includegraphics[width=0.3\textwidth]{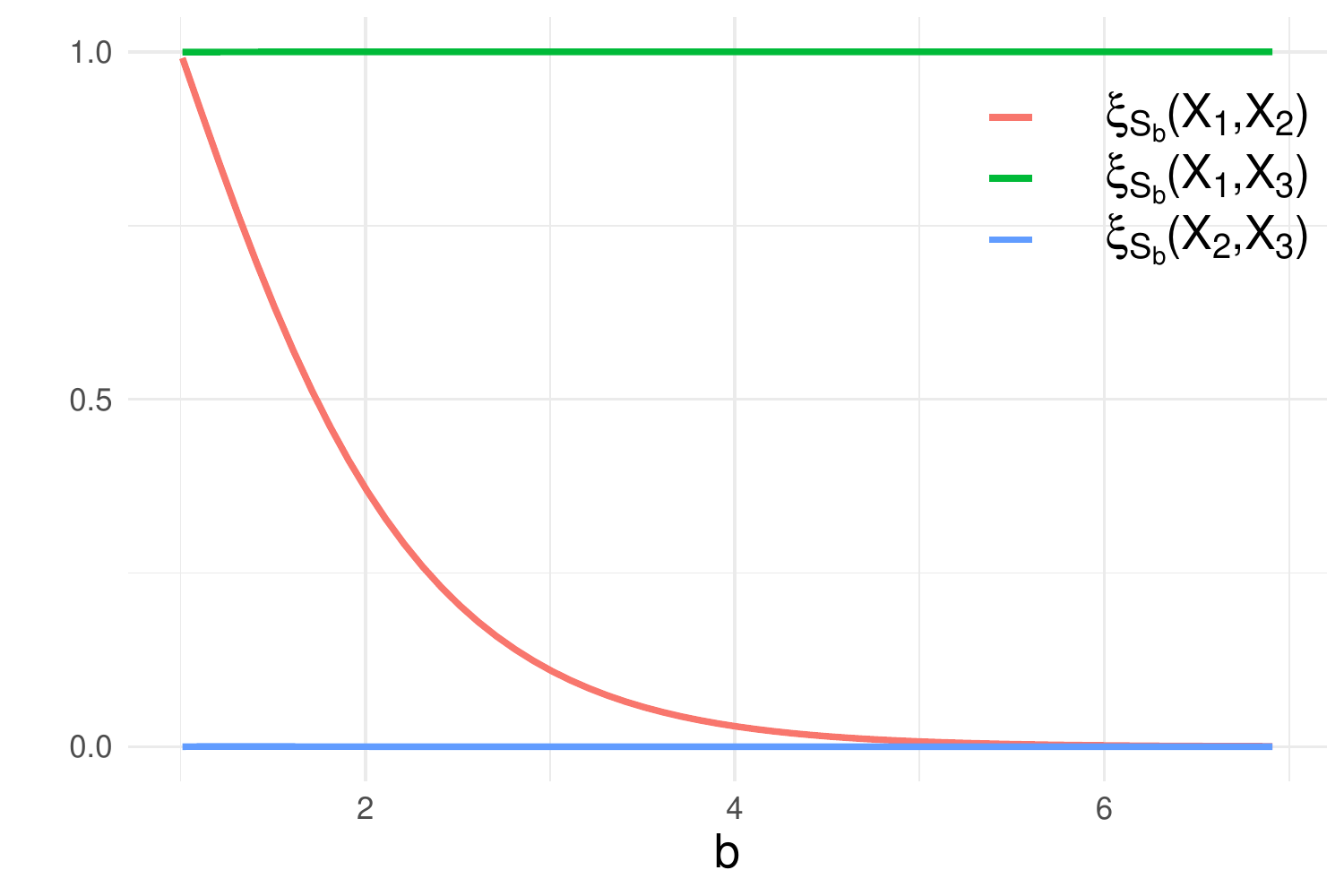}   
    \hspace{0.5em}
    \includegraphics[width=0.3\textwidth]{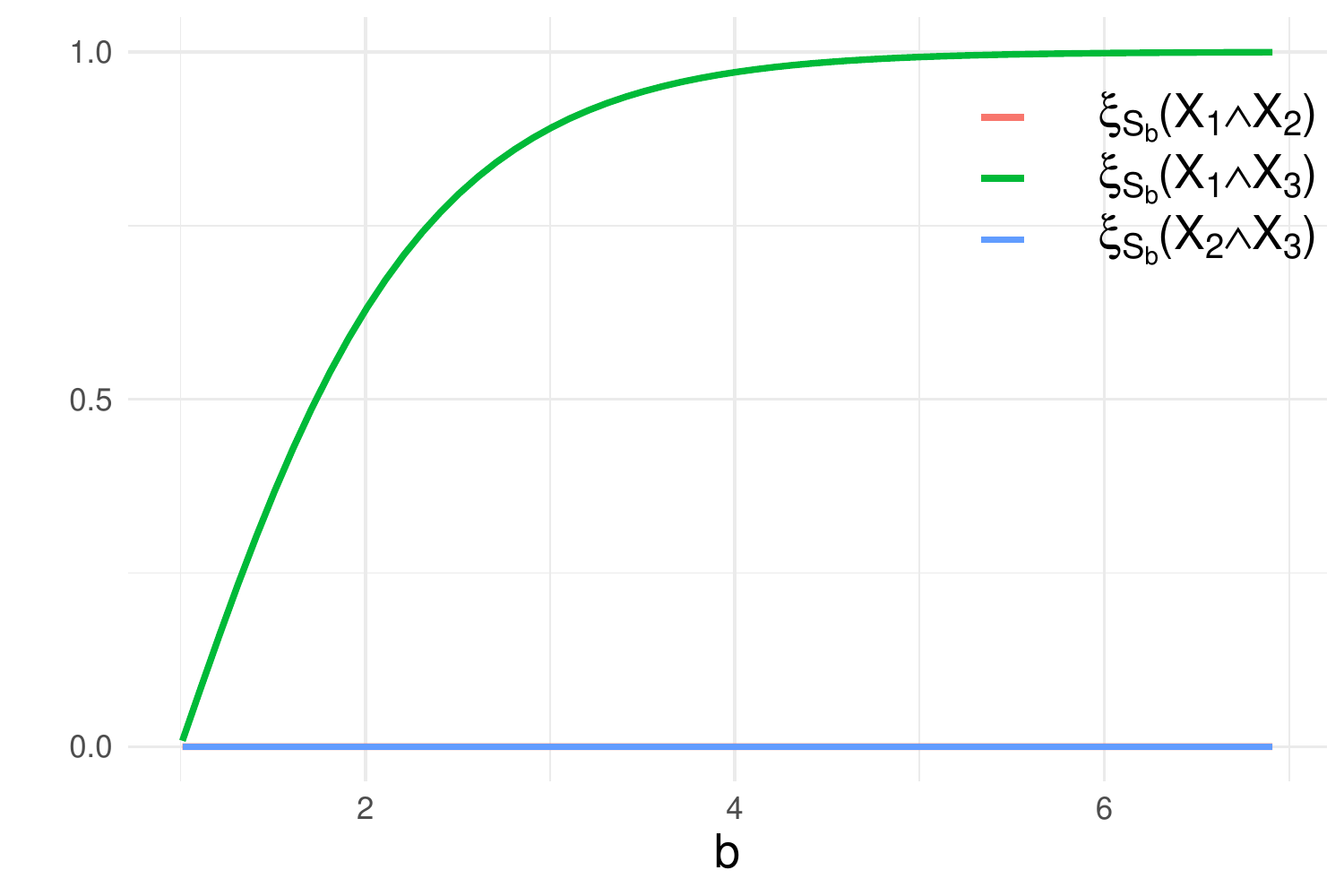}
    \caption{Top panels: Elementary (top panels) and homogeneous (bottom panels) score-based sensitivities for the mean functional and the Ishigami--Homma function with $a_1 = 1$ and $a_2 = 2$ . Left: sensitivities to a single risk factor. Middle: sensitivity to two risk facors. Right: interaction sensitivities. The sensitivity to $X_2$, i.e., $\xi_{S_\theta}(Y;X_2)$ and $\xi_{S_b}(Y;X_2)$, are numerically equal to the blue line, that is 0.}
    \label{fig:Ishigami-mean}
\end{figure}

Figure \ref{fig:Ishigami-mean} displays the score-based Murphy diagrams for the mean functional and for the choices $a_1 = 1$ and $a_2 = 2$. 
The top panels correspond to the Murphy diagrams using the elementary scoring functions and the bottom panels correspond to homogeneous scoring functions. 
Both the homogeneous and elementary Murphy diagrams describe a similar pattern. Indeed, we observe that risk factor $X_1$ is most influential and $X_2$ and $X_3$ are non-influential with a sensitivity for all elementary and homogeneous scores close to zero. 
The interaction term of $X_1$ and $X_3$, however, is equal to 1 for all scoring functions, reflecting the findings in \cite{Saltelli2008Wiley}. 
Note that the sensitivity for the homogeneous score with $b = 2$ are equal to the Sobol indices.
We obtain that the Sobol indices of $Y$ to $X_1$, $\xi_{S_2}(Y;X_1)$, is equal to $0.37$.
Similarly, $\xi_{S_2}(Y;X_2 )= 5.7 \times 10^{-5}$ and $\xi_{S_2}(Y;X_3 )=0$. We refer to \cite{Saltelli2008Wiley}, Equation (4.34), for the analytical derivation of the Sobol indices of the Ishigami--Homma function, and  to \cite{Pianosi2015EMS} and \cite{Baroni2020EMS} for a recent discussion on the Ishigami--Homma function.

\begin{figure}
    \centering
    \includegraphics[width=0.3\textwidth]{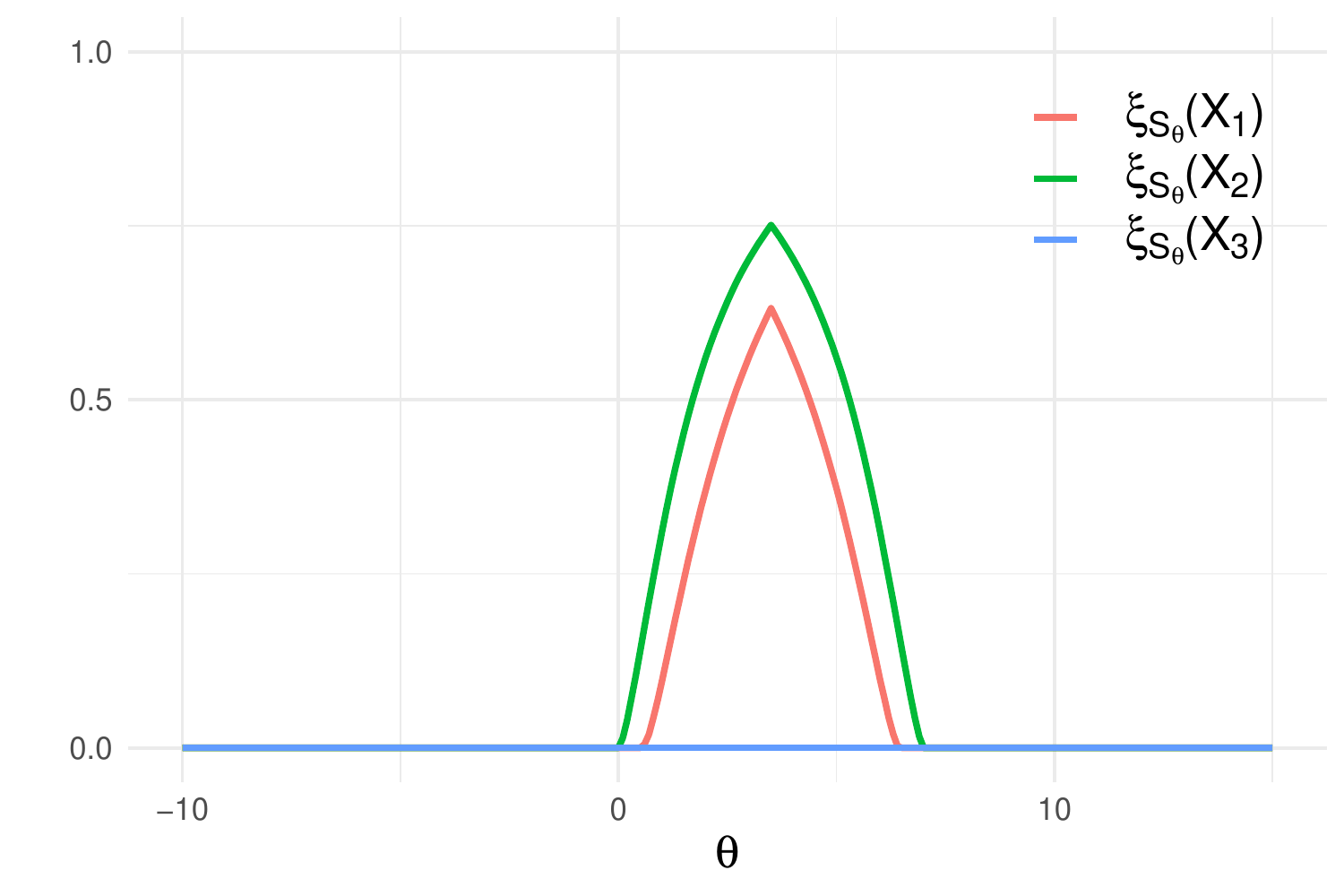}
    \hspace{0.5em}
    \includegraphics[width=0.3\textwidth]{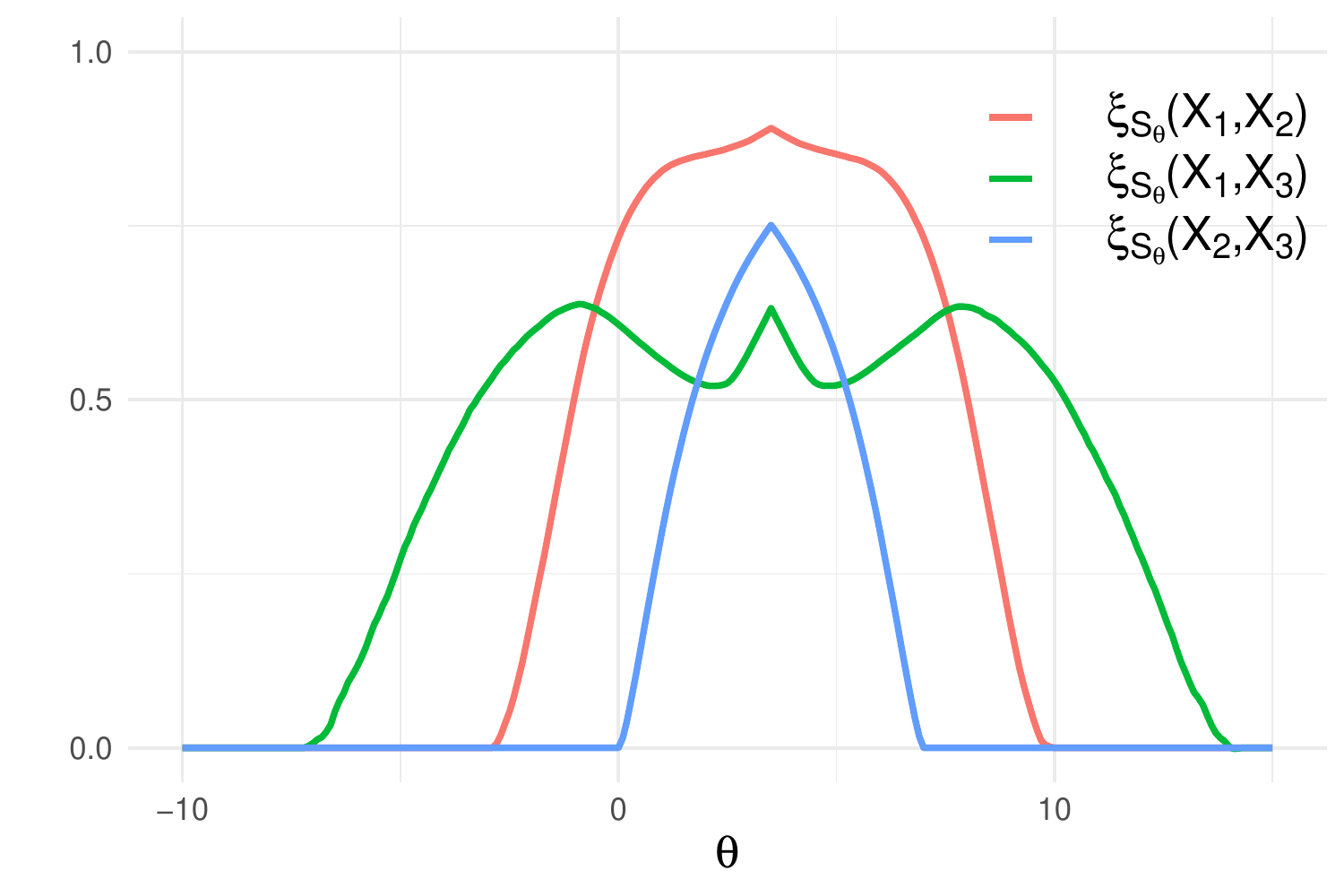}
    \hspace{0.5em}
    \includegraphics[width=0.3\textwidth]{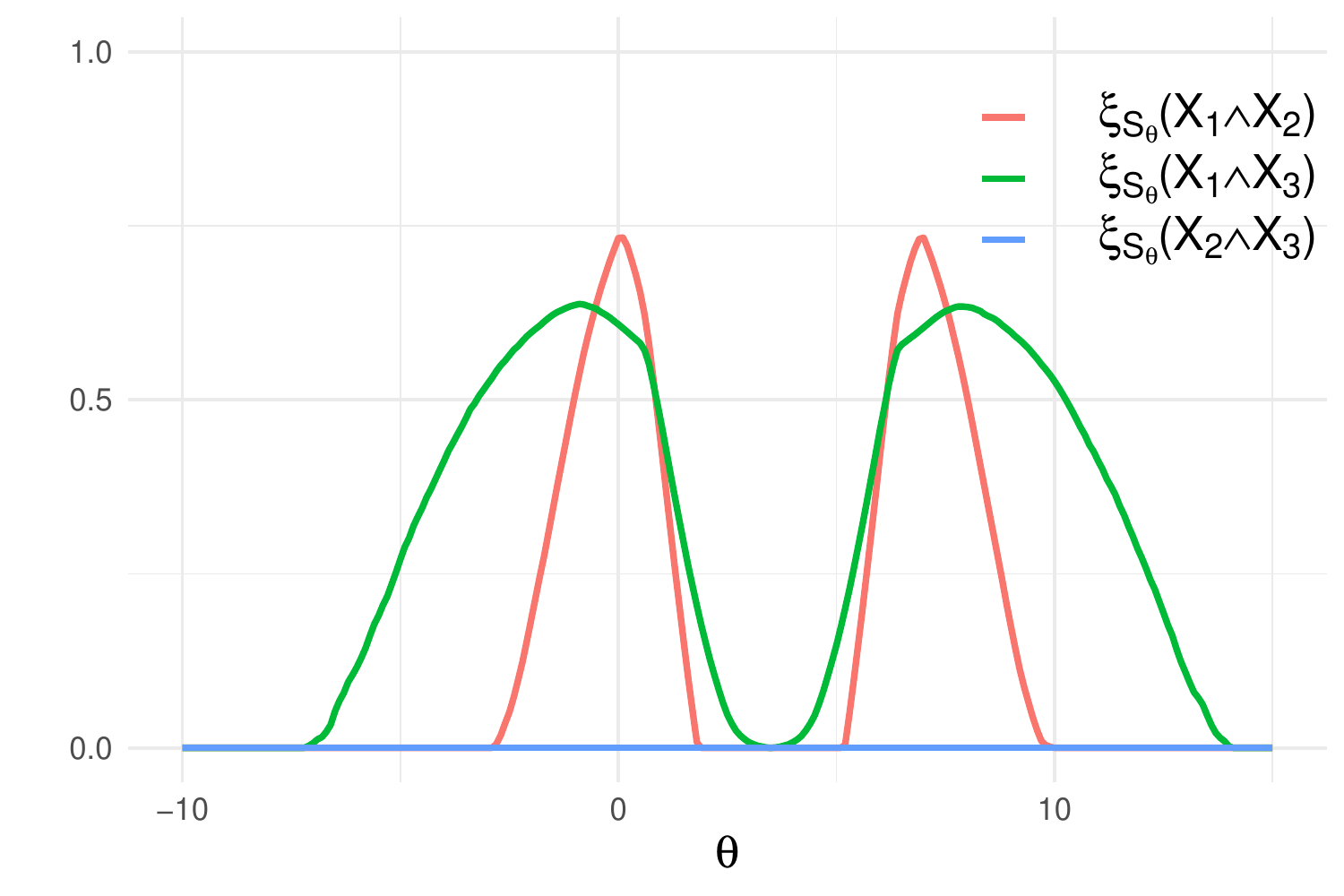}
    \\[0.5em]
    \includegraphics[width=0.3\textwidth]{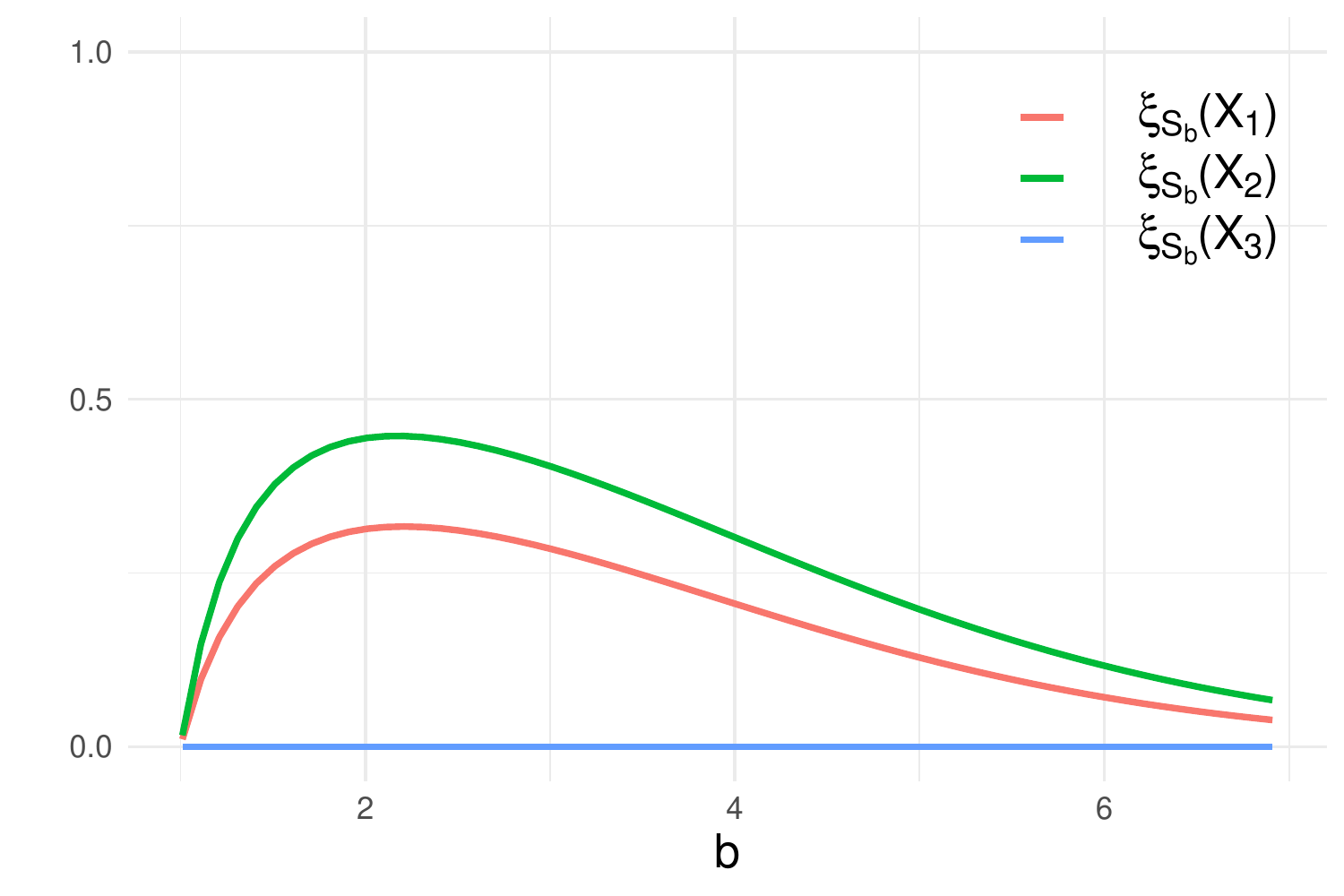}    \includegraphics[width=0.3\textwidth]{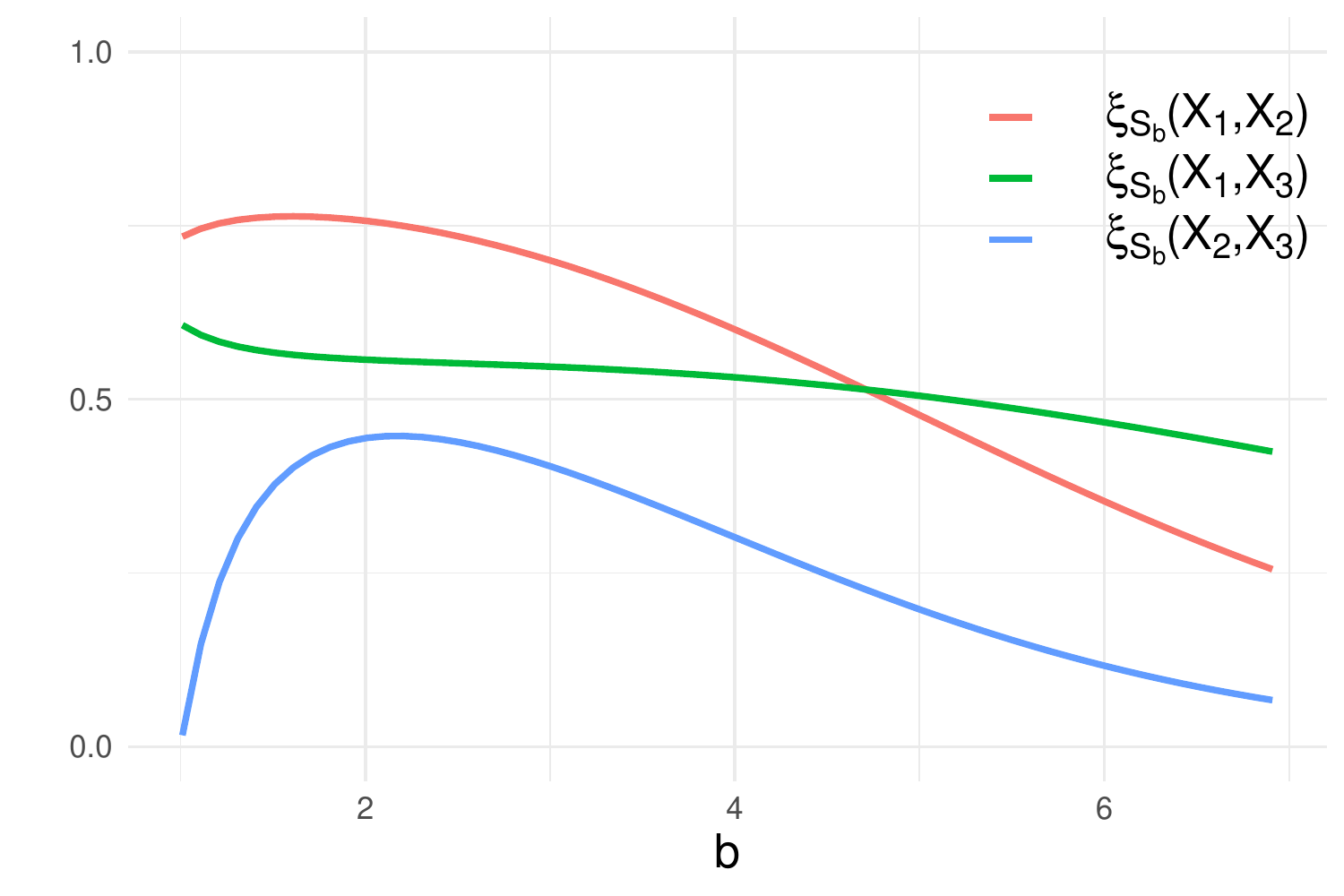}    \includegraphics[width=0.3\textwidth]{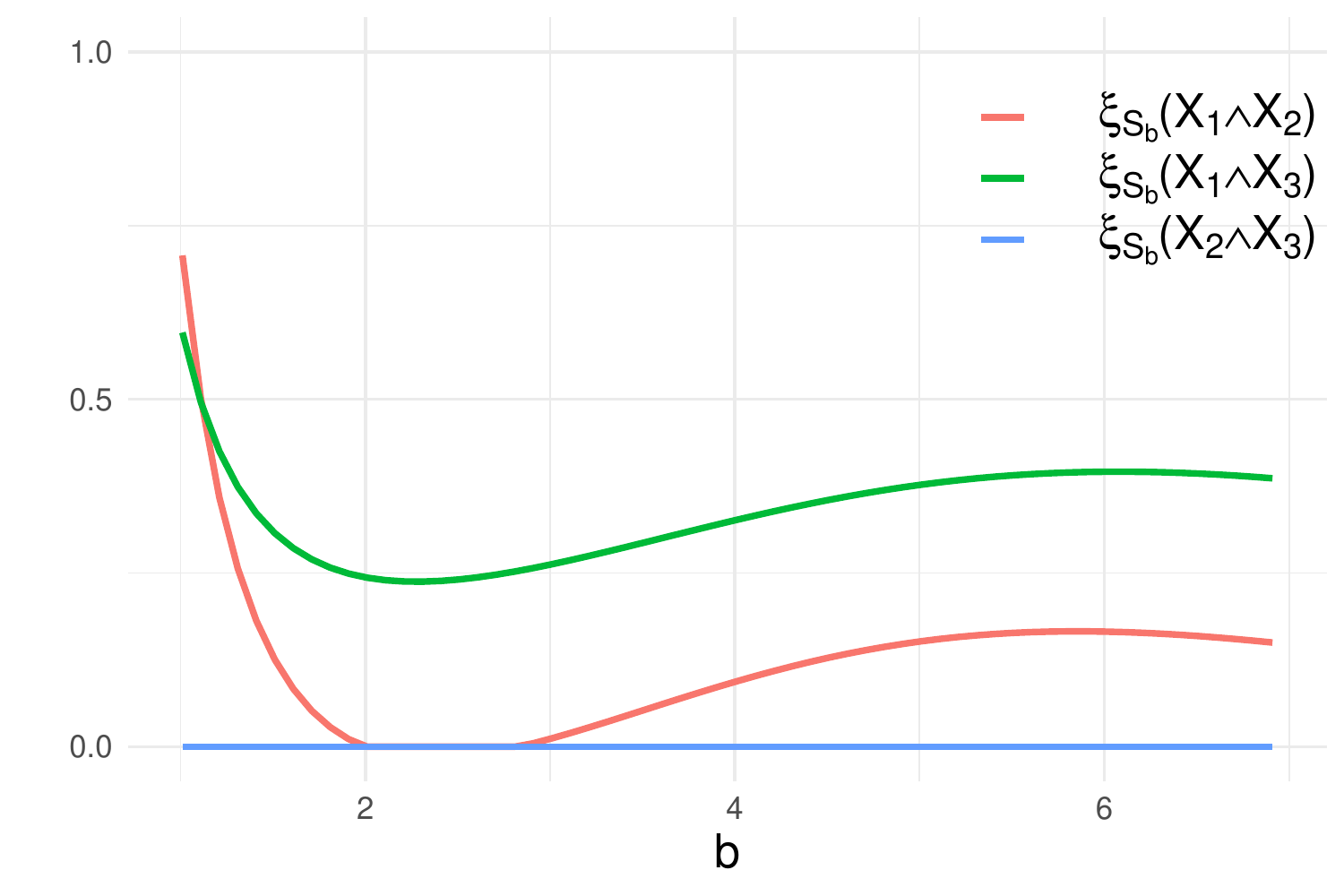}
    \caption{Top panels: Elementary (top panels) and homogeneous (bottom panels) score-based sensitivities for the mean functional and the Ishigami--Homma function with $a_1 = 7$ and $a_2 = 0.1$. Left: sensitivities to a single risk factor. Middle: sensitivity to two risk factors. Right: interaction sensitivities.}
    \label{fig:Ishigami-mean-a1_7}
\end{figure}

Next, we consider the Ishigami--Homma function with parameters $a_1 = 7$ and $a_2 = 0.1$ as in \cite{Marrel2009RESS}. The elementary and homogeneous score-based sensitivities for the mean functional are displayed in Figure \ref{fig:Ishigami-mean-a1_7}. We observe that for these choices of $a_1$ and $a_2$, the sensitivity to $X_2$ is larger than the sensitivity to $X_1$ for all strictly consistent scoring functions, see left panels in Figure \ref{fig:Ishigami-mean-a1_7}. 
The estimated Sobol indices for $X_1$, $X_2$, and $X_3$ are $0.31$, $0.44$, and $0$, respectively. 
The fact that the sensitivity to $X_2$ is larger than that to $X_1$ is in contrast to the choice $a_1 = 1$ and $a_2 = 2$, for which the sensitivity to $X_2$ is negligible. 
The sensitivity to $X_3$ is zero for both sets of parameters $a_1$ and $a_2$. 
For $a_1 = 7$ and $a_2 = 0.1$, the sensitivity to knowing two components, i.e. $(X_1, X_2)$, $(X_1, X_3)$, or $(X_2, X_3)$, strongly depends on the choice of the scoring function, as depicted in the right panels of Figure \ref{fig:Ishigami-mean-a1_7}.

It is well-known that for the Sobol indices, the interaction of $X_1$ and $X_3$ is non-zero, see Figures \ref{fig:Ishigami-mean} and  \ref{fig:Ishigami-mean-a1_7} bottom right panel. 
Interestingly, for the choice $a_1 = 1$ and $a_2 = 2$, this is the only interaction which is non-zero for all elementary scoring functions, see Figure \ref{fig:Ishigami-mean}. 
For homogeneous scores, however, the interaction between $X_1$ and $X_2$ becomes positive for large and small values of $b$. 
Thus, considering Murphy diagrams reveals a more wholesome picture on the individual sensitivities and their interactions.

\subsection{Non-linear Insurance Portfolio}

In this section we consider an insurance company with three different lines of business whose losses are  $X_1, X_2$, and $X_3$ that are subject to a multiplicative factor $X_4$, e.g., inflation. The insurance company has a reinsurance contract on the first two lines of business, $L = X_4(X_1 + X_2)$, with deductible $d$ and limit $l$. Thus, the insurance company's total loss is given by
\begin{equation*}\label{eq: total loss}
 Y = L - \min \{ (L - d)_+ , \;l\}  + X_3 X_4\,,
\end{equation*}
where $(x)_+ = \max\{x, 0\}$ denotes the positive part. 
The distributional assumptions are presented in Table \ref{tab: ex: insurance asm} and we set the deductible to $d = 380$ and the limit to $l = 30$. Furthermore, we assume the factors $(X_1, X_2, X_3, X_4)$ are dependent through a Gaussian copula with correlation matrix 
\begin{equation*}
    R = 
    \left(
    \quad
    \begin{array}{@{} S[table-format=1.3]S[table-format=1.3]S[table-format=1.3]S[table-format=-1.2] @{}}
    1  & 0.3 & 0 & 0.8\\
    0.3 & 1 & 0 & 0 \\
    0 & 0 & 1 & 0 \\
    0.8 & 0 & 0 & 1 \\
    \end{array}
    \quad
    \right)\,.
\end{equation*}

\begin{table}[t]
\centering
\caption{Distributional assumptions of risk factors of the non-linear insurance portfolio example.}\label{tab: ex: insurance asm}
\begin{tabular}{ c @{\hspace{2.5em}}  c @{\hspace{2em}} c c}
Risk factor & Distribution  & Mean & Std \\
\toprule 
\toprule 
$X_1$ &   Log-Normal$\, \left(4.98, ~ 0.23^2\right)$ & \multicolumn{1}{d{3.2}}{150} &  \multicolumn{1}{d{3.2}}{35.0}\\[0.5em]
$X_2$ &   Log-Normal$\, \left(4.98, ~ 0.23^2\right)$ & \multicolumn{1}{d{3.2}}{150} &   \multicolumn{1}{d{3.2}}{35.0}\\[0.5em]
$X_3$ &   Gamma$\, \left(100,~ 1\right)$ &	\multicolumn{1}{d{3.2}}{100} &   \multicolumn{1}{d{3.2}}{10.0}\\[0.5em]
$X_4$ &   Log-Normal$\, \left(-0.005, ~0.1^2\right)$ & \multicolumn{1}{d{3.2}}{1} &   \multicolumn{1}{d{3.2}}{0.1}\\[0.5em]
\bottomrule
\bottomrule
\end{tabular}
\end{table}
First, we calculate the score-based sensitivity for the $\VaR_\alpha$ at level $\alpha\in (0,1)$. For simplicity, we choose the pinball loss 
\begin{equation}
\label{eq:insurance-score}
    S_\VaR(z,y) = \big(\Id_{\{y\le z\}} - \alpha \big)\big(z - y\big)\,, \quad y,z \in \R\,,
\end{equation}
obtained with $g(x) = x$ in \eqref{eq:GPL}.

We calculate the score-based sensitivities to risk factors $\X_{\I}$, where $\I\subset\{1,2,3,4\}$ and $|\I|\in\{1,2\}$ is an index set of cardinality one or two such that $\X_{\I}$ is an at most two-dimensional subvector of $(X_1, \ldots, X_4)$. For this, we first calculate the conditional $\VaR_\alpha$ of the aggregate portfolio loss $Y$ given $\X_{\I}$, i.e., $T(Y|\X_{\I}) = \VaR_\alpha(Y|\X_{\I})$. Denote by $q_{\I}(\x_{\I}) = \VaR_\alpha(Y|\X_{\I} = \x_{\I})$, $\x_{\I}\in \R^{|\I|}$, the conditional quantile of $Y$ given $\X_{\I} = \x_{\I}$ viewed, for fixed $\alpha$, as a function of $\x_{\I}\in \R^{|\I|}$.
For each choice of $\I$, we approximate the function $q_\I$ via a neural network (NN) denoted by $G_{\vartheta_\I}\colon \R^{|\I|}\to\R$, where $\vartheta_\I$ are the network parameters. For relevant literature on estimating conditional VaR via NN -- albeit in a different context -- we refer the reader to \cite{Fissler2021SSRN}.
The network parameters $\vartheta_\I$ are learnt by iteratively minimising the expected score \eqref{eq:insurance-score}
\begin{equation}\label{eq:insurance-nn-param}
    \widehat \vartheta_\I = \argmin_\vartheta \;\frac{1}{N} \sum_{k = 1}^N S_\VaR\left(G_{\vartheta}(\x_{\I}^{(k)}), y^{(k)}\right)\,,
\end{equation}
over independently simulated mini-batches $(\x_\I^{(k)},y^{(k)})$, $k = 1, \ldots, N$. That is, for each iteration we independently simulate a mini-batch of $N$ i.i.d. samples of $(\X_\I, Y)$. We denote the learnt NN by $G_{\widehat{\vartheta}_\I}$.
After learning the NN, we estimate the sensitivity to $\X_{\I}$ out-of-sample, that is, using an independent sample $(\x_\I^{(l)},y^{(l)})$, $l = 1 \ldots, M$, of $(\X_\I, Y)$, via
\begin{equation}
\label{eq:insurance-sensitivity-estimate}
\widehat{\xi}_{S_\VaR} (Y;\X_{\I}) = 1 - 
\frac{\displaystyle\sum_{l = 1}^M S_\VaR\left(G_{\widehat \vartheta_\I}(\x_{\I}^{(l)}), y^{(l)}\right)}{\displaystyle\sum_{l = 1}^M S_\VaR\left(\widehat{\VaR}_{\alpha}(Y), y^{(l)}\right)}\,,
\end{equation}
where $\widehat{\VaR}_{\alpha}(Y)$ is the sample quantile of $Y$.

\begin{figure}[t]
    \centering
    \includegraphics[width = 0.78\textwidth]{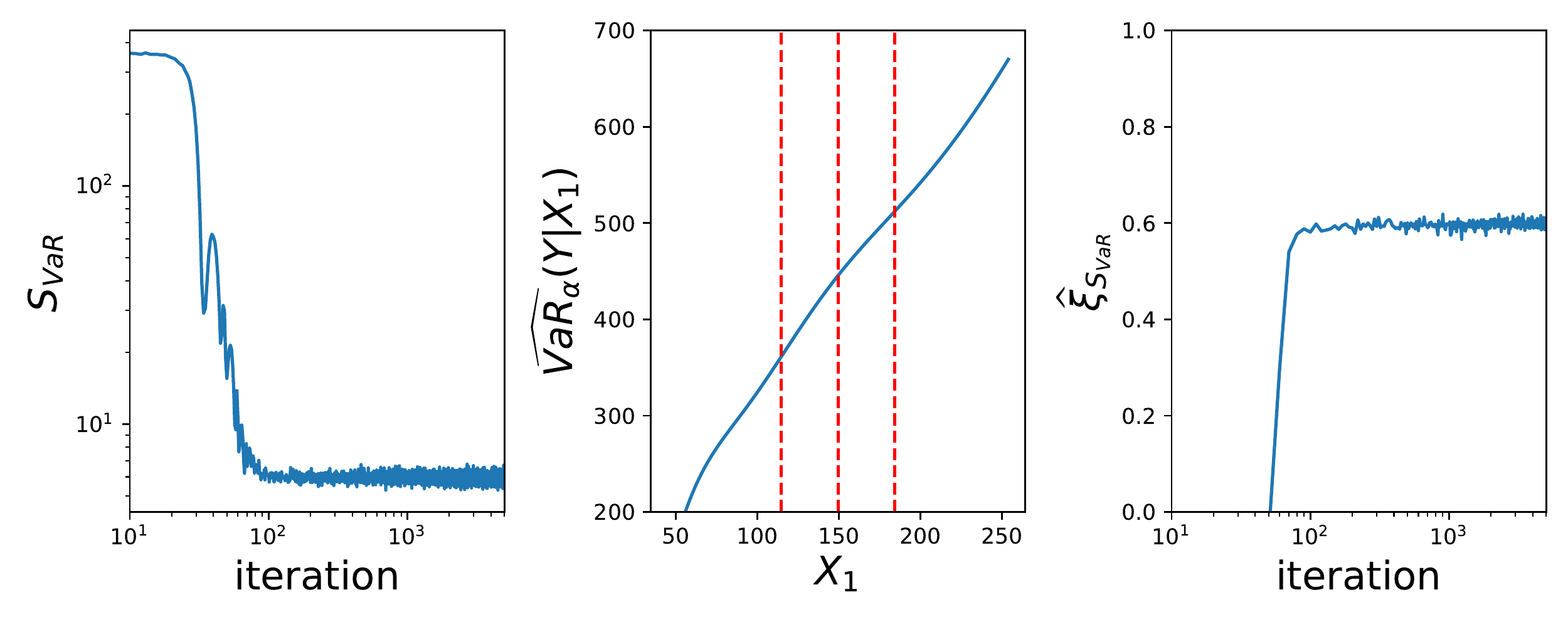}
    \\
    \includegraphics[width = 0.78\textwidth]{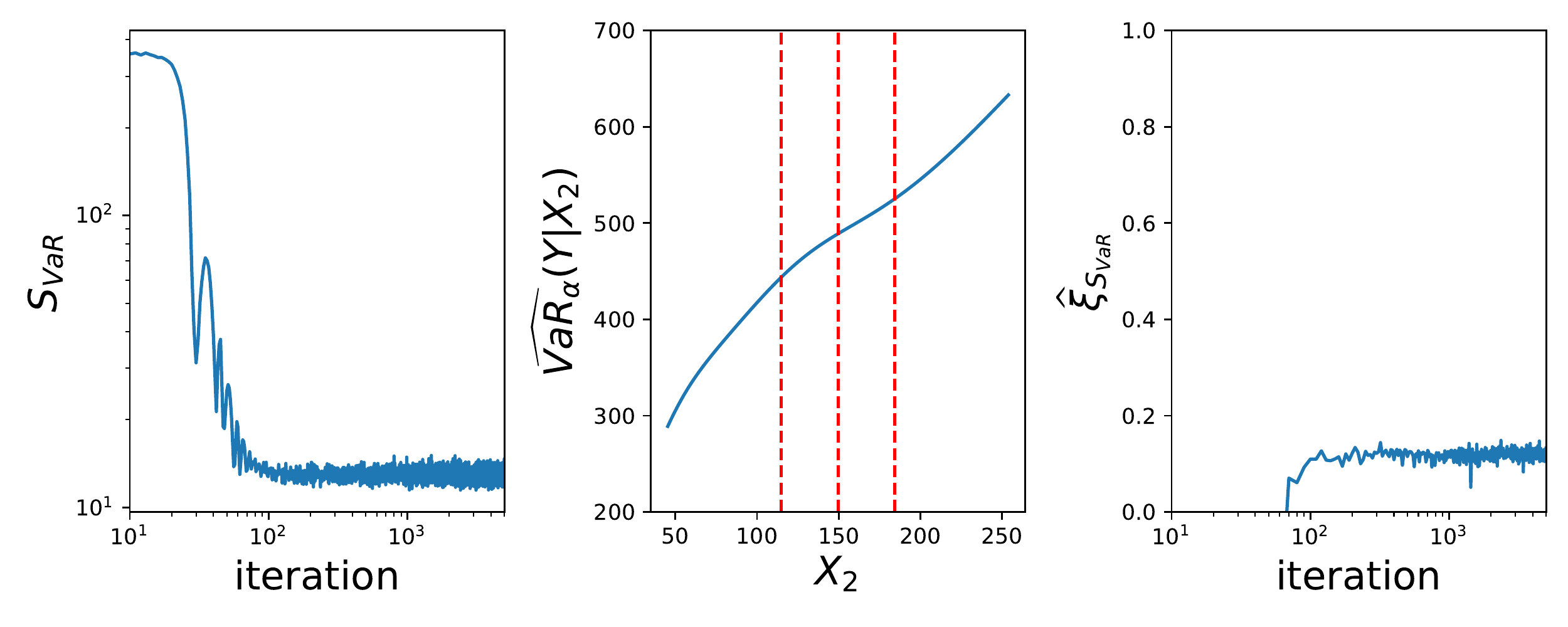}
    \\
    \includegraphics[width = 0.78\textwidth]{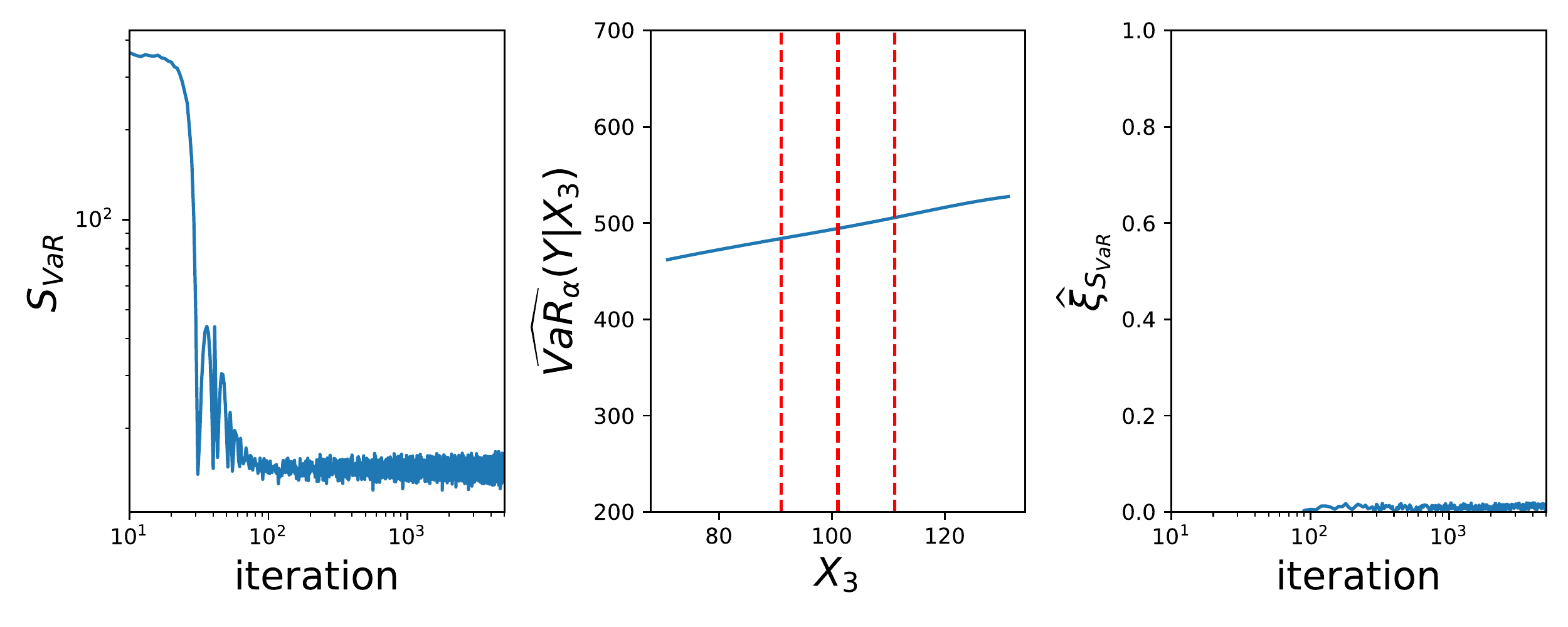}
    \\
    \includegraphics[width = 0.78\textwidth]{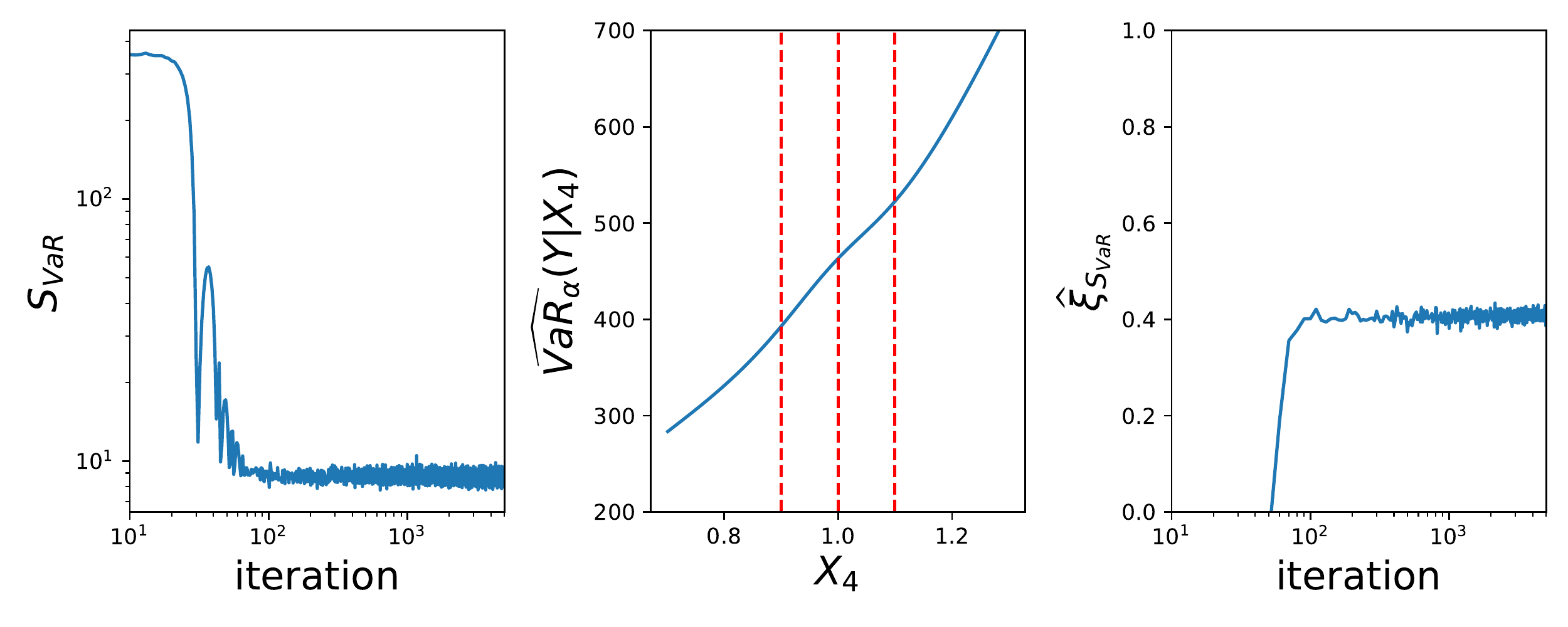}
    \caption{Score-based sensitivities for $\VaR_{0.9}$. Left: scoring function by iteration of learning the model. Middle: $\VaR_{0.9}(Y|X_i)$ of the learnt NN. Right: score-based sensitivities by iteration of learning the model.}
    \label{fig:ex-Insurance-VaR-single}
\end{figure}

For each choice of $\I$, the NN structure consists of 6 hidden layers with 20 neurons per layer and SiLU activation functions apart from the last layer. The networks are learnt via independently simulated mini-batches of size $N = 10^3$, i.e. every single iteration uses an independently simulated i.i.d. samples of $(\X_\I,Y)$ of size $N = 10^3$. Figure \ref{fig:ex-Insurance-VaR-single} displays the learning of the NN and the score-based sensitivities to single risk factors from $X_1$ (top row) to $X_4$ (bottom row), with $\alpha=0.9$. The left panel displays the average score for each mini-batch (Equation \eqref{eq:insurance-nn-param} with $N = 10^3$ per mini-batch) against the iterations of learning the NN. 
The right panel displays the out-of-sample estimates of the score-based sensitivity (Equation \eqref{eq:insurance-sensitivity-estimate} with $M = 10^4$) against the iterations of learning the model. 
The middle panels display the learnt NN, $G_{\widehat{\vartheta}_{i}} $, that is an estimate of $\VaR_{0.9}(Y | X_i)$ as a function of $X_i$, for $i= 1, \ldots, 4$ from top to bottom. 
The red dotted lines correspond to the mean of $X_i$ and the first standard deviations from its mean. 
We observe that risk factor $X_1$ has the largest sensitivity. 
The sensitivity to $X_1$ is significantly larger than that to $X_2$, even though $X_1$ and $X_2$ have the same distribution and the aggregate portfolio loss $Y$ is symmetric in $X_1$ and $X_2$. Thus, the difference in the sensitivities solely stems from the dependence structure of the portfolio; recall that $X_2$ is independent of $X_4$ while $X_1$ is highly correlated with $X_4$. Moreover, the sensitivity to $X_2$ (the independent business line) is negligible, while, the sensitivity to $X_4$, the multiplicative factor, is the second largest sensitivity.

Next, we estimate the score-based sensitivities  to two risk factors, i.e., $\xi_{S_{\VaR_{0.9}}}(Y;\X_{\I})$, for $|\I|=2$. We proceed similarly to estimating the score-based sensitivities to one risk factor in that we learn for each  $\I = \{i,j\}$ with $i \neq j$, a NN using Equation \eqref{eq:insurance-nn-param} and independently simulated mini-batches. The corresponding sensitivities to $\X_{\I}$ are estimated using the learnt NN out-of-sample via Equation \eqref{eq:insurance-sensitivity-estimate} with $M = 10^4$. The estimates of the sensitivities to one and two risk factors for the $\VaR_\alpha$ with level $\alpha = 0.9$ are reported in Table \ref{tab:ex-VaR-2D}. 
We observe that the sensitivities $\xi_{S_{\VaR_{0.9}}}(Y; X_1, X_i) $ with $i \in \{2, 3,4\}$ are close to the sensitivity to $X_1$, thus knowing $X_1$ and then learning an additional risk factor does not considerably increase the sensitivity. This is in contrast to the sensitivity to the pair $(X_2, X_4)$ which is equal to 0.740, providing a sensitivity that is substantially larger than the sum of the sensitivities to $X_2$ and to $X_4$. 

\begin{table}[t]
\centering
\caption{Score-based sensitivities for the insurance model and the $\VaR_\alpha$ at level $\alpha = 0.9$ with respect to one and two risk factors. The 90\% confidence interval (assuming the learned NNs are correct) are $(\pm 0.001)$ for all estimates.}\label{tab:ex-VaR-2D}
\begin{tabular}{ c c c c c}
 $\xi_{S_{\VaR_{0.9}}}(Y; X_i, X_j) $ & $X_1$ & $X_2$ & $X_3$ & $X_4$\\[0.5em]
\toprule
\toprule 
$X_1$ &   $0.601 $ & $0.707 $ & $0.616 $ & $0.610 $
\\[0.5em]
$X_2$ &   & $0.123 $  & $0.131 $ & $ 0.740 $
\\[0.5em]
$X_3$ &   &  & $0.014 $  & $0.410 $
\\[0.5em]
$X_4$ &   &  &  & $0.410 $ 
\\[0.5em]
\bottomrule
\bottomrule
\end{tabular}
\end{table}

\begin{figure}[h]
    \centering
    \includegraphics[width = 0.45\textwidth]{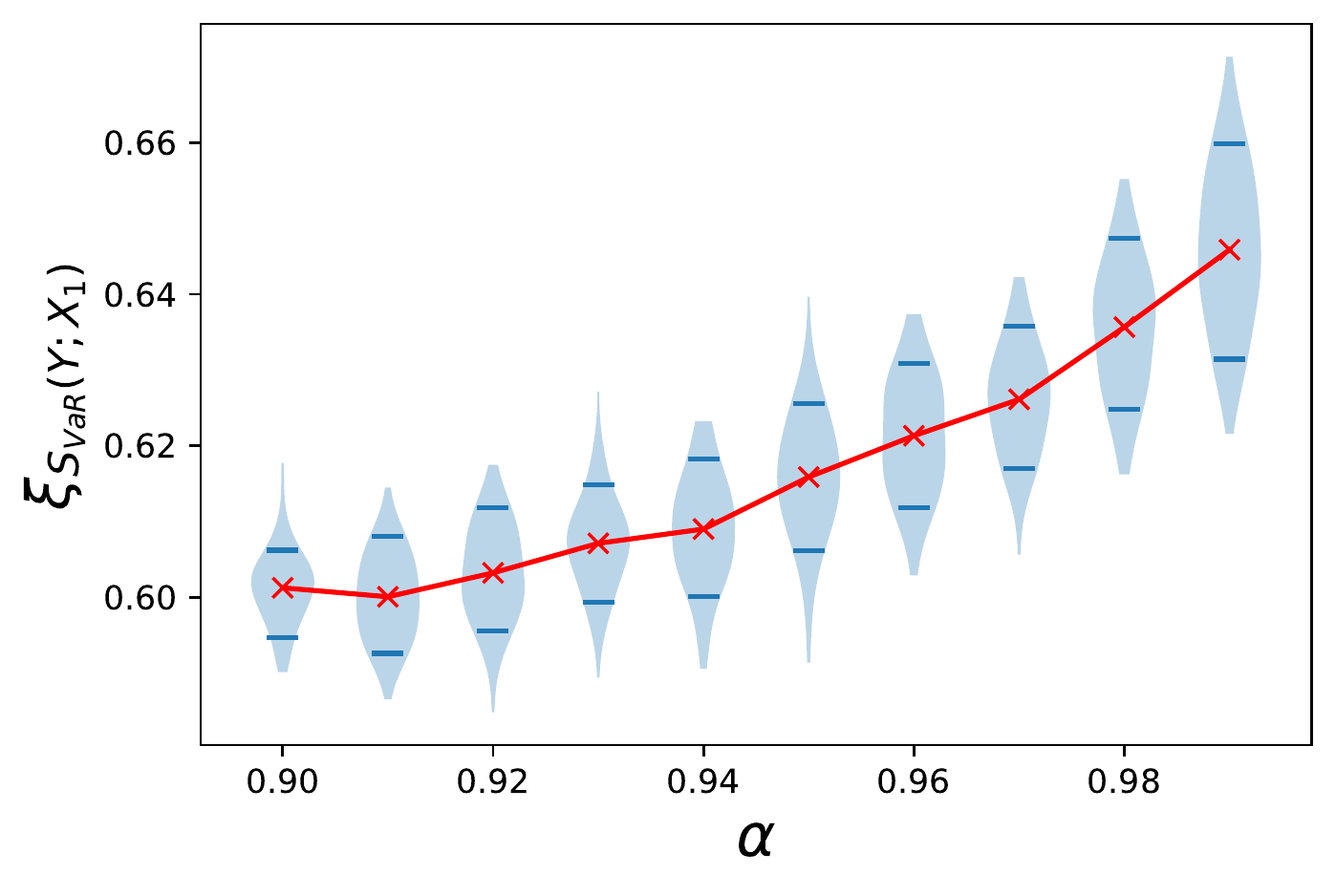}
    \includegraphics[width = 0.45\textwidth]{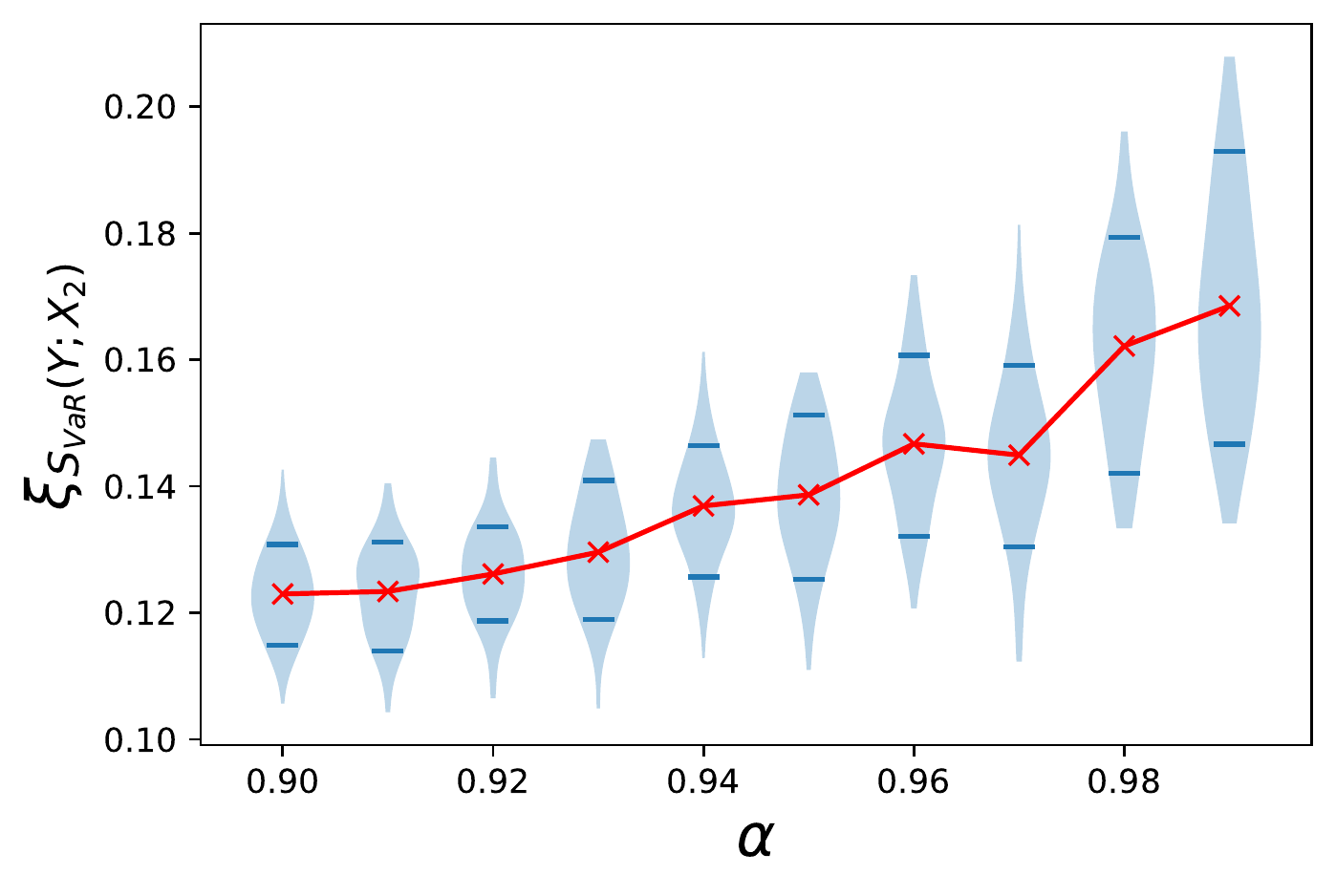}
    \\
    \includegraphics[width = 0.45\textwidth]{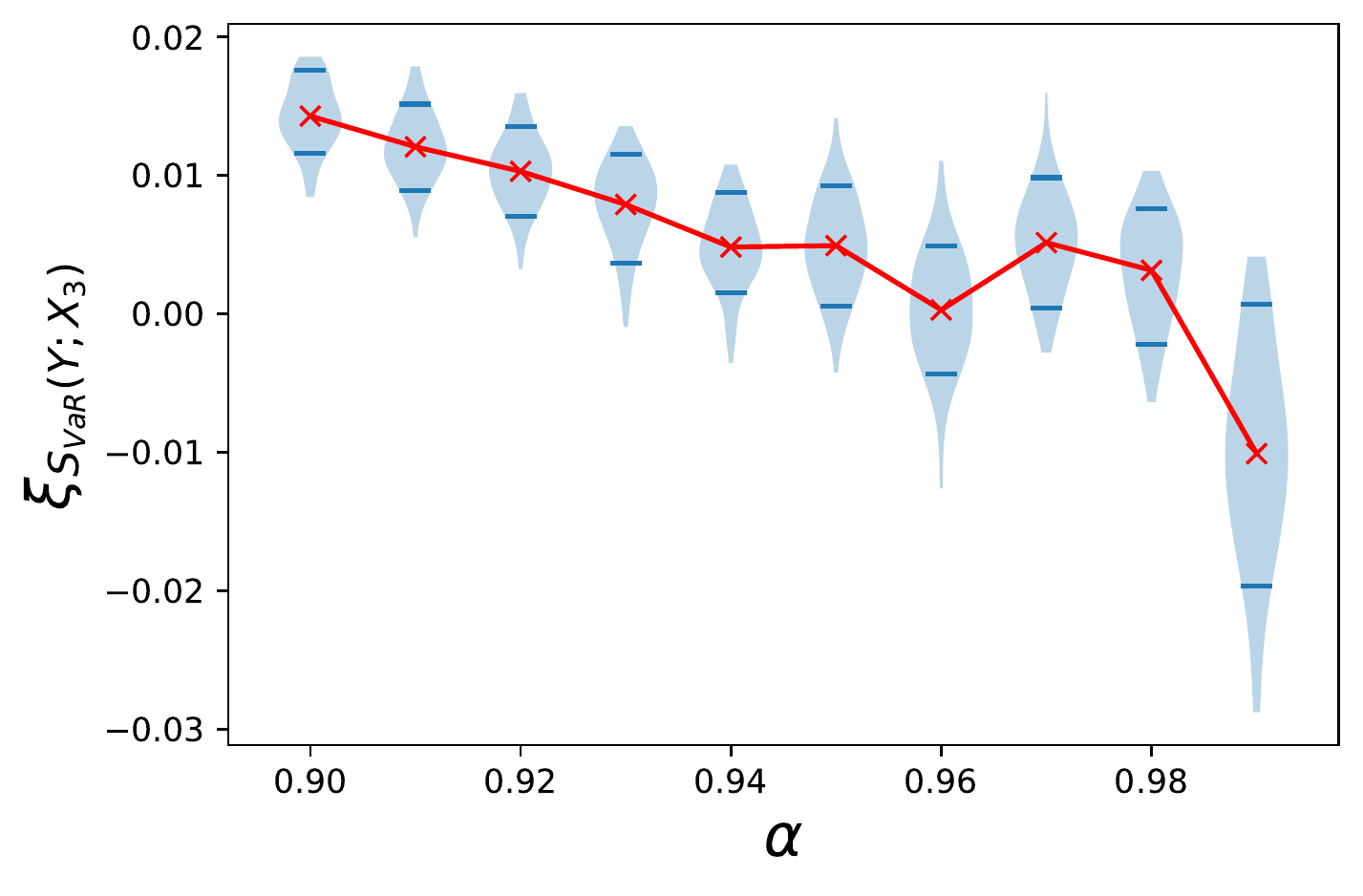}
    \includegraphics[width = 0.45\textwidth]{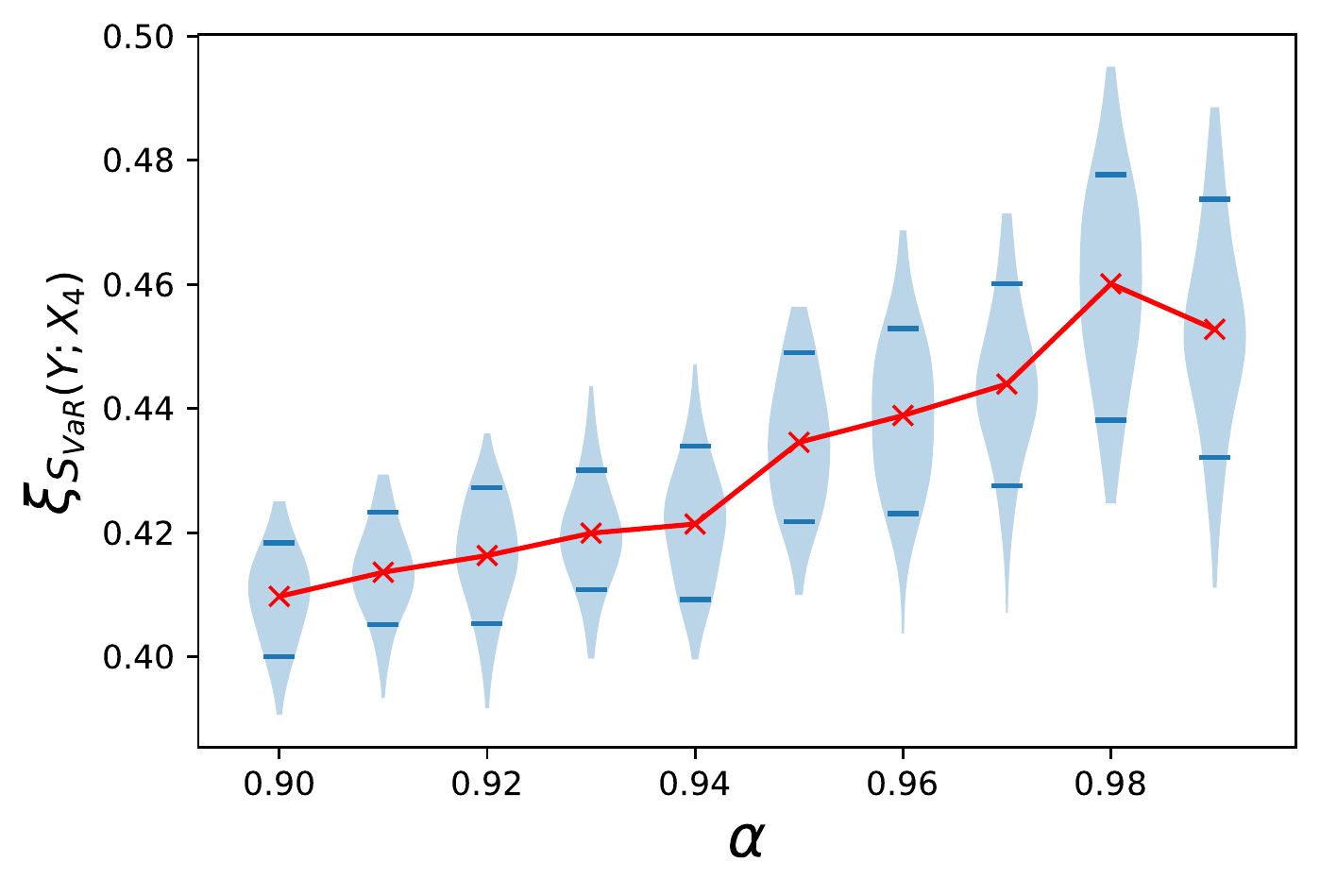}
    \caption{Score-based sensitivities for the $\VaR_{\alpha}$ with $\alpha$ ranging from 0.9 to 0.99. The violin plot displays the average sensitivity (red cross) and the 10\% and the 90\% quantile (blue lines), based on 100 estimates from the learnt NN. }
    \label{fig:es-Insurance-VaR-single-alpha}
\end{figure}

In Figure \ref{fig:es-Insurance-VaR-single-alpha} we compare the score-based sensitivities to single risk factors for different levels of $\VaR_\alpha$, that is from $\alpha  = 0.9$ to $\alpha = 0.99$. Specifically, we learn for each $\alpha$ and risk factor a NN using the procedure prescribed above. Figure \ref{fig:es-Insurance-VaR-single-alpha} displays violin plots of $\xi_{S_{\VaR}}(Y; X_i)$ for $i = 1, \ldots, 4$ and $\alpha = 0.9, \ldots, 0.99$ calculated using the learnt NNs. We observe that the sensitivities to $X_1, X_2$, and $X_4$ are increasing in $\alpha$, whereas the sensitivity to $X_3$ is decreasing. The bottom left panel, depicting the sensitivity to $X_3$, contains negative values for the sensitivities. These stem from statistical errors, as the sensitivities by definition should be non-negative. Note, however, that zero is contained in all 90\% confidence intervals (displayed via the blue lines).

Next, we calculate the sensitivity of the insurance portfolio for the pair $(\VaR_\alpha, \ES_\alpha)$. 
Recall that $\ES_\alpha$ is not elicitable on its own, but is jointly elicitable with $\VaR_\alpha$ \citep{FisslerZiegel2016}. We consider the 0-homogeneous strictly consistent scoring function given by
\begin{align} \label{ex:insurance-S-ES}
S_{\VaR, \ES}(z_1,z_2,y)
= \frac{1}{z_2\, (1 - \alpha)} \; \left(z_1 \left(\Id_{\{y \le z_1\}} -\alpha\right)+ y \Id_{\{y > z_1\}}\right) -1 + \log\left(\frac{z_2}{y}\right)\,,
\quad z_1,z_2,y>0\,,
\end{align}
which arises from the general family of scores in \eqref{eq:S ES} with $g(x) = 0$ and $\phi(x) = - \log(x)$.
(Note that all risk factors and also $Y$ are positive almost surely.)
We proceed similarly to estimating the score-based sensitivities for $\VaR_\alpha$, in that we use NNs to estimate the conditional functionals. Let $\X_{\I}$ be a sub-vector of the risk factors $(X_1, \ldots, X_4)$ with dimension 1 or 2, and denote by 
$q_{\I}(\x_{\I}) = \VaR_\alpha(Y|\X_{\I} = \x_{\I}) $ and $e_{\I}(\x_{\I}) = \ES_\alpha(Y|\X_{\I} = \x_{\I})$ the conditional VaR and ES, viewed, for fixed $\alpha$, as a function of $\x_{\I} \in \R^{|\I|}$. Then, we jointly estimate $q_\I$ and $e_\I$ via two neural nets $G_{\vartheta_{\I}}$ and $H_{\varsigma_{\I}}$, respectively. 
The network parameters are learnt by iteratively minimising the score \eqref{ex:insurance-S-ES}
\begin{equation}\label{eq:insurance-nn-param-ES}
\left(\hat{\vartheta}_{\I},\; \hat{\varsigma}_{\I}\right)
=
\argmin_{\vartheta,\; \varsigma} \frac{1}{N} \sum_{k  =1}^N  S_{\VaR, \ES} \left(G_{ \vartheta}(x_{\I}^{(k)}), \; H_{ \varsigma}(x_{\I}^{(k)}),\; y^{(k)}\right)\,,
\end{equation}
over independently simulated mini-batches $(\x_\I^{(k)}, y^{(k)})$, $k = 1, \ldots, N$, of $(\X_\I, Y)$. We denote the learnt NNs by $G_{\widehat{\vartheta}_{\I}}$ and $H_{\widehat \varsigma_{\I}}$, respectively, and estimate the sensitivity to $\X_{\I}$ using the learnt NNs out-of-sample, that is using an independent sample $(\x_\I^{(l)},y^{(l)})$, $l = 1 \ldots, M$, of $(\X_\I, Y)$, by
\begin{equation}\label{eq:insurance-sensitivity-estimate-ES}
\widehat{\xi}_{S_{\VaR, \ES}}(Y;\X_{\I}) = 1 - 
\frac{\displaystyle\sum_{l = 1}^M S_{\VaR, \ES}\left(G_{\widehat \vartheta_{\I}}(\x_{\I}^{(l)}), H_{\widehat \varsigma_{\I}}(\x_{\I}^{(l)}), y^{(l)}\right)}{
\displaystyle\sum_{l = 1}^M S_{\VaR, \ES}\left(\widehat{\VaR}_\alpha(Y), \widehat{\ES}_\alpha(Y), y^{(l)}\right)}\,,
\end{equation}
where $\widehat{\VaR}_\alpha(Y)$ and $\widehat{\ES}_\alpha(Y)$ are the sample quantile and sample ES of $Y$, respectively.

Since for all $\x_{\I}$ it holds that $\ES_\alpha(Y|\X_{\I} = \x_{\I})\ge \VaR_\alpha(Y|\X_{\I} = \x_{\I})$, we require that $ H_{ \varsigma_{\I}}(\x_{\I}) \ge G_{\vartheta_{\I}} (\x_{\I})$, for all $\x_{\I}$. Thus, we define the NN for estimating the conditional ES by $H_{ \varsigma_{\I}}(\x_{\I}) = G_{ \vartheta_{\I}}(\x_{\I}) + \mathcal{H}_{ \eta_{\I}}(\x_{\I})$, where $\mathcal{H}_{ \eta_{\I}}(\x_{\I})$ is constrained to be non-negative for all $\x_{\I}$; modelled using a softplus activation function on the last node. For each choice of $\I$, we choose a NN structure consists of 6 hidden layers with 20 neurons per layer. The activation functions are SiLU for each layer apart from the last layer which uses a softplus activation function. 

The estimates of the score-based sensitivities for $(\VaR_{0.9}, \,\ES_{0.9})$ are reported in Table \ref{tab:ex-VaR-ES-2D}. We observe that the sensitivities are similar in magnitude to the sensitivities for the $\VaR_{0.9}$, see also Table \ref{tab:ex-VaR-2D}. Moreover, the ranking of the risk factors stays the same. 

\begin{table}[t]
\centering
\caption{Score-based sensitivities for  $\VaR_\alpha$ and $\ES_\alpha$ at level $\alpha = 0.9$ with respect to one and two risk factors. The 90\% confidence interval (assuming the learned NNs are correct) are $(\pm 0.001)$ for all estimates.}\label{tab:ex-VaR-ES-2D}
\begin{tabular}{ c c c c c}
 $\xi_{S_{\VaR_{0.9}, \, \ES_{0.9}}}(Y; X_i, X_j) $ & $X_1$ & $X_2$ & $X_3$ & $X_4$\\[0.5em]
\toprule
\toprule 
$X_1$ &   $0.565 $ & $0.689 $ & $0.589 $ & $0.586 $
\\[0.5em]
$X_2$ &   & $0.121 $  & $0.122 $ & $ 0.718 $
\\[0.5em]
$X_3$ &   &  & $0.002 $  & $0.393 $
\\[0.5em]
$X_4$ &   &  &  & $0.384 $ 
\\[0.5em]
\bottomrule
\bottomrule
\end{tabular}
\end{table}

\section{Conclusion}

This paper provides a comprehensive framework for constructing sensitivity measures induced by strictly consistent scoring functions for any elicitable target functional $T$. 
These score-based sensitivities naturally quantify the \emph{relative} information gain -- when using available information ideally -- to model the target functional.
Theorem \ref{thm:properties} establishes intuitive and desirable properties for these score-based sensitivities such as zero information gain and full information gain. 
Following \cite{GriessenbergerETAL2022}, these properties suggest that the sensitivities can also be considered as a dependence measure between an output $Y$ and an information set $\ssalg$.
We further define a sensitivity called interaction information, which quantifies the information gain when learning the interaction of risk factors. We show that sensitivities based on a positively homogeneous score are scale-invariant, making them attractive in applications. 
Using Murphy diagrams for score-based sensitivities, we illustrate how to inspect entire classes of sensitivities, thus providing a holistic impression and revealing yet unknown model characteristics.

We emphasise that our approach is general and works for sensitivities which focus on any elicitable functional. 
In particular, we discuss the entire family of score-based sensitivity measures for the mean functional (of which the Sobol indices are one special case) and construct sensitivities for the pair consisting of VaR and ES; both, to the authors' best knowledge, novel endeavours to the literature.

\section*{Acknowledgements}
The authors thank Johanna Ziegel and Timo Dimitriadis for stimulating discussions about the topic
as well as three anonymous referees whose comments helped to substantially improve the paper.
SP would like to acknowledge support from the Natural Sciences and Engineering Research Council of Canada (grants DGECR-2020-00333 and RGPIN-2020-04289).

\newpage
\begin{APPENDICES}
\section{Scoring Functions of Risk Functionals}
\label{app:scores}

This appendix provides an overview of known characterisation results of (strictly) consistent scoring functions for popular functionals.
For precise technical conditions and statements see original references. 

\subsection{Entropic Risk Measures}

Entropic risk measures, also called exponential premium principle in insurance, \citep{Gerber1974ASTIN} are defined for a random variable $Y$ and parameter $\gamma>0$ by
\begin{equation*}
    \rho^\gamma(Y) = \frac{1}{\gamma} \log\, \E\left[\exp(\gamma Y)\right]\,.
\end{equation*}
Invoking the revelation principle \citep{Osband1985, Gneiting2011}, we can exploit the characterisation result \eqref{eq:Bregman} and obtain that, subject to mild regularity conditions, any (strictly) consistent score satisfying $S(y,y)=0$ is of the form
\begin{equation*}
    S(z,y) = \phi\big(\exp(\gamma y)\big) - \phi\big(\exp(\gamma z)\big) + \phi'\big(\exp(\gamma z)\big)(\exp(\gamma z)-\exp(\gamma y)) ,\qquad z,y\in\R,
\end{equation*}
where $\phi\colon\R\to\R$ is (strictly) convex with subgradient $\phi'$  such that $\int|\phi\big(\exp(\gamma y)\big) |\,\mathrm{d}F(y)<\infty$ for all $F\in\M$.

\subsection{Expectiles}

For $\tau\in(0,1)$, the $\tau$-expectile of a probability distribution $F$ is the unique solution of the equation 
\[
\tau \int_z^\infty (y-z)\,\mathrm{d}F(Y)
=(1-\tau) \int_{-\infty}^z (z-y)\,\mathrm{d}F(Y).
\]
\cite{Gneiting2011} characterises the class of (strictly) $\M$-consistent scoring functions for the $\tau$-expectile satisfying $S(y,y)=0$ as
\begin{equation*}
    \label{eq:S expectile}
    S(z,y) = |\Id_{\{y \le x\}} - \tau|\big(\phi(y) - \phi(z) + \phi'(z)(z-y)\big),\qquad z,y\in\R,
\end{equation*}
where $\phi\colon\R\to\R$ is (strictly) convex with subgradient $\phi'$ such that $\int | \phi(y)|\mathrm{d}F(y)<\infty$ for all $F\in \M$.

\subsection{Mode}

The mode is generally not elicitable on classes of absolutely continuous distributions \citep{Heinrich2014, HeinrichFissler2021}. 
However, on distributions with countably many outcomes, the mode is elicitable. \cite{Gneiting2017} showed that the zero-one loss $S(z,y)= \Id_{\{z\neq y\}}$ is basically the only strictly consistent score such that $S(y,y)=0$.
This yields the score-based sensitivity
\[
\sensA = \frac{(1-p_{\max}) - (1-\E[p_{\max|\ssalg}])}{1-p_{\max}},
\]
where $p_{\max}$ is the maximum of the counting density of $Y$ and $p_{\max|\ssalg}$ is the maximum of the conditional counting density of $Y$ given $\ssalg$.

\subsection{Mean and Variance}

We can again invoke the revelation principle to characterise the class of (strictly) $\M$-consistent scoring functions for the pair (mean, variance) satisfying Assumption \ref{asm:dirac-score} (i), such that $S(y,0,y)=0$.
Following \cite{FisslerZiegel2019}
they take the form
\begin{equation*}
    \label{eq:S mean var}
    S(z_1,z_2,y)
    = \phi(y,y^2) - \phi(z_1,z_2 + z_1^2) + \nabla \phi(z_1,z_2 + z_1^2) \begin{pmatrix}
    z_1 - y\\
    z_2 + z_1^2 - y^2
    \end{pmatrix}, \qquad z_1,y\in\R, \ z_2\ge0,
\end{equation*}
where $\phi\colon \R\times[0, \infty)\to \R$ is (strictly) convex with subgradient $\nabla \phi$, interpreted as a row vector, such that $\int|\phi(y,y^2)  + a(y)|\,\mathrm{d}F(y)<\infty$ for all $F\in\M$.

\subsection{VaR and ES}\label{}
ES generally fails to be elicitable \citep{Osband1985, Gneiting2011, Weber2006}. 
However, the pair $\VaR_\alpha$ together with $\ES_\alpha$, at the same probability level $\alpha\in(0,1)$, is elicitable, subject to mild conditions \citep{AcerbiSzekely2014, FisslerZiegel2016}. 
Note that different sign conventions are used in the literature. 
Upon using the one introduced in \ref{eq:def ES}, the (strictly) $\M$-consistent scoring functions for $(\VaR_\alpha, \ES_\alpha)$ satisfying Assumption \ref{asm:dirac-score} (i) such that $S(y,y,y)=0$ are of the form
\begin{align}\label{eq:S ES}
S(z_1,z_2,y)
&=  \big(\Id_{\{y\le z_1\}} - \alpha\big)\big(g(z_1) -g(y)\big)  \\ \nonumber
&+ \phi'(z_2)\Big(z_2 - \tfrac{1}{1 - \alpha} S^+_\alpha(z_1,y)\Big)- \phi(z_2) + \phi(y),
\quad z_1,z_2,y\in\R,
\end{align}
where $S^+_\alpha(z_1,y) = (1 - \alpha - \Id_{\{y > z_1\}})z_1 + \Id_{\{y> z_1\}}y = (\Id_{\{y\le z_1\}} - \alpha)z_1 - \Id_{\{y\le z_1\}}y + y$,
$\phi\colon\R\to\R$ is (strictly) convex with subgradient $\phi'$ and $g\colon\R\to\R$ is such that for all $z_2\in\R$
\begin{align*}
    &z_1\mapsto g(z_1) - z_1\phi'(z_2)/(1 - \alpha)
\end{align*}
is (strictly) increasing.
One convenient way to achieve this latter condition is to choose $\phi$ such that $\phi'<0$ and $g$ to be increasing.

To achieve strict consistency, the requirements on $\M$ are that the expected scores are finite at their minimum (amounting to the fact that $\int |\kappa(y)|\,\mathrm{d}F(y)<\infty$ for all $F\in\M$ and for $\kappa$ being $g$, $\phi$, or the identity), and that for all $F\in\M$, $F(\VaR_\alpha(F)+\epsilon)>\alpha$ for all $\epsilon>0$, see Proposition \ref{prop:sens mean quantile} (ii) and \cite{FisslerZiegel2016} for details.

\subsection{RVaR together with two VaR components}

The Range Value-at-Risk (RVaR) at levels $0< \alpha < \beta < 1$ for a random variable has been introduced by \cite{ContDeguestETAL2010} and is defined as
\begin{align*}
    \RVaR_{\alpha, \beta}(Y) = \frac{1}{\beta-\alpha} \int_\alpha^\beta \VaR_u(Y)\, \mathrm{d}u\,.
\end{align*}
RVaR fails to be elicitable on its own in general \citep{WangWei2020}.
However, \cite{FisslerZiegel2019_RVaR} show that the triplet $(\VaR_\alpha, \VaR_\beta, \RVaR_{\alpha, \beta})$ is elicitable subject to mild conditions. The (strictly) $\M$-consistent scoring functions satisfying Assumption \ref{asm:dirac-score} (i) such that $S(y,y,y,y)=0$ are of the form
\begin{align*}\label{eq:S RVaR}
S(z_1,z_2,z_3,y)
&=  \big(\Id_{\{y\le z_1\}} - \alpha\big)\big(g_1(z_1) -g_1(y)\big)  +\big(\Id_{\{y\le z_2\}} - \beta\big)\big(g_2(z_2) -g_2(y)\big)\\ \nonumber
&+ \phi'(z_3)\Big(z_3 + \frac{1}{\beta - \alpha} \big(S_\beta(z_2,y) - S_\alpha(z_1,y) \big)\Big)- \phi(z_3) + \phi(y),
\quad z_1,z_2,z_3,y\in\R\,,
\end{align*}
where $S_\gamma(z,y) = (\Id_{\{y\le z\}} - \gamma)z - \Id_{\{y\le z\}}y$ for $\gamma\in\{\alpha,\beta\}$,
$\phi\colon\R\to\R$ is (strictly) convex with subgradient $\phi'$ and $g_1,g_2\colon\R\to\R$ are such that for all $z_3\in\R$ the functions
\begin{align*}
    &z_1\mapsto g_1(z_1) - z_1\phi'(z_3)/(\beta - \alpha) \quad \text{and}\\
    &z_2\mapsto g_2(z_2) + z_2\phi'(z_3)/(\beta - \alpha)
\end{align*}
are (strictly) increasing.
We refer to \cite{FisslerZiegel2019_RVaR} for examples of such scores as well as to the detailed conditions on $\M$.

\end{APPENDICES}

\bibliographystyle{informs2014}
\bibliography{references.bib} 

\end{document}